\documentclass{aastex631}
\usepackage{amsmath}
\usepackage{amssymb}
\usepackage{epigraph}
\usepackage{bm}

\shorttitle{Binary Black Hole Mergers in AGN Accretion Disks with ZTF}
\shortauthors{Graham et al.}
\graphicspath{{./}{figures/}}

\begin{document}

\title{A light in the dark: searching for electromagnetic counterparts to black hole-black hole mergers\\ in LIGO/Virgo O3 with the Zwicky Transient Facility}

\correspondingauthor{Matthew Graham}
\email{mjg@caltech.edu}

\author[0000-0002-3168-0139]{Matthew J. Graham}
\affiliation{California Institute of Technology, 1200 E. California Blvd, Pasadena, CA 91125, USA}

\author[0000-0002-9726-0508]{Barry McKernan}
\affiliation{Department of Science, CUNY Borough of Manhattan Community College, 199 Chambers Street, New York, NY 10007, USA}
\affiliation{Department of Astrophysics, American Museum of Natural History, Central Park West, New York, NY 10028, USA}
\affiliation{Physics Program, CUNY Graduate Center, 365 5th Avenue, New York, NY 10016, USA}
\affiliation{Center for Computational Astrophysics, Flatiron Institute, New York, NY 10010, USA}

\author[0000-0002-5956-851X]{K.~E.~Saavik Ford}
\affiliation{Department of Science, CUNY Borough of Manhattan Community College, 199 Chambers Street, New York, NY 10007, USA}
\affiliation{Department of Astrophysics, American Museum of Natural History, Central Park West, New York, NY 10028, USA}
\affiliation{Physics Program, CUNY Graduate Center, 365 5th Avenue, New York, NY 10016, USA}
\affiliation{Center for Computational Astrophysics, Flatiron Institute, New York, NY 10010, USA}

\author[0000-0003-2686-9241]{Daniel Stern}
\affiliation{Jet Propulsion Laboratory, California Institute of Technology, 4800 Oak Grove Drive, Pasadena,
CA 91109, USA}

\author[0000-0002-0603-3087]{S.~G.~Djorgovski}
\affiliation{California Institute of Technology, 1200 E. California Blvd, Pasadena, CA 91125, USA}

\author[0000-0002-8262-2924]{Michael Coughlin}
\affiliation{School of Physics and Astronomy, University of Minnesota, Minneapolis, Minnesota 55455, USA}

\author[0000-0002-7226-836X]{Kevin B. Burdge}
\affiliation{Department of Physics, Massachusetts Institute of Technology, Cambridge, MA 02139, USA}
\affiliation{Kavli Institute for Astrophysics and Space Research, Massachusetts Institute of Technology, Cambridge, MA 02139, USA}

\author[0000-0001-8018-5348]{Eric C. Bellm}
\affiliation{DIRAC Institute, Department of Astronomy, University of Washington, 3910 15th Avenue NE, Seattle, WA 98195, USA}

\author{George Helou}
\affiliation{IPAC, California Institute of Technology, 1200 E. California Blvd, Pasadena, CA 91125, USA}

\author[0000-0003-2242-0244]{Ashish~A.~Mahabal}
\affiliation{Division of Physics, Mathematics and Astronomy, California Institute of Technology, Pasadena, CA 91125, USA}
\affiliation{Center for Data Driven Discovery, California Institute of Technology, Pasadena, CA 91125, USA}

\author[0000-0002-8532-9395]{Frank J. Masci}
\affiliation{IPAC, California Institute of Technology, 1200 E. California Blvd, Pasadena, CA 91125, USA}

\author{Josiah Purdum}
\affiliation{Caltech Optical Observatories, California Institute of Technology, Pasadena, CA 91125, USA}

\author{Philippe Rosnet}
\affiliation{Universit\'{e} Clermont Auvergne, CNRS/IN2P3, LPC, F-63000 Clermont-Ferrand, France}

\author[0000-0001-7648-4142]{Ben Rusholme}
\affiliation{IPAC, California Institute of Technology, 1200 E. California Blvd, Pasadena, CA 91125, USA}



\begin{abstract}

The accretion disks of active galactic nuclei (AGN) are promising locations for the merger of compact objects detected by gravitational wave (GW) observatories. Embedded within a baryon-rich, high density environment, mergers within AGN are the only GW channel where an electromagnetic (EM) counterpart must occur (whether detectable or not).  Considering AGN with unusual flaring activity observed by the Zwicky Transient Facility (ZTF), we describe a search for candidate EM counterparts to binary black hole (BBH) mergers detected by LIGO/Virgo in O3. After removing probable false positives, we find nine candidate counterparts to BBH mergers during O3 (seven in O3a, two in O3b) with a $p$-value of 0.0019. Based on ZTF sky coverage, AGN geometry, and merger geometry, we expect $\approx 3(N_{\rm BBH}/83)(f_{\rm AGN}/0.5)$ potentially detectable EM counterparts from O3, where $N_{\rm BBH}$ is the total number of observed BBH mergers and $f_{\rm AGN}$ is the fraction originating in AGN. Further modeling of breakout and flaring phenomena in AGN disks is required to reduce our false positive rate. Two of the events are also associated with mergers with total masses $> 100M_\odot$, which is the expected rate for O3 if hierarchical (large mass) mergers occur in the AGN channel. Candidate EM counterparts in future GW observing runs can be better constrained by coverage of the Southern sky as well as spectral monitoring of unusual AGN flaring events in LIGO/Virgo alert volumes. A future set of reliable AGN EM counterparts to BBH mergers will yield an independent means of measuring cosmic expansion ($H_0$) as a function of redshift.

\end{abstract}





\section{Introduction} \label{sec:intro}

\epigraph{`Great black holes have little black holes in their disks a-mergin', \\
And with ZTF we can see the light that comes a-splurgin''}{Anonymous (with apologies to A. de Morgan)}


The gravitational wave (GW) detectors Advanced LIGO \citep{2015CQGra..32g4001L} and Advanced Virgo \citep{2015CQGra..32b4001A} (hereafter referred to as LIGO/Virgo) detected binary black hole (BBH) mergers in the local ($z<1$) Universe at a rate of about once per week during O3 \citep[O3a:  2019 March 1 - 2019 September 30; O3b:  2019 November 1 - 2020 March 30;][]{Abbott19}. BBH mergers can come from two broad classes of channels \citep[for a recent review, see][]{Mapelli21}: field binary origin \citep[i.e., from the evolution of a field binary system consisting of two massive stars; e.g.,][]{Belczynski10, deMink16} and dynamical origin. Among dynamical BBH mergers, sub-channels include mergers in globular clusters \citep[e.g.,][]{Rodriquez16a, Rodriquez16b}, mergers in quiescent galactic nuclei \citep[e.g.,][]{Antonini14, AntoniniRasio16, Fragione19}, and mergers in the accretion disks of active galaxies \citep[e.g.,][]{McK19a, Graham20}. Due to pair instability supernovae which leave no compact remnant, the explosive deaths of massive stars are not thought to produce black holes (BHs) in the ``upper mass gap'' range of $\sim 50-130M_{\odot}$ \citep{Woosley17}.  There is also a ``lower mass gap'' range of $\sim 3-5M_{\odot}$, corresponding to the observed absence of compact objects with masses between the most massive neutron stars (NSs) and the least massive BHs. Massive BBH merger progenitors detected by LIGO/Virgo in the upper end of the upper mass gap range strongly imply a hierarchical (i.e., dynamical) merger origin. Since BHs can receive a strong kick at merger \citep[e.g.,][]{Varma22}, hierarchical mergers are more easily retained in deep gravitational potentials, such as in the nuclei of galaxies \citep[e.g.,][]{Gerosa19, GerosaFishbach21}. 


A promising location for hierarchical mergers are active galactic nuclei (AGN) \citep[e.g.,][]{McK12, McK14, McK18, Bellovary16, Bartos17, Stone17, Secunda19, Secunda20, Yang19, Tagawa19, Tagawa21, Samsing22}; \citet{Graham20} presented the first candidate counterpart for such an event. Merger kicks, even of large magnitude \citep{Varma22}, are insufficient to escape an AGN environment where the Keplerian orbital velocity is $\mathcal{O}(10^{4}){\rm km\, s^{-1}}(R/10^{3}r_{g})$ at disk radius $R$, where $r_{g}=GM_{\rm SMBH}/c^{2}$ is the supermassive black hole (SMBH) gravitational radius and $M_{\rm SMBH}$ is the SMBH mass.  This makes AGN ideal for retaining and growing BHs via hierarchical mergers. AGN are expected to dominate the rate of mergers in the deep potential wells of gas-free galactic nuclei \citep{FMcK22}. Besides BBH mergers at the upper end of the mass gap, other pointers to a significant contribution to observed BH mergers from the AGN channel include significantly asymmetric mass ratio BH mergers and the observed anti-correlation between BH mass ratio and BBH effective spin \citep{Callister21}, which at present can only be explained in the context of the AGN channel \citep{qX21,Yihan21}. 

Unlike all other BH merger channels, significant detectable EM counterparts may develop due to compact object mergers in AGN \citep[e.g.,][]{McK19a, Graham20, Perna21, WangJ21, Kimura21}. Furthermore, identified counterparts to BH mergers in AGN accretion disks provide a test of the dynamics of the merger, as well as a probe of fundamental AGN disk properties \citep{Vajpeyi22}. In addition, if we can confidently associate particular GW mergers with specific AGN at identified redshifts \citep[e.g.,][]{Juan21, Palmese21, Ashton21}, the GW signal becomes a standard siren that provides a new, independent measurement of the Hubble constant $H_0$ as a function of redshift \citep{Chen22, Mukherjee20}. 

If we are optimistic about identifying EM counterparts to BH mergers in AGN disks, this approach promises to produce the merger locations of GW sources, an important new probe of AGN accretion disks, and a powerful technique to measure the expansion history of our Universe over a critical redshift range. However, one might also be pessimistic about identifying EM counterparts in AGN disks. Even if AGN are responsible for most of the GW-detected BH mergers, we might not detect EM counterparts due to either the muffling of embedded EM signatures by optically thick disks, or the emerging EM flare might be too faint to detect against a bright, variable quasar disk. Even in the pessimistic case, however, a search for unusual AGN flares is valuable as a test of extreme variability mechanisms in AGN disks.  We also note that even if EM counterparts are never confidently detected from AGN, a significant AGN contribution to the merger rate can still be estimated from a statistical approach \citep[see, e.g.,][]{Imre17, Veronesi22}.

The EM signature of a merged BBH in an AGN disk depends on the reaction of the surrounding disk gas to the merger. In general, a merged BBH in an AGN disk moves away from the merger site with recoil kick velocity $v_k$, which is a function of binary mass asymmetry and spin orientation. Gas that was gravitationally bound to the BBH will attempt to follow the merged product, but collides  with surrounding AGN disk gas, heating it and producing a bright shock in the disk, possibly detectable in the optical/UV waveband in a thin or relatively dim disk \citep{McK19a}. An even brighter EM signature may result from the continued onward progression of the recoiling BH through the disk as surrounding gas produces a Bondi drag `tail' behind the kicked BH. This can generate a significant, detectable luminosity at super-Eddington accretion rates \citep{Graham20}, as long as a jetted outflow allows radiation to escape and emerge \citep{McK19a, WangJ21}. EM counterparts can emerge on the side of the AGN disk facing towards, or away from, the observer's sightline. 



The search for EM counterparts to isolated NS-NS or BH-NS mergers typically requires rapid followup with coverage of as much of the LIGO/Virgo localization map as possible \citep{coughlin19, anand21, kasliwal20}. In contrast, the EM signal for BBH mergers in AGN disks only emerges days to weeks after the merger event \citep{McK19a}. The detection strategy is therefore different from the non-AGN case and requires regular monitoring of the AGN population within the LIGO/Virgo error volume rather than rapid scanning. Modern time domain surveys which observe large areas of sky with day-to-week cadences therefore present an ideal data set for identifying possible counterpart events. However, this also illustrates the importance of correctly updating public localization maps: if the parameterization of a GW event changes over time, the two-dimensional and three-dimensional event localization can change significantly. 


In this paper, we present a search with the Zwicky Transient Facility \citep[ZTF;][]{Bellm19,Graham19} for EM counterparts in AGN disks to all BBH merger detections by LIGO/Virgo during the O3 run. Over this period, ZTF covered the visible sky above Dec = -30$^{\circ}$ from Palomar Observatory every three nights in $g$- and $r$-bands to $\sim$20.5 magnitude ($5 \sigma$ detection limit). This provides a data set with the required large spatial coverage and sampling cadence to detect an association between a BBH merger and an AGN flare. The paper is structured as follows: in \S~2, we consider how a BBH merger in an AGN disk could generate an EM signal; in \S~3, we consider other events that could produce an equivalent signal; \S~4 describes our search procedure; and \S~5 presents our results. We discuss the implications of our results in \S~6 and detail our conclusions in \S~7.  Throughout, magnitudes are reported in the AB system and we adopt the Planck 2015 cosmology, $H_0 = 67.7\, {\rm km}\, {\rm s}^{-1}\, {\rm Mpc}^{-1}$, $\Lambda = 0.693$, and $\Omega_m = 0.307$ \citep{Planck15}.  

\section{EM counterparts to BBH mergers in AGN disks}
BBH mergers in AGN disks occur in the presence of gas and must always produce some EM radiation, though the detectability of the resulting EM signature depends on three basic factors: \\

\noindent
{\em 1. Can the EM counterpart escape from the midplane of a dense, optically thick disk on sufficient timescales?}

\noindent
If a BBH merger occurs in (and remains in) the midplane of a disk that is both optically and geometrically thick, the diffusion timescale for radiation produced at the midplane is many years and the signature is not detectable \citep{McK19a}. If the merger happens in the midplane of a razor-thin disk, an EM signature could emerge promptly. However, razor-thin disks are disfavoured since models of pressure-supported disks tend to generate modest disk aspect ratios \citep[e.g.,][]{SirkoGoodman03,Thompson05}. A BH remnant that is strongly kicked at merger could quickly emerge from an optically thick midplane into a diffuse, optically thin, disk atmosphere. There is, however, also a geometrical problem: an EM counterpart to a merger in an edge-on AGN or a kick that pushes the remnant to the far side of the disk with respect to the observer will be obscured.\\

\noindent
{\em 2. Is the change in brightness relative to the already bright AGN disk detectable?}

\noindent
The brighter the AGN, the less likely it is that we can identify a flare associated with the BBH merger. The brightness of the flare also depends on how the EM emission emerges from the merger. A shock is usually too dim to show up against bright AGN and will only be detectable against lower luminosity AGN \citep{McK19a, Graham20}. However, a jet from a kicked, rapidly spinning and accreting merger product may be sufficiently bright \citep{WangJ21}, or may be powerful enough to mechanically clear a low optical depth path out from the midplane, unless, e.g., pre-merger outflows have excavated a local bubble \citep{Kimura21} or the accretion rate is insufficient to generate high luminosity \citep{Pan21}.\\

\noindent
{\em 3. Can we distinguish a resulting flare from false positives?} 

\noindent
Even if the flare is bright enough to be observable, we still need to be able to distinguish the flare from other AGN variability events or known false positives. This requires models of intrinsic extreme AGN variability \citep{Graham17} as well as an understanding of light curve and color evolution from embedded disk eruptions such as supernovae (SNe) or tidal disruption events (TDEs) in the presence of an AGN accretion disk \citep{Chan19, Yang21}, as well as microlensing events \citep[e.g.,][]{Lawrence16}.\\

In this section, we address the first two of these points and review our model for a prompt EM counterpart to a BBH merger in an AGN disk \citep{McK19a}. This underpinned our reporting of the first plausible candidate EM counterpart to a GW BBH merger trigger \citep{Graham20}. In particular, we consider what the model implies for the properties of the associated flare, including the characteristic timescales, luminosity, and flare shape. We discuss the third point, namely the statistical uniqueness of the EM signature, in \S~3 and 4.

\subsection{Parameters from the initial GW trigger}

The initial notification of a candidate GW event from LIGO/Virgo reports the luminosity distance ($d_{L}$), the confidence interval for the luminosity distance ($\Delta d_L$), and the 90\% confidence interval for the sky localization area ($A_{90}$) of the event \citep{Singer16, singer16b, Veitch15}. The probability distribution for $d_{L}$ is a convolution of source mass, detector orientation, and source angle to the observer line-of-sight.  The localization area is proportional to the signal-to-noise ratio (SNR) of the event detection by LIGO/Virgo, $A_{90} \propto$ SNR$^{-2}$ \citep{Berry15}, where SNR $\propto M_c^{5/6}\, d_{L}^{-1}$ \citep{FinnChernoff93}, and $M_c$ is the chirp mass\footnote{Defined as $M_c \equiv \mu^{3/5}\, M_{\rm BBH}^{2/5}$, where the $M_{\rm BHH} = m_1 + m_2$ is the total binary mass and $\mu = m_1 m_2 / M_{\rm BBH}$.}. Thus, for any given GW event trigger at time $t=0$, we can estimate the approximate source frame BBH mass, $M_{\rm BBH}$, and we have a search volume for AGN EM counterparts in the volume given by $A_{90}\times \Delta d_{L}$.
 
\subsection{Parameters from the bound gas shock}
At merger, the new BH recoils with a kick velocity, $v_k$. In an AGN disk, gas at distance $R_{\rm bound} < GM_{\rm BBH}/v_k^{2}$ is bound to the merged BBH and attempts to follow the kicked merger product. In doing so, it collides with surrounding disk gas and a shock luminosity emerges on a timescale $t_{\rm bound}=R_{\rm bound}/v_{k}=GM_{\rm BBH}/v_{\rm k}^{3}$ \citep{McK19a}, which can be parameterized as
\begin{equation}
    t_{\rm bound} \sim 20\, {\rm day}\left(\frac{M_{\rm BBH}}{100\, M_{\odot}}\right)\left(\frac{v_{\rm k}}{200\, {\rm km}\, {\rm s}^{-1}}\right)^{-3}.
\end{equation} 

\noindent
This is a low luminosity effect ($\mathcal{O}(10^{42})$ erg s$^{-1}$) compared to other mechanisms discussed here and so we do not consider it any further. We note, however,  that if $v_k$ is very small ($<50$ km s$^{-1}$) then this low luminosity prompt flare is likely the only EM counterpart.

\subsection{Parameters from the Bondi drag accretion and shock}
Once the kicked BH leaves behind originally bound gas, the disk gas it passes through is accelerated around the BH, producing a shocked Bondi tail \citep[e.g.,][]{Ostriker99, Antoni19}. This tail both acts as a drag on the BH and accretes onto it. We assume the associated  Bondi-Hoyle-Lyttleton (BHL) luminosity is $L_{\rm BHL}=\eta \dot{M}_{\rm BHL}c^{2}$, where $\eta$ is the radiative efficiency and the mass accretion rate is
\begin{equation}
    \dot{M}_{\rm BHL}=\frac{4\pi G^{2}M_{\rm BBH}^{2} \rho}{v_{\rm rel}^{3}},
\end{equation}
where $\rho$ is the local disk gas density, $v_{\rm rel}=v_{k}+c_{s}$, and $c_{s}$ is the gas sound speed, assumed to be $c_{s} \sim 50\, {\rm km}\, {\rm s}^{-1}$ \citep{Graham20}. Then we can parameterize $L_{\rm BHL}$ as
\begin{eqnarray}
    L_{\rm BHL}&=&2.5 \times 10^{45}{\rm erg}\, {\rm s}^{-1} \left( \frac{\eta}{0.1}\right)\left( \frac{M_{\rm BBH}}{100\, M_{\odot}}\right)^{2} \left(\frac{v_{rel}}{200\, {\rm km}\, {\rm s}^{-1}} \right)^{-3} \left( \frac{\rho}{10^{-10}\, {\rm g}\, {\rm cm^{-3}}}\right).
    \label{eq:bhl}
\end{eqnarray}
Following \citet{Graham20}, the dynamical time in the source-frame associated with the ram pressure shock (or the time for the merger remnant to cross the sphere of bound gas) is $t_{\rm ram} = G\, M_{\rm BBH} / v_k^3 \sim 20~{\rm day}\, (M_{\rm BBH}/100~M_\odot)\, (v_k / 200\, {\rm km}\, {\rm s}^{-1})^{-3}$.  We assume that the luminosity of the flare rises linearly to $L_{\rm BHL}$ from $t=[t_{\rm ram},2t_{\rm ram}]$ as the disk gas rearranges itself around the kicked BH. At $t=2t_{\rm ram}$, the flare luminosity is assumed constant at $L_{\rm BHL}$. Note that this represents hyper-Eddington accretion as parameterized and it is an open question as to whether enough radiation could escape from a hyper-Eddington accretion rate BH to justify the choice of $\eta \sim 0.1$. Simulations of super-Eddington accretion that reach up to $1500 \times$ Eddington imply that $\eta \rightarrow 0.01$ in this context \citep{YanFei19}. In order for enough radiation to escape to produce for a bright flare against a quasar, jetted or collimated outflows are required. In this study we have no constraints on very high energy emission (X-rays) that we should expect from such outflows. We recommend that future work on simulations of hyper-Eddington accretion establish whether there is an upper limit to accretion which can choke off jets. This will help establish luminosity upper limits on any flares that emerge from kicked mergers in AGN disks.

Bondi drag slows down the kicked BH from an initial kinetic energy of $1/2\, M_{\rm BBH}\, v_k^{2}$. The drag force is $\dot{M}_{\rm BHL}v_k$ and is equal to $M_{\rm BBH}v_k/\tau_{\rm dec}$, where $\tau_{\rm dec}$ is the source frame deceleration timescale. $\tau_{\rm dec}$ is $\mathcal{O}(10^{2})$yrs for plausible BBH merger and disk parameters \citep{Graham20}, and kicks are likely not exactly aligned with the (relatively thin) AGN disk. Therefore, it is most likely that a modest inclination kick (i.e., $\theta \neq 0^{\circ}$, where $\theta=0^{\circ}$ is the disk mid-plane) lets the kicked BH exit the disk on a timescale $t_{\rm exit} \ll \tau_{\rm dec}$. We define $t_{\rm exit}$ as the time for the remnant to reach the $\tau=1$ optical depth surface from the merger point and we assume a Gaussian atmosphere with scale height $H$, i.e. $\rho=\rho_0 \exp(-z^2/2H^2)$, where $z$ is the height above the mid-plane, $\rho$ is the disk gas density at height $z$, and $\rho_0$ is the mid-plane density. The most rapid exit will be for a vertical kick directly out of the mid-plane, i.e. $\theta=90^{\circ}$; if the inclination angle is (as is likely) smaller, then the relevant velocity is just the vertical component of $v_k$, i.e. $v_k~ \rm{sin}(\theta)$. In the absence of strong constraints on $\theta$, we adopt the fastest exit time (see Appendix~\ref{app:derivation} for a full derivation):
\begin{eqnarray}
    t_{\rm exit}&=&\frac{H \sqrt{2 \ln(\tau_{mp})}}{v_k},
    \label{eqn:t_exit}
\end{eqnarray}
where $\tau_{\rm mp}$ is the merger point optical depth (we have used the fact that $\rho \propto \tau$). The peak flare luminosity will occur on a timescale $\approx t_{\rm exit}$ and we can use this calculation to constrain the range of kick velocities we are sensitive to based on the temporal search window.  Our ZTF search, described in \S~4, assumes a flare peak $<200~$days post-merger.

Assuming a gas-pressure supported disk (which should be true for $a > 10^2~r_g$, where $a$ is the BH orbit semi-major axis in units of $r_g$), we expect $H=c_s/\Omega$, where $\Omega=v_{\rm orb}/a$, and so
\begin{eqnarray}
    v_{k} &>&  \left(\frac{c_s a}{v_{\rm orb}~ t_{\rm exit}} \right) \sqrt{2 \ln \left(\tau_{mp} \right)}.
\label{eq:vklower}
\end{eqnarray}
In general, for remnants kicked at $\theta=90^{\circ}$, the minimum kick velocity could be as small as a few km/s; however, for more realistic parameters, we expect $v_{k}> \rm{O}(100~{\rm km/s})$.

On the other hand, larger $v_{k}$ will reduce $L_{\rm BHL}$, so we can also establish an upper limit on $v_{k}$ by approximating the total energy of the flare, $E_{\rm tot}$, as the luminosity times the flare duration, $t_{\rm flare}=t_{\rm exit}-t_{\rm st}$. Here $t_{\rm st}$ is the start time of the EM flare, and not the merger time. Rearranging, we find

\begin{eqnarray}
    v_k &<& t_{\rm flare}^{1/3} \left( \frac{\eta}{0.1}\right)^{1/3}\left( \frac{M_{\rm BBH}}{100\, M_{\odot}}\right)^{2/3} 
    \left(\frac{E_{\rm tot}}{2 \times 10^{45}\, {\rm erg}\, {\rm s}^{-1}} \right)^{-1/3} \left( \frac{\rho}{10^{-10}\, {\rm g}\, {\rm cm^{-3}}}\right)^{1/3}.
    \label{eq:vkupper}
\end{eqnarray}

Thus we can reject candidate counterparts in cases where these two limits are mutually exclusive.





\subsection{Flare form as a function of observer orientation}
For kicked BBH mergers in AGN disks, which are assumed to be face-on (i.e., within $\sim 45^{\circ}$ to the observer), we expect around half of all kicks to be directed away from the observer, out the far side of the disk, and thus not detectable.  A merger that occurs off the disk midplane and directed away from the observer should appear to become redder and diluted on an increasing diffusion timescale $t_{\rm diff}$ as the source is carried through a deeper scattering screen with an increasing optical depth.  Once a merger remnant crosses the midplane we consider it undetectable due the optical depth of the accretion disk. 



\section{False positives from AGN disks}
\label{falseposother}
AGN are intrinsically variable \citep[e.g.,][]{Matthews63}. Most of the optical signatures that ZTF will detect from AGN will therefore be false positives. Here we summarize properties of different kinds of known optical variability that may occur in AGN.

\subsection{Disk variability}
Optical/UV variability in AGN is typically $\mathcal{O}$(10$\%$) over a few months \citep[e.g.,][]{Krolik99, Kasliwal15} and is generally associated with the behaviour of the accretion disk. Larger scale optical/UV variability can also occur in quasars \citep[e.g.,][]{Graham17, Ross18, Stern18}. There are a very large number of possible causes of accretion disk variability including (but not limited to): disk inhomogeneities, different instabilities, fronts and  magnetic reconnection. \citet{Stern18} conveniently parameterizes timescales for orbital ($T_{\rm orb}$), thermal ($T_{\rm thermal}$), heating/cooling front propagation ($T_{\rm cool}$) and viscous transport ($T_{\nu}$) as a function of disk radius ($R$), aspect ratio (height over radius, $h \equiv H/R$), viscous parameter ($\alpha$) and SMBH mass ($M_{\rm SMBH}$). Following \citet{Stern18}, we write:
\begin{equation}
    T_{\rm orb} \sim 100\, {\rm day}\, M_{8}\, \left(\frac{R}{750r_{g}}\right)^{3/2}
\label{eq:orb}
\end{equation}
\begin{equation}
    T_{\rm thermal} \sim 100\, {\rm day}\, M_{8}\, \alpha_{0.03}^{-1}\, \left(\frac{R}{65r_{g}}\right)^{3/2}
\label{eq:thermal}
\end{equation}
\begin{equation}
    T_{\rm front} \sim 100\, {\rm day}\, M_{8}\, h_{0.1}^{-1}\, \alpha_{0.03}^{-1}\, \left(\frac{R}{15r_{g}}\right)^{3/2}
\label{eq:front}
\end{equation}
\begin{equation}
    T_{\rm \nu} \sim 100\, {\rm day}\, M_{8}\, h_{0.2}^{-2}\, \alpha_{0.03}^{-1}\, \left( \frac{R}{6r_{g}}\right)^{3/2}
    \label{eq:viscous}
\end{equation}
where $M_{8}=M_{\rm SMBH}/10^{8}M_{\odot}$, $h_{0.1} = h / 0.1$, and $\alpha_{0.03}$ corresponds to a viscous parameter $\alpha=0.03$. From equations~(\ref{eq:orb})-(\ref{eq:viscous}), we can associate variability on timescales of $<100$ days in a given AGN disk to: for example, an embedded or interacting orbiter at $<10^{3}r_{g}M_{8}$, thermal variability in the disk at $\leq 65r_{g}M_{8}$, front propagation at $\leq 15r_{g}M_8$ of the SMBH, or viscous changes near the innermost, stable circular orbit (ISCO; $\approx 6\, r_{g}$), assuming a moderately puffed-up and viscous inner disk.


\subsection{Supernovae and kilonovae}
Supernovae (SNe) are expected to occur in AGN accretion disks, but the expected rate is small, $>2 \times 10^{-7}\, {\rm AGN}^{-1}\, {\rm yr}^{-1}$ in the {\it WISE} sample \citep[e.g.,][]{Assef18}. SNe will also occur in the host galaxy, and appear consistent with an origin in the galactic nucleus if their separation is less than the survey angular resolution. For unobscured SNe we expect rise times of $\mathcal{O}(20-50)$~days and a decay time or plateau of $\sim 100 - 200$~days \citep{Kasen10} and we also expect an evolution in color over time \citep{Foley11}.  We were able to rule out a SN origin for the flare reported in \citet{Graham20} based on its much shorter flare timescale and lack of color evolution.

\subsection{Tidal disruption events}
Tidal disruption events (TDEs) also occur in AGN \citep[e.g.,][]{Chan19, Ricci20, Starfall21}. Main sequence star disruptions can occur around the central SMBH in a galaxy, but only for $M_{\rm SMBH} \leq 10^{8} \,M_{\odot}$ \citep[for a non-spinning SMBH;][]{rees88,Taeho20}. These are typically characterized by a fast rise (i.e., several weeks), $\sim t^{-5/3}$ decay signatures, and will be false positives in our search for EM counterparts to any O3 GW triggers. 

In addition, TDEs can also occur around small BHs in AGN disks, as neutron star (NS) or white dwarf (WD) disruptions by stellar origin BHs \citep[e.g.,][]{Yang21}. Thus, for BH-NS mergers, where the BH is $\leq 7-10M_{\odot}$ (depending on BH spin), the expected EM counterpart corresponds to a NS tidal disruption emerging from inside an AGN disk. We expect the rate of such events at $z<0.5$ should span $\sim [4,113]\, (f_{\rm AGN}/0.1)\, {\rm yr^{-1}}$, where $f_{\rm AGN}$ is the fraction of BBH mergers expected from the AGN channel \citep{McK20R}.  The expected integrated total energy of such events is $\mathcal{O}(10^{52}\, {\rm erg})$ \citep{Cannizzaro20}.  BH-WD disruptions lead to underluminous Type Ia SN with integrated energy $10^{49-51}{\rm erg}$ \citep{rosswog09}.

\subsection{Microlensing}
Microlensing is uniform in color at restframe UV/optical bands and is expected for AGN with an expected rate of $\mathcal{O}(10^{-4})$ per AGN \citep{Lawrence16}. However, the expected characteristic timescale for microlensing is $\mathcal{O}$(yrs) \citep{Lawrence16}, which generally is far longer than the flaring timescales considered here. The Einstein ring radius of a gravitational lens is
\begin{equation}
r_{E}=\sqrt{\frac{4GM_{\ast}}{c^2} \frac{D_{L}D_{LS}}{D_{S}}}
\end{equation}
where $GM_{\ast}/c^2=1.5(M_{\ast}/M_{\odot})$km, $D_{S}$ is the distance to the source, $D_{L}$ is the distance to the lens, and $D_{LS}$ is the distance from the lens to the source. If the lens is in the AGN host $D_{L} \sim D_{S}$. The duration of a microlensing event is $t_{\mu}=r_E/v_{\rm rel}$ where $v_{\rm rel}$ is the relative velocity of the lens. The resulting magnification is 
\begin{equation}
{\cal{M}}=\frac{u^{2}+2}{u \sqrt{(u^{2}+4)}}
\end{equation}
where $u=b/r_{E}$ is the impact parameter ($b \sim 1$) of the lensing event in units of $r_{E}$. Assuming the stellar orbits at distance $D_{LS}$ in the bulge of the host galaxy are evenly distributed in the hemisphere facing the observer, they cover an area $A=\pi D_{LS}^{2}$ and each star sweeps out a microlensing area of $A_{\mu,\ast}=\pi D_{LS} r_{E}$ in half its Keplerian orbital time. 

A fiducial $M_{\odot}$ lens in the source galaxy with  $D_{LS} \sim 1\, {\rm kpc}$ and $v_{\rm rel} \sim 200\, {\rm km \, s^{-1}}$ gives us a timescale of $\sim 2 \times 10^6\, {\rm s}$ (i.e., $\sim 3$~weeks) with magnification depending on the choice of $u(r_{E})$. Assuming a population of $\mathcal{O}(10^{10})$ stars in random orbits, geometric considerations produce a rate of $\mathcal{O}(10^{-5}) \, {\rm events \, yr^{-1} \, AGN^{-1}}$. We also note that the flare in \citet{Graham20} has an noticeable asymmetry that is not expected for microlensing and therefore disfavors this explanation for that flare.

\section{Method and Data Sets}
\label{sec:method}

\subsection{Zwicky Transient Facility (ZTF)}
\label{sec:ztf}
ZTF is a state-of-the-art time-domain survey employing a 47 deg$^2$ field-of-view camera on the Palomar 48-inch Samuel Oschin Schmidt telescope \citep{Bellm19,Graham19}. Since March 2018, it has operated a number of observing programs, including a public survey covering the visible northern sky every 2 to 3 nights in the $g$- and $r$-bands to $\sim 20.5$ magnitude, as well as boutique partnership programs such as higher cadence coverage of specific regions and use of an $i$-band filter \citep{Bellm19b}. Each ZTF observation is processed by an image differencing pipeline \citep{Masci19}, which generates real-time alerts for all 5$\sigma$ detections of point-source transient events \citep{Patterson19}. Each observation is also processed by a PSF photometry pipeline which produces a single epoch catalog covering {\em all} identified sources. These are archived and used to create light curves for all detected objects in ZTF data releases.

ZTF DR5\footnote{https://www.ztf.caltech.edu/page/dr5}, released on 2021 June 6, provides data from public surveys thru 2021 January 31 (as well as partnership data thru 2019 December 31) and therefore covers the entire LIGO/Virgo O3 data release. A forced photometry data set has been produced based on all difference images available for DR5 (Mroz et al., in prep.) and source positions from the PS1 DR1 catalog \citep{Chambers16}. 

  

\subsection{AGN matching}
\label{sec:catalog}
We employed the set of spectroscopically confirmed AGNs and high probability AGN candidates from the Million Quasar Catalog \citep[v7.3,][MQC]{Flesch19} as the primary AGN catalog. Given the sensitivity of LIGO/Virgo in O3, we excluded sources at redshift $z > 1.2$ from this analysis, as well as known blazars. The catalog was crossmatched using a 3\arcsec\ matching radius against the set of forced photometry ZTF light curves, giving a data set of 524,666  sources.

For each LIGO/Virgo event, we identified all ZTF sources from the light curve data set that were located within the 90\% credible volume of the event using the \texttt{crossmatch} method from the \texttt{ligo.skymap} Python package\footnote{https://lscsoft.docs.ligo.org/ligo.skymap/}. Table~\ref{tab:alert_summary} gives the waveform employed for each LIGO/Virgo event but, in summary, the NRSur7dq4 waveform was used for O3a alerts when available and the SEOBNRv4PHM waveform as an alternate. The IMRPhenomXPHM waveform was used for all O3b events.



Each ZTF source may have a $g$- or $r$-band light curve and we fitted each (in flux space) with a Bayesian block representation \citep{scargle13}. This provides an optimal segmentation of the data in terms of a set of discontinuous piecewise constant components and makes significant local variations more easily detectable. We identified candidate flares using the hill-climbing procedure proposed by \cite{meyer2019}: peaks are identified as blocks that are higher than both previous and subsequent blocks and then extended in both directions as long as succeeding blocks are lower. The data contained within such a set of blocks is then characterized by a Gaussian rise-exponential decay form:
\begin{eqnarray}
y(t) & = r_0 + A \exp\left(-\frac{(t - t_0)^2}{2 t_g^2}\right),  & t \le t_0 \nonumber \\
& = r_0 + A \exp\left(-\frac{(t - t_0)}{t_e}\right),& t > t_0 
\label{eq:flare}
\end{eqnarray}

\noindent
with rise and decay times, $t_g$ and $t_e$, respectively. The flare peaks at time $t_0$ with an amplitude $A$ above a background flux of $r_0$. We added an additional random noise term $f$ in quadrature to the model covariance to account for potential systematic errors in the model. We assumed uniform priors for each parameter (see Table~\ref{tab:model_priors}) and maximized the posterior likelihood for the flare model.
 
Sources were rejected when the time between the LIGO/Virgo event and a flare peak is greater than 200 days or less than the Gaussian rise time from the model fit. Random noise in a ZTF light curve may be misidentified as a flare, particularly in low signal-to-noise data, so a LIGO/Virgo event must have a corresponding flare in both filters (coincident in time) and all single filter flares were rejected. Similiarly, the amplitude of the flare must be at least 10\% greater than the background flux level ($r_0$).


\subsection{Astrophysical implications of our search assumptions}
The search criteria in Table~\ref{tab:model_priors} correspond to assumptions about the mergers. In particular, given $M_{\rm SMBH}$, a radial location in the disk of the merger $a$, and a local disk scale height (or, equivalently, local sound speed $c_s$), the search timescales correspond to a minimum $v_{k}$. For each source, we can measure $M_{\rm SMBH}$, but can only make plausible assumptions for a range of $a$ and $c_{s}$. The total energy of the flare (which is related to the amplitude) sets a maximum $v_{k}$, given $M_{\rm BBH}$, $\eta$, and the local disk density $\rho$. As with the minimum $v_{k}$, we can use the GW measurements to find $M_{\rm BBH}$, but $\eta$ and $\rho$ are free parameters. Note that for some combinations of candidate flare and GW source, there may be no plausible sets of $a$, $c_{s}$, $\eta$, and $\rho$ which yield consistent min/max $v_{k}$, given the observed $M_{\rm SMBH}$ and $M_{\rm BBH}$. For most pairs, we can construct a parameter set that yields an internally consistent scenario. However, some of those may be implausible astrophysically.

In particular, given our search window of only 200 days post-merger, our search tends to require small $a$ and $c_{s}$, such that the remnant can escape the optically thick part of the disk sufficiently quickly to be observed, yet still have a sufficiently small $v_{k}$ that it can produce a detectable flare via BHL accretion. Similarly, to generate sufficiently bright flares that they will be detectable above the background AGN luminosity, we are pushed to large $\eta$ and $\rho$; we are also more likely to detect larger $M_{\rm BBH}$ and find them more easily around smaller $M_{\rm SMBH}$. Large $\eta$ is expected for thin disk accretion onto a highly spinning BH (which would be expected for a BBH merger remnant); however simulations imply that $\eta$ decreases for highly super-Eddington accretion, as required if these flares are associated with the GW events. Large $\rho$ may indeed be expected in the innermost regions of an AGN disk, where we are more sensitive to detecting mergers (due to the shorter length scales and shorter timescales); however, there are several cases where only a maximal density can produce a consistent pair of optical flare and GW signal.

We find the lowest minimum $v_{k}$ assuming $a=100~r_g$ and $c_s=10~{\rm km~s^{-1}}$. For $M_{\rm SMBH}=10^8~M_{\odot}$ and a 200 day search window, we are sensitive to $v_{k}>1~{\rm km}\, {\rm s}^{-1}$ (with lower velocities probed around lower $M_{\rm SMBH}$). We probe the largest maximum $v_{k}$ assuming $\eta=0.3$ and $\rho=10^{-9}~{\rm g~cm^{-3}}$, and larger flare energies imply smaller $v_{k}$ at fixed $M_{\rm SMBH}$ and $M_{\rm BBH}$. We note that $a=1000~r_g$ and $c_s=50~{\rm km~s^{-1}}$ are more consistent with our chosen density \citep{SirkoGoodman03}, and our choice of $\eta$ may be larger than warranted. 

\subsection{Discriminating false positive signals}
\label{sec:discrim_other}

\noindent
In Section~\ref{falseposother}, we considered other potential sources of flaring activity from an AGN that might be misidentified as an EM counterpart to a GW event.
Normal TDEs, SNe, and AGN flares can be distinguished on the basis of rise and fade timescales, $g-r$ color, and the rate of color evolution \citep{vanVelzen21}. The total observed energy of the event can also be added to this list. Since instantaneous colors are not available for ZTF light curves and data sometimes exist for only one filter for a few nights, we model the multiband light curve for each source as a 2D surface with time and wavelength (passband) as the two axes. A suitable interpolation scheme over the irregularly-sampled surface then allows fluxes to be predicted in each passband at each observed time. 

\begin{figure*}
   \centering
   \includegraphics[width = 0.425\textwidth]{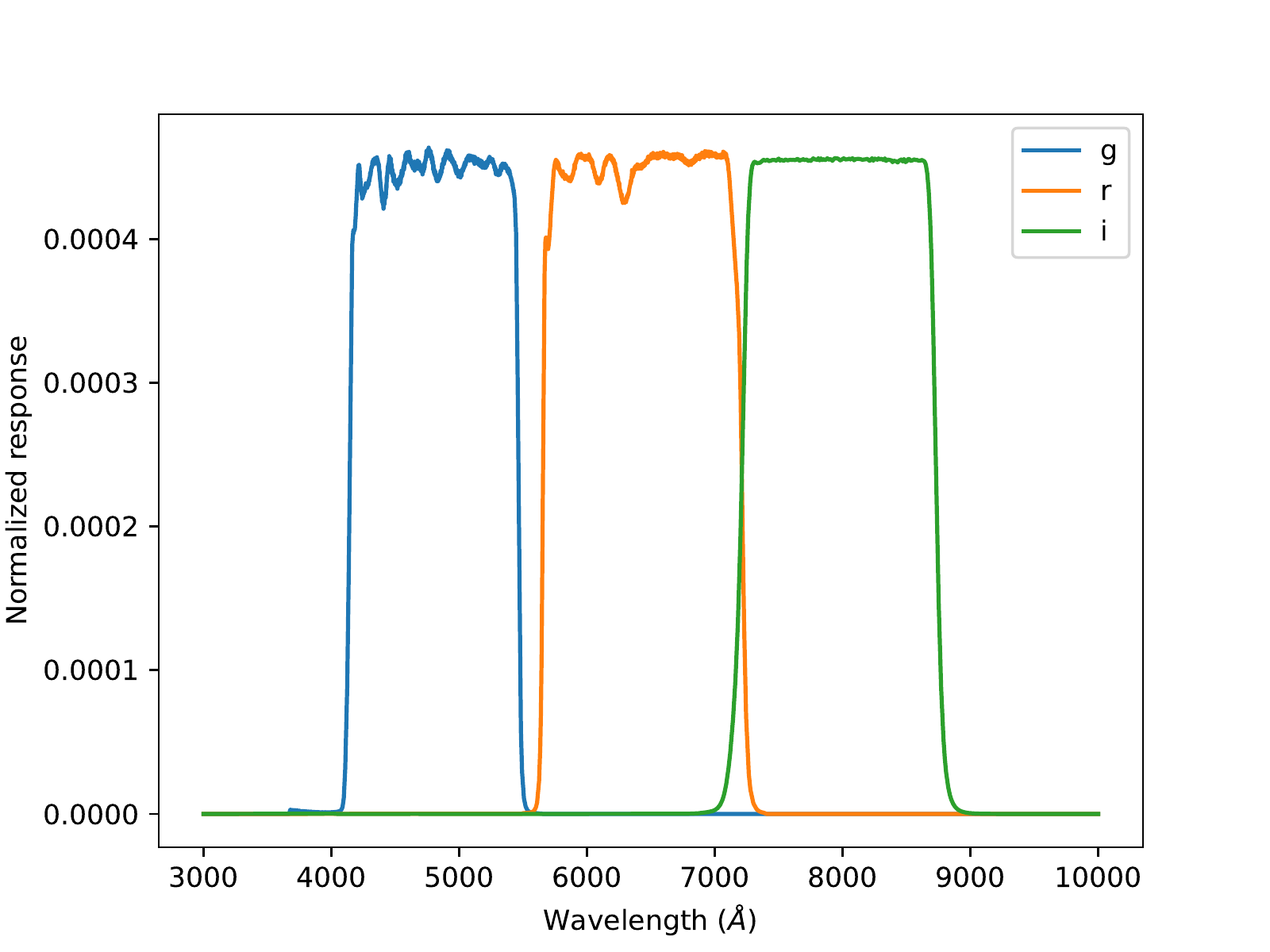}
   \includegraphics[width = 0.565\textwidth]{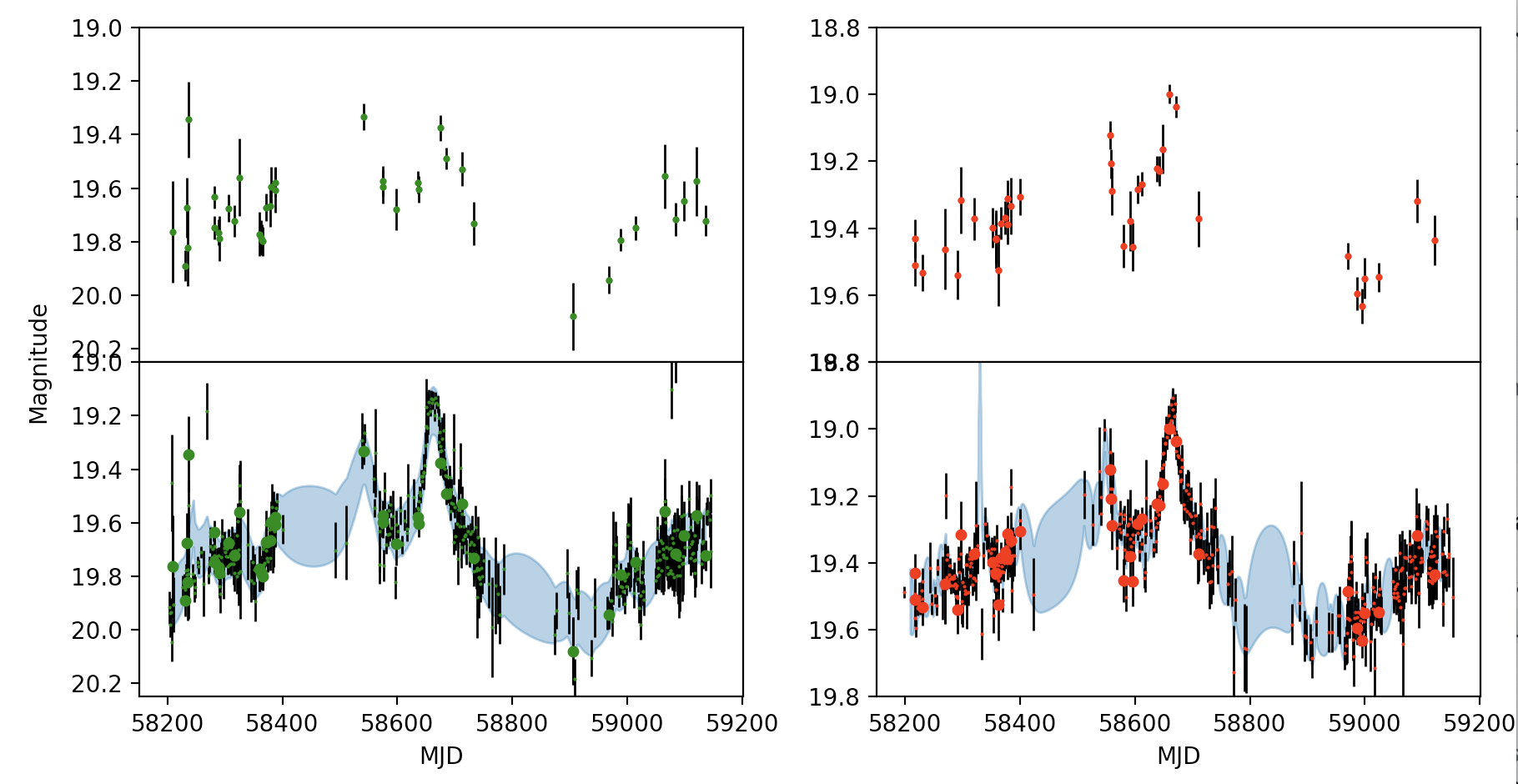}
   \caption{(left) The $g$-, $r$-, and $i$-band normalized filter response curves for ZTF (Bellm et al. 2018). (right) The upper plots show $g$- and $r$-band subsampled light curves of an AGN; the lower plots show the interpolated light curves using a 2D GP fit (blue shaded region). The larger points indicate the points used in the regression and the smaller points are from the full light curve.}
\label{fig:filters}
\end{figure*}

Thin-plate splines have been used for similar purposes \cite[e.g., fitting stellar spectral energy distributions as a function of temperature and extinction;][]{BailerJones2011} but a more probabilistic approach is provided by a Gaussian process (GP), as demonstrated by \cite{Boone19} and \citet[][V20]{Villar20}. Note that the thin-plate spline and GP interpolation schemes can be shown to be equivalent with a particular choice of kernel function. Following V20, we use a composite kernel function for the covariance between observations at times $t_i$ and $t_j$ in filters $f_i$ and $f_j$, respectively: 
\[
K(t_i, t_j, f_i, f_j) = \sigma^2 l_t \exp \left[ -\frac{|t_i - t_j|}{l_t} \right] \exp \left[-\frac{d(f_i, f_j)^2}{2l_f^2} \right] 
\]
 
\noindent
where $\sigma^2$ is the variance, and $l_t$ and $l_f$ are the respective length scales along each axis. This describes crossband information via a squared exponential with a Wasserstein distance metric, $d(f_i, f_j)$, between each filter's normalized transmission curve (see Fig.~\ref{fig:filters}). V20 also describe the temporal variability with a squared exponential function; however, this gave inadequate fits to ZTF data, particularly during the seasonal observation gaps of a few months. Since AGN variability can be described as a damped random walk (DRW) process, an Ornstein-Uhlenbeck (Mat\'{e}rn-1/2) kernel was found to be a better choice. Note, however, that perfect correlation between bands is assumed in this model with no wavelength dependency in the variability amplitude or characteristic timescale. 

\begin{figure*}
   \centering
   \includegraphics[width = \textwidth]{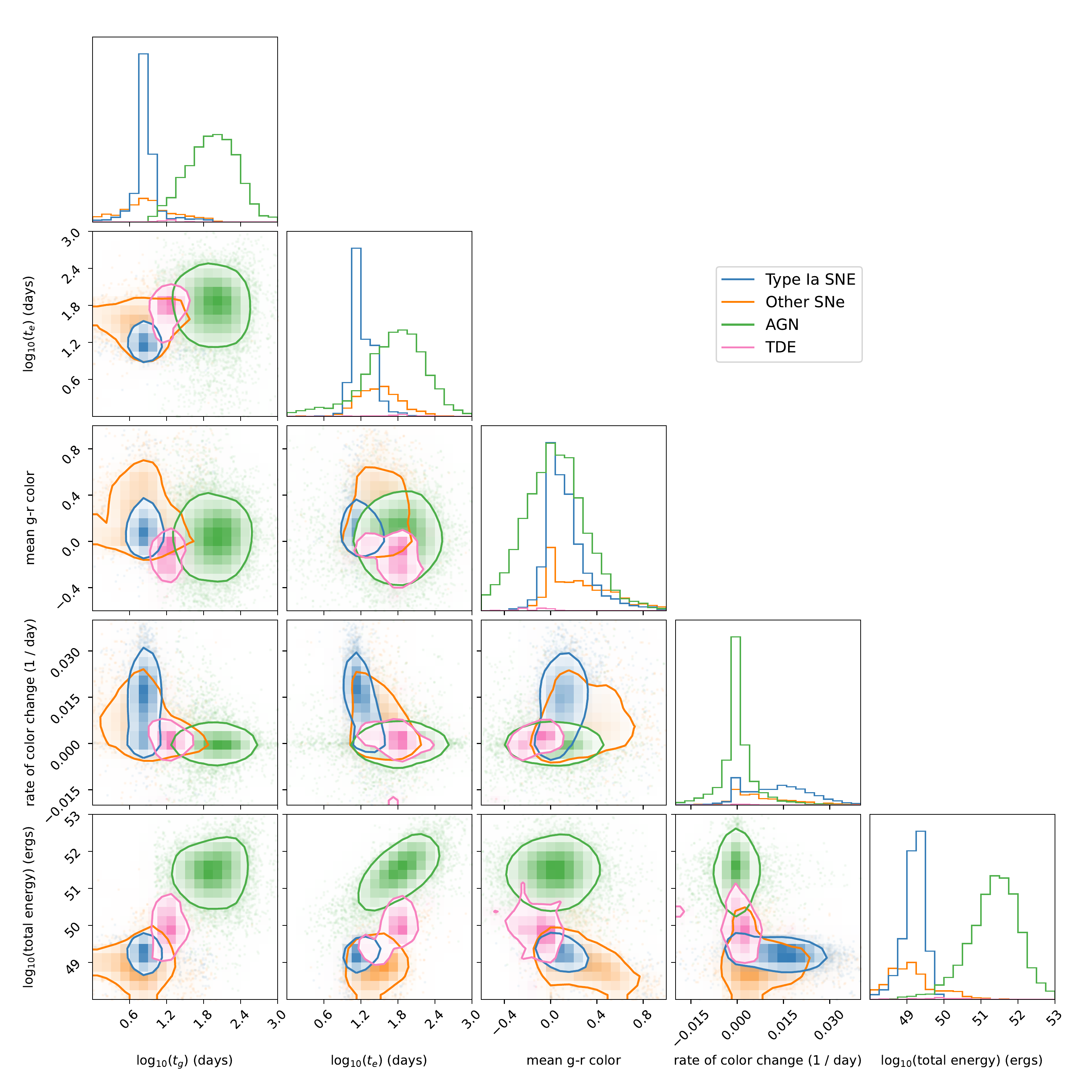}
   \caption{The distributions of rise time ($t_g$), decay time ($t_e$), mean $g-r$ color, rate of $g-r$ color change, and total flare energy for type Ia SNe (blue) and other SNe (orange) from the ZTF Bright Transient Survey \citep{Fremling20}, optically selected TDEs in ZTF (pink) \citep{vanVelzen21, Hammerstein21}, and a general sample of AGN flares (green) (see text for details). The timescales are measured in the restframe of the source. The contours show one sigma levels.}
\label{fig:tgte}
\end{figure*}

Fig.~\ref{fig:tgte} shows projections of the discriminating space and the respective distributions of ZTF optically selected TDEs  \citep{vanVelzen21, Hammerstein21}, SNe from the ZTF Bright Transient Survey \citep{Fremling20, Perley20}, and a general sample of flares in AGN identified by applying the flare model to AGN light curves without a LIGO/Virgo timing constraint. A balanced random forest classifier [BalRF04] was trained on this data, achieving $\sim$90\% accuracy, and applied to candidate flares to determine whether they might be a TDE or a SNe. Note that these could be associated with an off-nucleus event rather than a TDE or SN embedded in the accretion disk of an AGN. If a source is not clearly classified then we do not reject it. Finally, we test whether the time series in both bands is better fit ($\Delta BIC > 10$) with a microlensing profile than the flare (eqn.~\ref{eq:flare}). 

\subsection{Discriminating from AGN activity}
\label{sec:discrim_agn}
Variability is an inherent property of AGN, often exhibited on short timescales, and the most likely origin of a false positive signal, i.e., an identified flare is intrinsic behavior related to general gas physics and accretion disk activity rather than associated with a particular event in the accretion disk. Change point detection is an aspect of statistical time series analysis that tries to identify when the probability distribution of a stochastic process changes, for example, with the addition of a secondary temporary signal from an accompanying process.

We employed a GP-based algorithm (see Appendix~\ref{gpchange} for details) to determine whether the flare seen in a given AGN is consistent with the type and level of intrinsic variability exhibited by the AGN as described with a DRW model \citep[see][and references therein]{Moreno19}. We used a sliding window with a 50 day width to detect variations from the DRW and rejected all flares where the probability that the flare is just intrinsic AGN variability is greater than 0.5\% in either filter. 
We note, however, that an incorrect choice of model for the intrinsic AGN behaviour could be a source of false positives, i.e., intrinsic activity mistaken for significant flares. While a discussion of the correct model for AGN variability is outside the scope of this paper, the DRW model is seen as an adequate statistical description of AGN behaviour and the fidelity of the data used here is such that higher order autoregressive models are indistinguishable from it. A more promising avenue would be to use a multi-survey data set with extended temporal baselines to better constrain the models but again there is work to be done on multivariate AGN models.
The change-point algorithm, though, will work with any model provided.




\section{Results}

\begin{figure*}
   \centering
   \includegraphics[width = 0.497\textwidth]{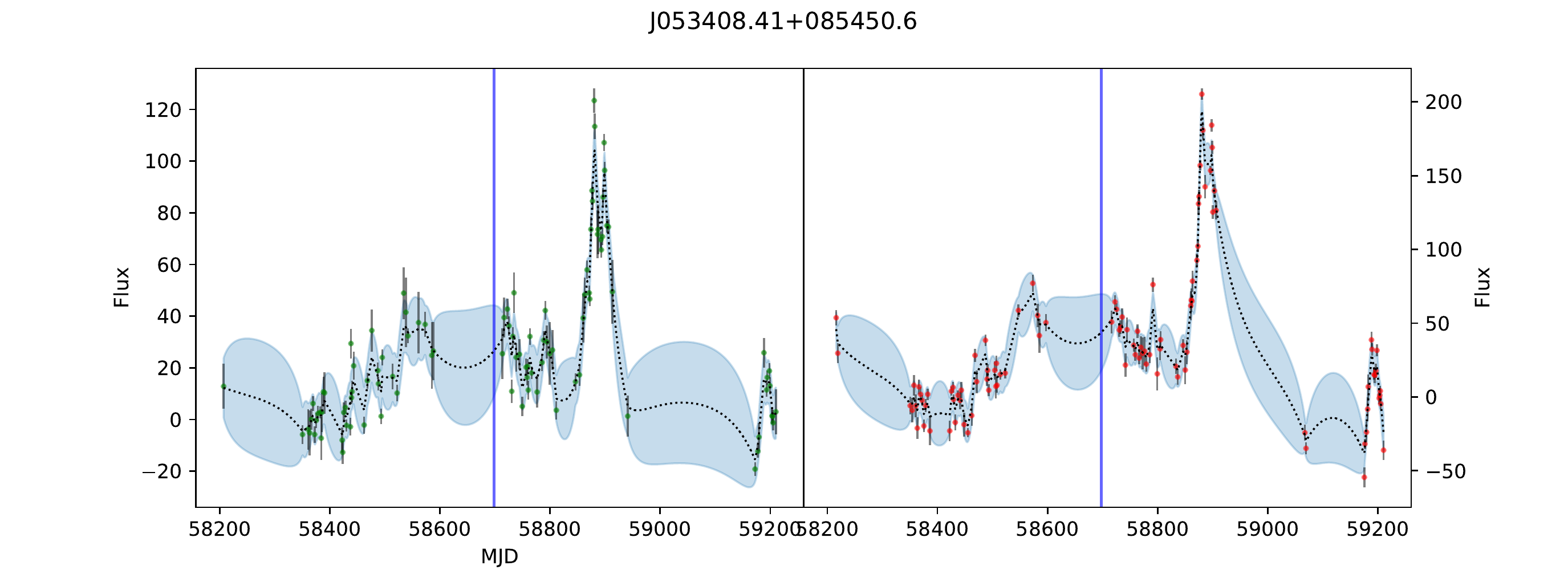}
   \includegraphics[width = 0.497\textwidth]{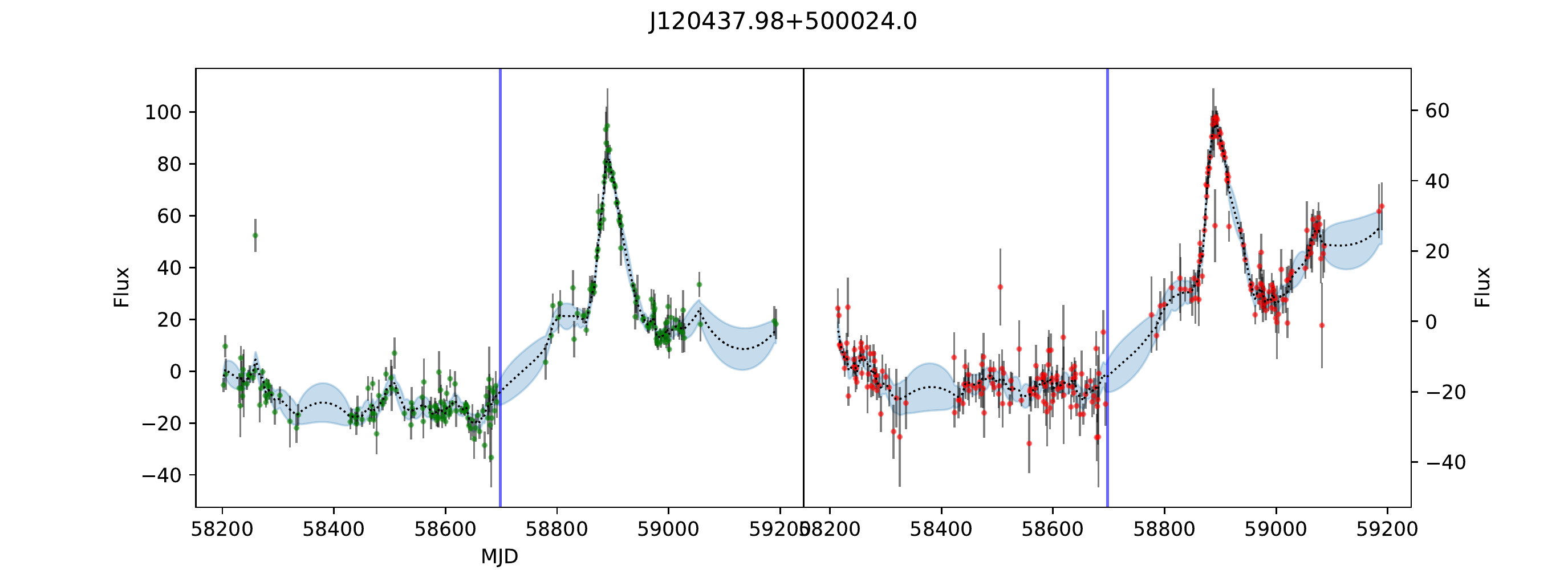}
   \includegraphics[width = 0.497\textwidth]{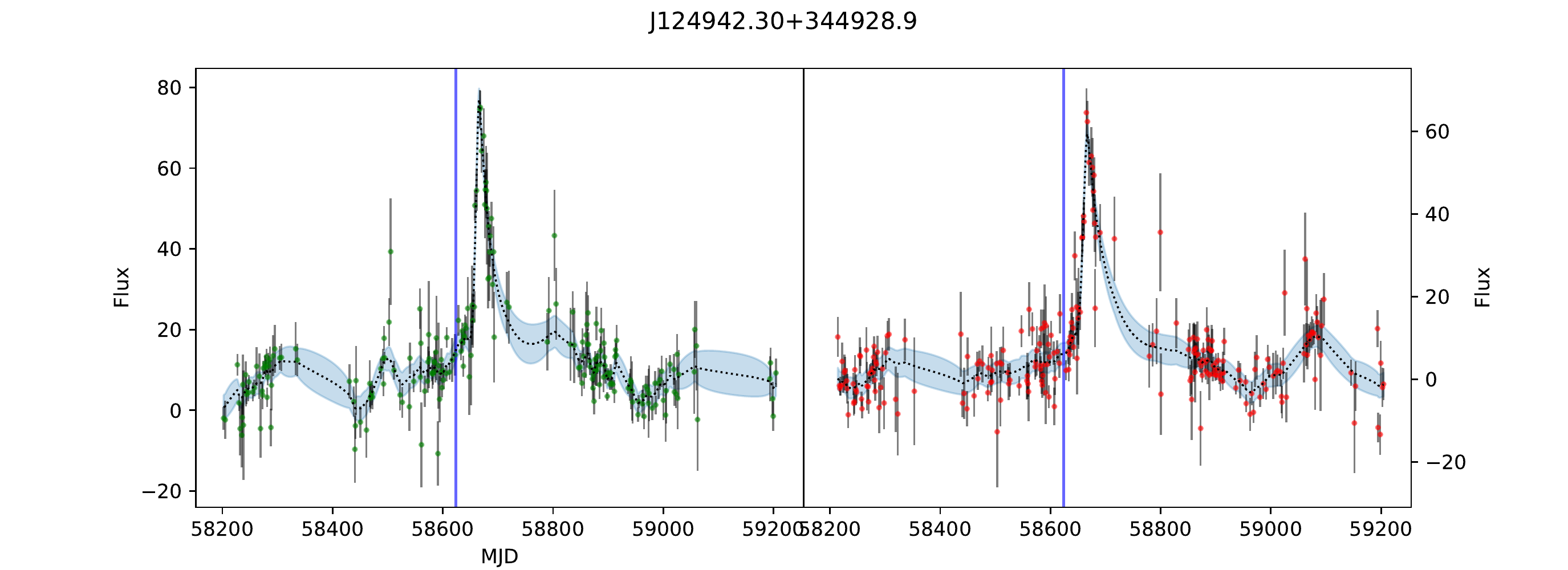}
   \includegraphics[width = 0.497\textwidth]{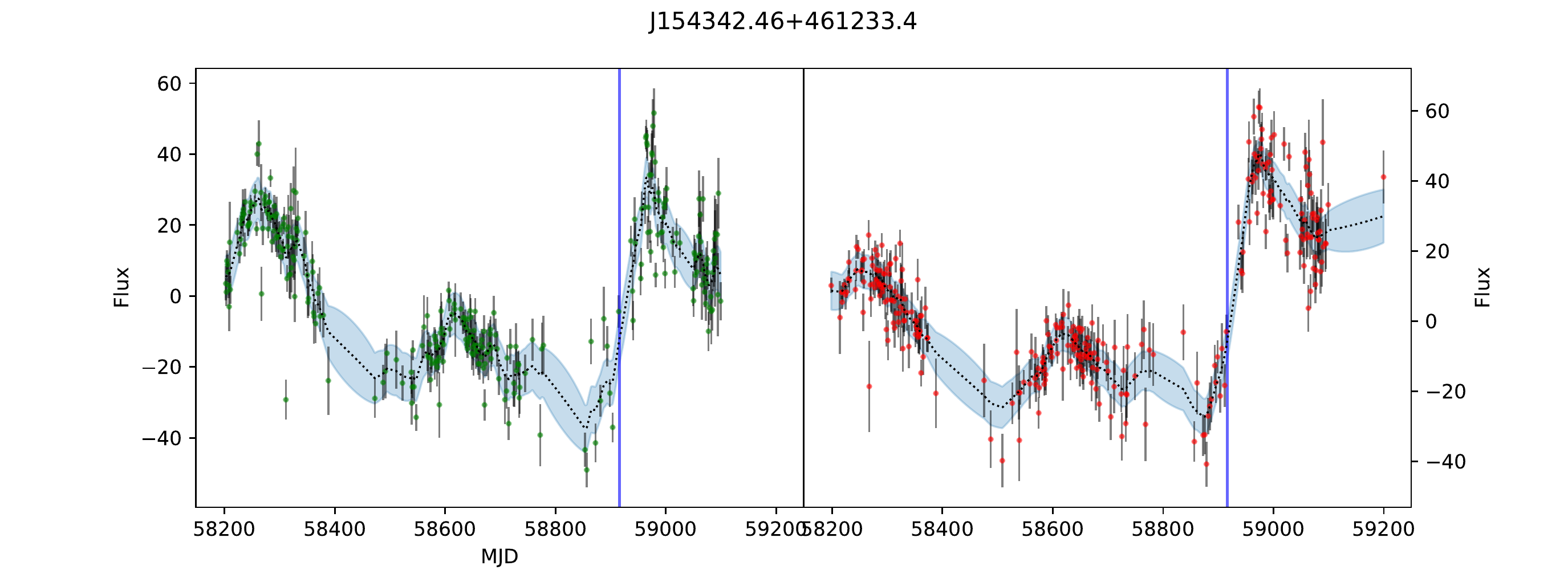}
   \includegraphics[width = 0.497\textwidth]{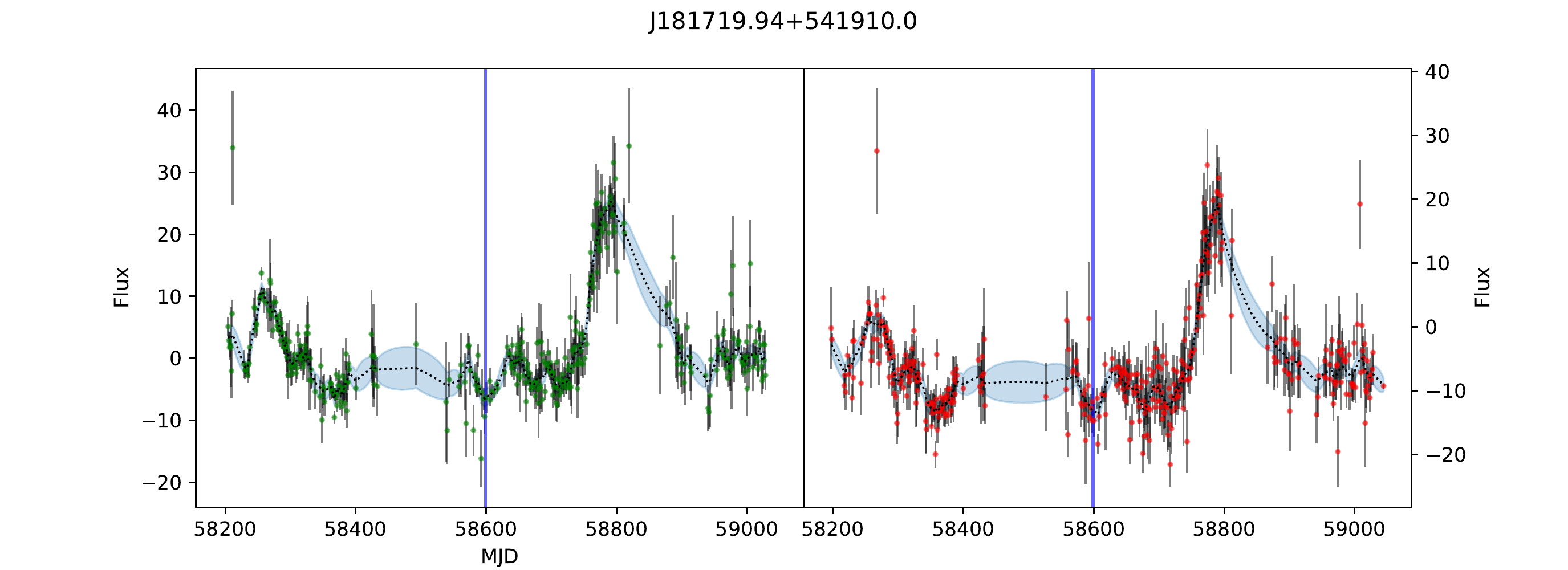}
   \includegraphics[width = 0.497\textwidth]{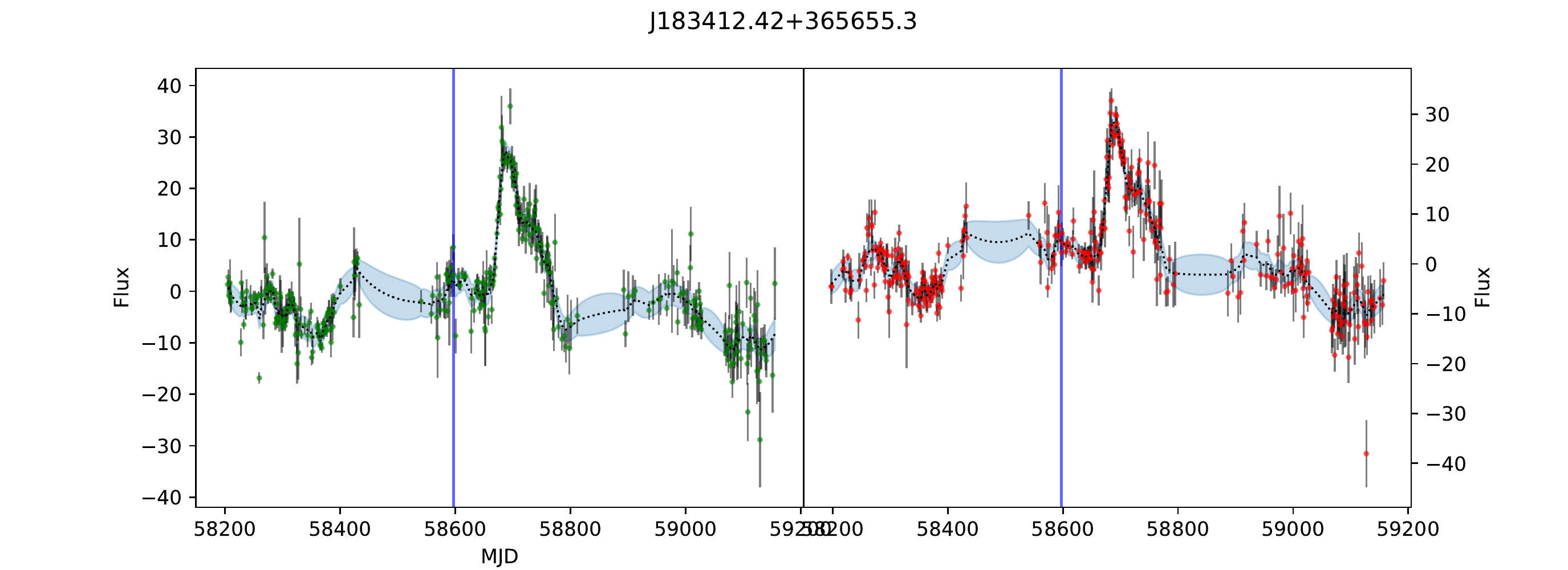}
   \includegraphics[width = 0.497\textwidth]{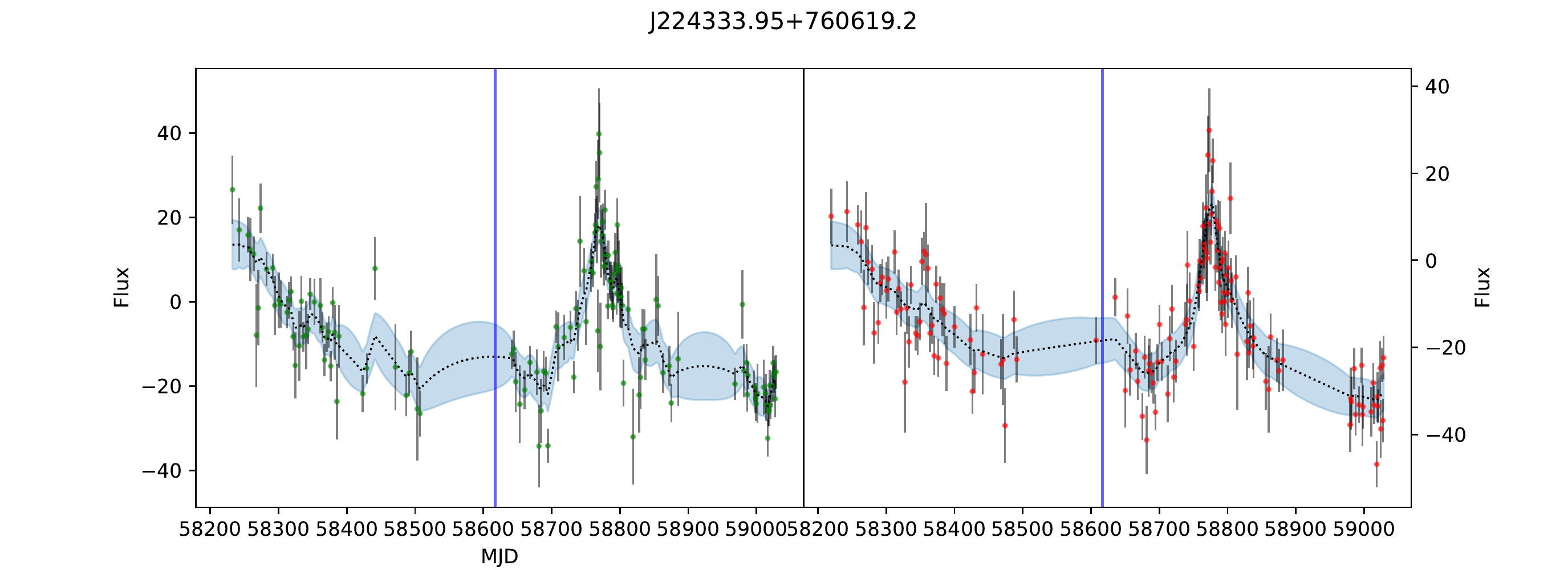}
   \caption{The ZTF $g$- and $r$-band light curves for the flares associated with LIGO/Virgo events. A GP fit to the data combining a mean flare function and a damped random walk kernel is shown with its predicted uncertainties (blue shaded region).}
\label{fig:magdiff}
\end{figure*}

From the 83 LIGO/Virgo BBH and lower mass gap merger alerts, we find 7 AGN flares that are statistically associated with one or more of 9 LIGO/Virgo events (see Table~\ref{tab:flare_summary} for details and Fig.~\ref{fig:magdiff} for their light curves). The chance coincidence of this can be computed by considering that there are 20 AGN flares in the full ZTF data set that meet our selection criteria: i.e., they have the correct morphology, correct energetics, acceptable color evolution, and are inconsistent with being a TDE, SNe, or regular AGN activity. The total comoving volume probed by ZTF to $z = 1.2$ is $1.643 \times 10^{11}$ Mpc$^3$. Assuming a mean flare lifetime of 100 days and 1000 days of ZTF data, the effective source density of flares in the survey volume at any given time is $1.321 \times 10^{-11}$ Mpc$^{-3}$. The expectation value for the number of flares from O3 is the product of the effective density and the combined localization volume covered by ZTF ($2.144 \times 10^{11}$ Mpc$^3$), implying 2.83 expected random events. Using a Poisson distribution, the chance coincidence of finding 9 matches is $p = 1.90 \times 10^{-3}$.  

One of the sources, J053408.41+085450.6, has a quoted photometric redshift of $z = 0.5$ from MQC and a quoted spectroscopic redshift of $z = 1.62$ from GAIA DR3 \citep{GaiaDR3}, though the latter has flags indicative of an ambiguous redshift identification (i.e., {\tt FLAGS\_QSOC} = 13).  The higher redshift would place this source outside the LIGO/Virgo detection limit, though we still consider this candidate in our analysis given the low score of that redshift solution.  We obtained a 600s spectrum of the source from Keck Observatory (see Appendix~\ref{app:spectra}), finding a relatively featureless spectrum and an inconclusive redshift determination. The source is also associated with a 32.1 mJy radio source at 21 cm from the NRAO VLA Sky Survey \citep{Condon1998}, which suggests that this might actually be a blazar, although previous optical activity of the object is fairly quiet and it does not appear in any blazar catalogs.

As a comparison, we consider the original 45 BAYESTAR \citep{Singer16} or LALInference \citep{Veitch15} LIGO/Virgo alerts for BBH and lower mass gap events.  This is a subset of the 83 such alerts eventually published from O3, and generally correspond to the higher signal-to-noise ratio events.  We use the original released skymaps rather than the final published skymaps, and find that three flares are associated with three LIGO/Virgo events. Of these, only J124942.3+344928.9 and its association with GW190521 is common to both sets of input. The other two alerts, GW190519\_153544 and GW200112\_155838, fall outside the 90\% LIGO/Virgo confidence volumes of the final published skymaps, which reduced in volume by $\sim$15 -- 20\% during further processing of the GW signal. This highlights the importance of providing accurate and revised volume localizations in O4 (preferably on the timescale of days after an event) to improve our ability to identify an EM counterpart to a BBH merger in an AGN disk.


The implied kick velocity constraint of each event is also broadly consistent with what LIGO/Virgo reported and so we cannot rule out any event on this basis here.

\begin{deluxetable*}{cccccccc}
\tabletypesize{\scriptsize}
\tablenum{1}
\tablecaption{A summary of the $83$ LIGO/Virgo BBH and lower mass gap merger alerts with associated ZTF AGN flares. $f_{\rm cov}$ is the fraction of the LIGO/Virgo 90\% confidence area covered by ZTF with at least 15 observations in the 200 days following the LIGO alert. $n_s$ is the number of Million Quasar Catalog (v7.3) sources within the LIGO/Virgo 90\% confidence volume. The divisions indicate events from O3a (GWTC 2.0), O3a (GWTC 2.1), and O3b. 
The NRSur7dq4 waveform is used for O3a detections except for those marked with an asterisk where the SEOBNRv4PHM waveform is used. The IMRPhenomXPHM waveform is used for all O3b detections. SNR are the published values and are derived as follows: in O3a, from GstLAL (including Virgo) where possible and PYCBC BBH otherwise; in O3b, from IMRPhenomXPHM; and in GWTC 2.1, from PYCBC BBH where possible and GstLAL otherwise. LIGO/Virgo considers signals with SNR $>8$ to be confidently real.\label{tab:alert_summary}}
\tablewidth{0pt}
\tablehead{
\colhead{LIGO/Virgo alert ID}  & \colhead{SNR} & \colhead{50\% area} & \colhead{90\% area} & \colhead{Distance} & \colhead{$f_{\rm cover}$}  & \colhead{$n_s$} \\
  &   & \colhead{(deg$^2$)} & \colhead{(deg$^2$)} & \colhead{(Gpc)} & &   
}
\startdata
GW190408\_181802*  & 14.7 & 23 & 145  & $1.58^{+0.40}_{-0.59}$ & 1.000 & 332  \\ 
GW190412*  & 18.9 & 3 & 12 & $0.74^{+0.14}_{-0.17}$ & 1.000 &  34  \\
GW190413\_052954 & 8.6 & 331 & 1488 & $4.10^{+2.41}_{-1.89}$ &  0.522 &  33453 \\
GW190413\_134308  & 10.0& 54 & 589 & $5.15^{+2.44}_{-2.34}$ &  0.180 & 11961 \\
GW190421\_213856   & 10.6 & 296 & 1267 & $3.15^{+1.37}_{-1.42}$ & 0.000 &  8185 \\ 
GW190424\_180648  & 10.0 & 10754 & 28618 &  $2.55^{+1.56}_{-1.33}$ & 0.563 & 207206  \\
GW190425 & 13.0 & 2417 & 9958 & $0.16^{+0.07}_{-0.07}$ & 0.377 & \nodata \\
GW190426\_152155* & 10.1 & 289  & 1395 & $0.38^{+0.19}_{-0.16}$ & 0.496 & 620  \\
GW190503\_185404 &  12.1 & 26 & 94 & $1.52^{+0.71}_{-0.66}$ & 0.000 & 503 \\
GW190512\_180714* &  12.3 & 46 & 227  & $1.49^{+0.53}_{-0.21}$ & 0.233 & 367 \\ 
GW190513\_205428 &  12.3 & 112 & 404 & $2.16^{+0.94}_{-0.80}$ & 0.949 & 1151   \\
GW190514\_065416 &  8.3 & 444 & 2930 & $4.93^{+2.76}_{-2.41}$ & 0.652 & 64784  \\
GW190517\_055101 &   10.6 & 53 & 425  & $2.11^{+1.79}_{-1.00}$ & 0.322 & 1803  \\
GW190519\_153144 &  12.0 & 170 & 838 & $ 2.85^{+2.02}_{-1.14} $ & 0.542 & 7880  \\
GW190521 &  15.0 & 180 & 822 & $ 4.53^{+2.30}_{-2.13} $ & 0.384 & 20225  \\
GW190521\_074359 &  24.4 & 129 & 546 & $ 1.28^{+0.38}_{-0.57} $ & 0.836  & 797 \\
GW190527\_092055 &  8.9 & 1115 & 3628 & $3.10^{+4.85}_{-1.64}$ & 0.396 & 44299 \\
GW190602\_175927 &  12.1 & 180 & 687 & $ 2.99^{+2.02}_{-1.26} $ & 0.262 & 6173 \\
GW190620\_030421 &  10.9& 764 & 7570 & $3.16^{+1.67}_{-1.43}$ & 0.654 & 103774  \\
GW190630\_185205 &   15.6 & 204 & 1258 & $ 0.93^{+0.56}_{-0.40} $ & 0.465 & 5198 \\
GW190701\_203306 &   11.6 & 14 & 46 & $ 2.14^{+0.79}_{-0.73} $ & 0.891 &  652  \\
GW190706\_222641 &   12.3 & 111 & 770 & $ 5.07^{+2.57}_{-2.11} $ & 0.690 & 29332 \\
GW190707\_093326* &  13.0 & 248 & 1328 &  $0.80^{+0.37}_{-0.38}$ & 0.484 & 1390 \\
GW190708\_232457* &  13.1 & 2325 & 14420 & $0.90^{+0.33}_{-0.40}$ & 0.552 & 43181 \\
GW190719\_215514 &  8.0 & 219 & 874 & $4.61^{+2.84}_{-2.17}$ & 0.856 & 31178 \\
GW190720\_000836* &  11.7 & 63 & 574 & $ 0.81^{+0.71}_{-0.33} $ & 0.345  & 2131 \\
GW190727\_060333 &  12.3 & 140 & 880 & $ 3.60^{+1.56}_{-1.51} $ & 0.386 & 4941 \\ 
GW190728\_064510* &  13.6 & 62 & 574 & $ 0.89^{+0.25}_{-0.37} $ & 0.257 & 736 \\
GW190731\_140936 &  8.5 & 728 & 3203 & $3.97^{+2.56}_{-2.07}$ & 0.269 & 41160  \\
GW190803\_022701 &  9.0 & 364 & 1497 & $3.69^{+2.04}_{-1.69}$ & 0.910 & 16456 \\
GW190814* &  22.2 & 3 & 15 &  $0.24^{+0.04}_{-0.05}$ & 0.712 & 1 \\
GW190828\_063405 &  16.0 & 101 & 539 & $ 2.22^{+0.63}_{-0.95}$ & 0.461 & 1609 \\
GW190828\_065509* & 11.1 & 145 & 655 & $ 1.66^{+0.63}_{-0.61}$ &  0.254 & 1265 \\
GW190909\_114149* & 8.5 & 1355 & 4613 & $4.77^{+3.70}_{-2.66} $ & 0.570 & 79634 \\
GW190910\_112807 & 13.4 & 2786 & 10120 & $1.57^{+1.07}_{-0.64}$ & 0.516 & 35170 \\
GW190915\_235702* &  13.1 & 76 & 370 & $ 1.70^{+0.71}_{-0.64} $ &  0.863 & 3192 \\
GW190924\_021846* &  13.2 & 105 & 348 & $ 0.57^{+0.22}_{-0.22} $ & 0.895 & 1088 \\  
GW190929\_012149* &  9.9 & 524 & 2031  & $3.68^{+2.98}_{-1.68}$ & 0.522 & 13618 \\
GW190930\_133541* & 10.0 & 496 & 1723 & $ 0.78^{+0.37}_{-0.33}$ & 0.789 &  2406 \\ 
\hline
GW190403\_051529 & 8.0 & 1111 & 4250 & $8.28^{+6.72}_{-4.29}$ & 0.499 & 132911 \\
GW190426\_190642 & 9.6 & 2041 & 8031 & $4.58^{+3.40}_{-2.28}$ & 0.496 & 145368 \\
GW190725\_174728 & 8.8 & 325 & 2436 & $1.03^{+0.52}_{-0.43}$ & 0.684 & 5403 \\
GW190805\_211137 & 8.3 & 660 & 3089 & $6.13^{+3.72}_{-3.08}$ & 0.543 & 77630 \\
GW190916\_200658 & 7.9 & 986 & 3573 & $4.94^{+3.71}_{-2.38}$ & 0.539 & 116115 \\
GW190917\_114630 & 9.5 & 451 & 1801 & $0.72^{+0.30}_{-0.31}$ & 0.419 & 2876 \\
GW190925\_232845 & 9.9 & 112 & 955 & $0.93^{0.46}_{-0.35}$ & 0.195 & 2813 \\
GW190926\_050336 & 7.8 & 625 & 2212 & $3.28^{3.40}_{-1.73}$ & 0.461 & 32379 \\
\hline
\enddata
\end{deluxetable*}

\begin{deluxetable*}{cccccccc}
\tablenum{2}
\tablewidth{0pt}
\tablehead{
\colhead{LIGO/Virgo alert ID}   & \colhead{SNR} & \colhead{50\% area} & \colhead{90\% area} & \colhead{Distance} & \colhead{$f_{\rm cover}$}  & \colhead{$n_s$}  \\
  &   & \colhead{(deg$^2$)} & \colhead{(deg$^2$)} & \colhead{(Gpc)} & & 
}
\startdata
GW191103\_012549 & $8.9^{+0.3}_{-0.5}$& 667 & 2171 & $0.99^{+0.50}_{-0.47}$   & 0.616 & 7377 \\
GW191105\_143521 &  $9.97^{+0.3}_{-0.5}$& 100 & 641 & $1.15^{+0.43}_{-0.48}$   & 0.375 & 2622  \\
GW191109\_010717  &$17.3^{+0.5}_{-0.5}$ & 418 & 1649 & $1.29^{+1.13}_{-0.65}$   & 0.364 & 5826 \\
GW191113\_071753 &  $7.9^{+0.5}_{-1.1}$& 578 & 2483 & $1.37^{+1.15}_{-0.62}$   & 0.444 & 13157 \\
GW191126\_115259 &  $8.3^{+0.2}_{-0.5}$& 354 & 1378 & $1.62^{+0.74}_{-0.74}$   & 0.473 & 7214 \\
GW191127\_050227 &  $9.2^{+0.7}_{-0.6}$& 137 & 983 & $3.4^{+3.1}_{-1.9}$   & 0.460 & 15259 \\
GW191129\_134029 &  $13.1^{+0.2}_{-0.3}$& 208 & 856 & $0.79^{+0.26}_{-0.33}$   & 0.409 & 2387 \\
GW191204\_110529  &$8.8^{+0.4}_{-0.6}$ & 922 & 3675 & $1.8^{+1.7}_{-1.1}$   & 0.495 & 39638 \\
GW191204\_171526  &$17.5^{+0.2}_{-0.2}$ & 79 & 256 & $0.65^{+0.19}_{-0.25}$   & 0.286 & 161 \\
GW191215\_223052  &$11.2^{+0.3}_{-0.4}$ & 139 & 586 & $1.93^{+0.89}_{-0.86}$   & 0.297 & 2158 \\
GW191216\_213338 & $18.6^{+0.2}_{-0.2}$& 61 & 206 & $0.34^{+0.12}_{-0.13}$   & 0.592 & 59 \\
GW191219\_163120 & $9.1^{+0.5}_{-0.8}$ &  &  & $0.55^{+0.25}_{-0.16}$   & \nodata & 7120 \\ 
GW191222\_033537  & $12.5^{+0.2}_{-0.3}$& 503 & 2168 & $3.0^{+1.7}_{-1.7}$   & 0.451 & 24418 \\
GW191230\_180458  & $10.4^{+0.3}_{-0.4}$& 302 & 1086 & $4.3^{+2.1}_{-1.9}$   & 0.187  & 17057 \\ 
GW200105\_162426 & $13.7^{+0.2}_{-0.4}$& & & $0.27^{+0.12}_{-0.11}$   & \nodata & 596309 \\ 
GW200112\_155838 & $19.8^{+0.1}_{-0.2}$& 550 & 3200 & $1.25^{+0.43}_{-0.46}$   & 0.486 & 8558 \\
GW200115\_042309 & $11.3^{+0.3}_{-0.5}$& 56 & 388 & $0.29^{+0.15}_{-0.10}$   & \nodata & 174 \\
GW200128\_022011 & $10.6^{+0.3}_{-0.4}$ & 714 & 2415 & $3.4^{+2.1}_{-1.8}$   & 0.435 & 34995 \\ 
GW200129\_065458 &  $26.8^{+0.2}_{-0.2}$& 6 & 31 & $0.90^{+0.29}_{-0.38}$   & 0.968  & 41 \\
GW200202\_154313 &  $10.8^{+0.2}_{-0.4}$& 44 & 150 & $0.41^{+0.15}_{-0.16}$   & 0.886  & 295 \\
GW200208\_130117 & $10.8^{+0.3}_{-0.4}$ & 9 & 30 & $2.23^{+1.00}_{-0.85}$   & 0.000 & 80 \\
GW200208\_222617 &  $7.4^{+1.4}_{-1.2}$& 385 & 2040 & $4.1^{+4.4}_{-1.9}$   & 0.505 & 36630 \\
GW200209\_085452 &  $9.6^{+0.4}_{-0.5}$& 217 & 877 & $3.4^{+1.9}_{-1.8}$   & 0.643 & 15681 \\
GW200210\_092255 &$8.4^{+0.5}_{-0.7}$ & 290 & 1388 & $0.94^{+0.43}_{-0.34}$   & 0.473 & 5193 \\  
GW200216\_220804 &  $8.1^{+0.4}_{-0.5}$& 727 & 2924 & $3.8^{+3.0}_{-2.0}$   & 0.861 & 64883 \\
GW200219\_094415 &  $10.7^{+0.3}_{-0.5}$& 88 & 781 & $3.4^{+1.7}_{-1.5}$   & 0.392 & 14993 \\ 
GW200220\_061928 &  $7.2^{+0.4}_{-0.7}$& 1065 & 4477 & $6.0^{+4.8}_{-3.1}$  & 0.272 & 75639 \\
GW200220\_124850 &  $8.5^{+0.3}_{-0.2}$& 855 & 3129 & $4.0^{+2.8}_{-2.2}$   & 0.500 & 65013 \\
GW200224\_222234 &  $20.0^{+0.2}_{-0.2}$& 11 & 42 & $1.71^{+0.49}_{-0.64}$   & 0.000 & 178 \\
GW200225\_060421 &  $12.5^{+0.3}_{-0.4}$& 150 & 498 & $1.15^{+0.51}_{-0.53}$  & 0.511 & 1125 \\ 
GW200302\_015811 & $10.8^{+0.3}_{-0.4}$ & 1604 & 6016 & $1.48^{+1.02}_{-0.70}$   & 0.377 & 18761 \\
GW200306\_093714 &  $7.8^{+0.4}_{-0.6}$& 965 & 3907 & $2.1^{+1.7}_{-1.1}$   & 0.524 & 35435 \\
GW200308\_173609 &  $7.1^{+0.5}_{-0.5}$& 3671 & 25292 & $5.4^{+2.7}_{-2.6}$   & 0.400 & 596309 \\
GW200311\_115853 &  $17.8^{+0.2}_{-0.2}$& 10 & 35 & $1.17^{+0.28}_{-0.40}$   & 0.821 & 81 \\ 
GW200316\_215756 &  $10.3^{+0.4}_{-0.7}$& 12 & 187 & $1.12^{+0.47}_{-0.44}$   & 0.726 & 276 \\
GW200322\_091133 &  $6.0^{+1.7}_{-1.2}$& 6250 & 28703 & $3.6^{+7.0}_{-2.0}$   & 0.475 & 596309 \\    
\enddata
\end{deluxetable*}

\begin{table}
  \tablenum{2}
  \begin{tabular}{clll}
  \hline
  Parameter & Description & Prior range & Search range \\
  \hline
  $t_g$ & Gaussian rise time (days) & $0 \le t_g < 4000$ & $5 \le t_g \le 100 $ \\
  $t_e$ & Exponential decay time (days) & $0 \le t_e < 4000$ & $10 \le t_e \le 200$\\
  $t_0$ & Time of flare peak (MJD) & $58574 \le t_0 < 59000$ & $0 \le t_0 - \mathrm{MJD}_{\rm LIGO} \le 200$ \\
  $A$ & Flare amplitude (ADU) & $ 0 < A \leq 1000$ & $\log_{10}(A / r_0 + 1) > 0.04$ \\
  $r_0$ & Baseline flux (ADU) & $r_0 < 100$ & \nodata \\
  $f$ & Additional noise term & $-8 \le \log_{10}f \le 2$ & \nodata\\
  \hline
    \end{tabular}
    \caption{A summary of the parameter ranges used in the flare modeling.}
    \label{tab:model_priors}
\end{table}

\begin{deluxetable*}{cccccccccc}
\tablenum{3}
\tablecaption{A summary of the 7 ZTF AGN flares that match LIGO/Virgo events. The name used for each AGN is its position in sexagesimal format. BH masses are taken from PyQSOFit \citep{PyQSOSFit} fits to available spectra for the sources (see Appendix~\ref{app:spectra}) using the virial mass relationship of \cite{Ho15} for H$\beta$ and \cite{Shen12} for H$\alpha$ (asterisked values). The rise and decay time scales are measured in the restframe of the AGN. $min v_k$ and $max v_k$ are the minimum and maximum kick velocities for the merged black hole. Conf. limit is percentile confidence contour within the 90\% credible volume of the event at which the AGN is located. The redshift for J053408.41+085450.6 is a photometric redshift and taken from the literature. Its spectrum does not have any broad emission features to evaluate a virial mass so a fiducial mass of $\log_{10}(M_{BH}) = 8$ is used. \label{tab:flare_summary}}
\tablewidth{0pt}
\tablehead{
\colhead{LIGO/Virgo alert ID} & \colhead{Name} & \colhead{Redshift} &
\colhead{$\log_{10}(M_{\rm BH})$} & \colhead{Conf. limit} &
\colhead{$t_g$} & \colhead{$t_e$} & \colhead{$\log_{10}(E)$} & \colhead{$\min v_{k}$}
& \colhead{$\max v_{k}$} \\
& & & \colhead{($M_\odot$)} & &(days) & (days) & \colhead{(ergs)} &
\colhead{(km\, s$^{-1}$)} & \colhead{(km\, s$^{-1}$)}
}
\startdata
GW190403\_051519 & J124942.30$+$344928.9 & 0.438 & 8.6 & 0.606 & 11.7 & 45.3 & 51.6 & 5 & 990 \\ 
GW190403\_051519 & J183412.42$+$365655.3 & 0.419 & 9.1 & 0.864 & 12.7 & 41.0 & 50.5 & 15 & 2300 \\ 
GW190424\_180648 & J181719.94$+$541910.0 & 0.234 & 8.0 & 0.099 & 12.9 & 35.6 & 51.4 & 1 & 800 \\ 
GW190514\_065416 & J124942.30$+$344928.9 & 0.438 & 8.6 & 0.754 & 11.7 & 45.3 & 51.6  & 5 & 740 \\ 
GW190514\_065416 & J224333.95$+$760619.2 & 0.353 & 8.8 & 0.664 & 11.3 & 18.3 & 50.5 & 6 & 1400 \\ 
GW190521 & J124942.30$+$344928.9 & 0.438 & 8.6 & 0.596 & 11.7 & 45.3 & 51.6  & 5 & 1300 \\ 
GW190731\_140936 & J053408.41$+$085450.6 & 0.5 & (8.0) & 0.754 & 7.6 & 27.4 & 51.0 & 1 & 990 \\ 
GW190803\_022701 & J053408.41$+$085450.6 & 0.5 & (8.0) & 0.488 & 7.6 & 27.4 & 51.0 & 1 & 920 \\ 
GW190803\_022701 & J120437.98$+$500024.0 & 0.389 & 8.0* & 0.304 & 20.2 & 47.4 & 51.5 & 2 & 780 \\ 
GW190909\_114149 & J120437.98$+$500024.0 & 0.389 & 8.0* & 0.057 & 20.2 & 47.4 & 51.5 & 2 & 1100 \\ 
GW200216\_220804 & J154342.46$+$461233.4 & 0.599 & 9.3 & 0.699 & 12.0 & 123.4 & 51.4 & 24 & 1300 \\ 
GW200220\_124850 & J154342.46$+$461233.4 & 0.599 & 9.3 & 0.113 & 12.0 & 123.4 & 51.4 & 24 & 1100 \\ 
\hline
\enddata
\end{deluxetable*}

\begin{deluxetable*}{ccccccc}
\tablenum{4}
\tablecaption{A summary of the merger parameters for the 9 LIGO/Virgo events with associated flares. \label{tab:associated_alert_summary}}
\tablewidth{0pt}
\tablehead{
\colhead{LIGO/Virgo alert ID} & Data set & \colhead{Total mass} & \colhead{$M_1$} & \colhead{$M_2$} & \colhead{Chirp mass} & \colhead{$\chi_{\rm eff}$} \\
& & \colhead{($M_\odot$)} & \colhead{($M_\odot$)} &  \colhead{($M_\odot$)} & \colhead{($M_\odot$)} &
}
\startdata
GW190403\_051519 & GWTC 2.1 & 111 & 85.0 & 20.0 & 34.0 & 0.68  \\ 
GW190424\_180648 & GWTC 2 & 73 & 40.5 & 31.8 & 31.0 & 0.13 \\ 
GW190514\_065416 & GWTC 2 & 68 & 39.0 & 28.4 & 28.5  & -0.19 \\ 
GW190521 & GWTC 2 & 164 & 95.3 & 69.0 & 69.2 & 0.03 \\ 
GW190731\_140936 & GWTC 2 & 70.1 & 41.5 & 28.8 & 29.5 & 0.06 \\ 
GW190803\_022701 & GWTC 2 & 65 & 37.3 & 27.3 & 27.3 & -0.03 \\ 
GW190909\_114149 & GWTC 2 & 75 & 45.8 & 28.3 & 30.9 & -0.06 \\ 
GW200216\_220804 & GWTC 3 & 81 & 51 & 30 & 32.9 & 0.10  \\ 
GW200220\_124850 & GWTC 3 & 67 & 28 & 39 & 28.2 & 0.10 \\ 
\hline
\enddata
\end{deluxetable*}

\begin{figure*}
   \centering
   \includegraphics[width = 0.9\textwidth]{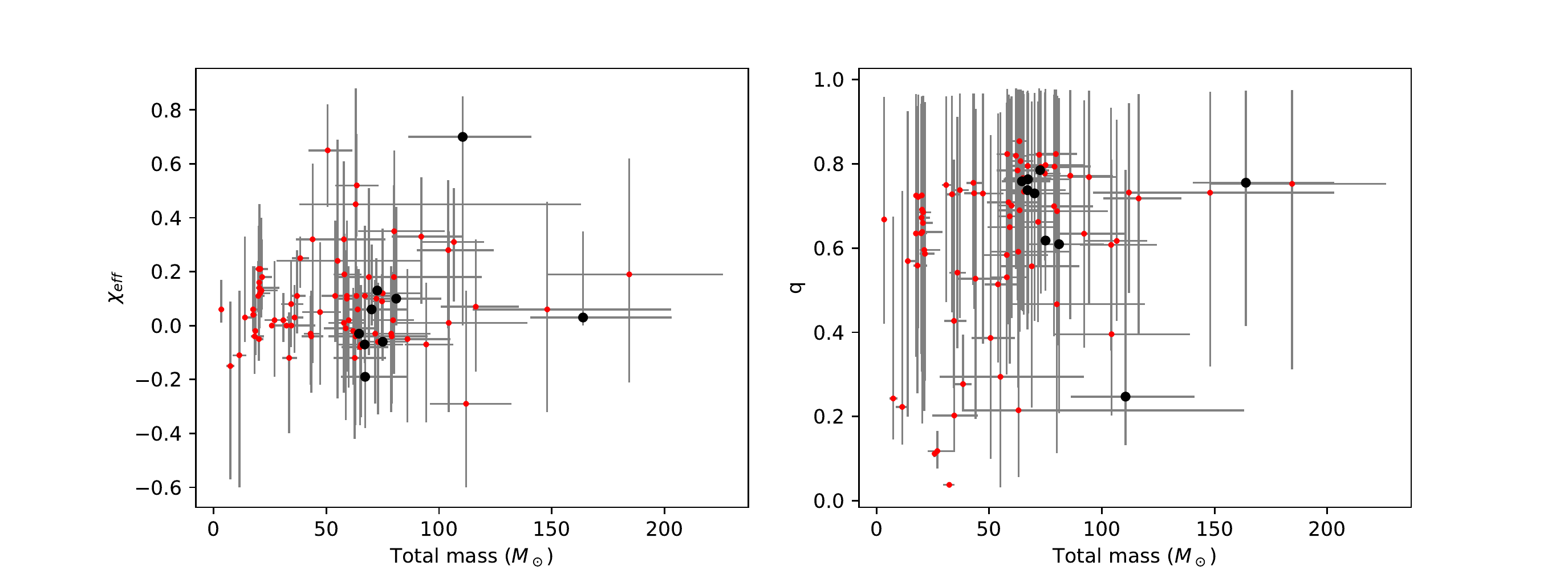}
   \caption{The distribution of merger parameters -- (left) $\chi_{\rm eff}$ vs. total mass and (right) mass ratio ($q$) vs. total mass -- for flare events (black points) and all O3 BBH mergers (red points).   }
\label{fig:popnstat}
\end{figure*}

To further assess the probability of finding a positive result, we have rerun our selection procedure with 1000 simulations of the full O3 LIGO/Virgo run. We apply a random rotation to the RA of each LIGO/Virgo error volume which allows for random spatial localization but also maintains latitude-averaged dependencies. We also assign a random time to each LIGO/Virgo event drawn from the respective ranges: MJD = 58574 -- 58756 for O3a events and MJD = 58788 -- 58930 for O3b events.  We note, however, that the localization volume of some events is large enough that a substantial range of rotations will still place a particular association found in the real data within the volume. Similarly, the post event time window of 200 days used to identify associated flares is of the order of the durations of O3a and O3b and so a temporal match is also still possible with any time shift. This makes particular associations more likely in the simulations and so we should consider these results as an upper bound on the results from the real data. Future studies could use the observing scenario skymaps simulated for LIGO O4 and O5 runs \citep{Petrov21} to create a larger background set.

From the simulations, we can determine the expected rate of an association between an AGN flare and a LIGO/Virgo detection
\[ \langle n \rangle \, = n_s \cdot p_{\rm flare} \cdot f_{\rm cover} \cdot f_{\rm prof} \cdot f_{\rm FP} \cdot  f_{\rm WISE} \cdot f_{K} \cdot f_{\rm CP},\]
where $n_s$ is the number of sources in the LIGO/Virgo error volume, $p_{\rm flare}$ is the probability that an AGN has a flare within a 200 day window, $f_{\rm cover}$ is the fraction that have at least 20 ZTF detections within 200 days of a LIGO/Virgo event, $f_{\rm prof}$ is the fraction that have the right flare profile, $f_{\rm FP}$ is the fraction where the flare is not consistent with a SNe or TDE, $f_{\rm WISE}$ is the fraction that have {\it WISE} colors consistent with an AGN, $f_{K}$ is the fraction that are known flaring sources, and $f_{\rm CP}$ is the fraction where the flare is not consistent with general activity in the source. Calculated values for these parameters are given in Table~\ref{tab:parameters}. The appropriate values for $n_s$ and $f_{\rm cover}$ for each LIGO/Virgo event are given in Table~\ref{tab:alert_summary}.

The expected number of associations from the simulated O3 LIGO/Virgo runs with 83 detections is 3.15. This is consistent with the rates-based estimate.

Another way of approaching the reality of the associations is to consider the distribution of merger parameters, e.g., chirp mass, mass ratio, etc., for the flare events against those of non-flare events. BBH mergers occurring in an AGN disk are expected to favor certain regions of merger parameter space, for example, higher masses or more extreme mass ratio ($q$) values, than mergers originating in other formation channels. We note that in this study we are dealing with small number statistics and this line of argument will have more power when the overall population statistics of BBH mergers are better determined. 

The relative distributions of total mass, mass ratio ($q = m_2 / m_1$), and $\chi_{\rm eff}$ are shown in Fig.~\ref{fig:popnstat} for both the flare events identified here and the total sample of merger events in O3. It can be seen that there is mass preference shown for the flare events with no flare event have a total mass less than 65 $M_{\odot}$ and an Anderson-Darling test between the two distributions gives a significance level of 0.005, indicating that a low likelihood that the populations come from the same distribution. The values of $q$ and $\chi_{\rm eff}$ are, however, not significantly different. If we consider 100,000 random subsamples of 9 events from the LIGO/Virgo O3 full sample then only 282 show similar characteristics with a mass preference but no difference in $q$ or $\chi_{\rm eff}$.

The association of the most likely merger event to originate in the AGN channel, namely GW190521, has already been discussed elsewhere \citep{Graham20} but we can also consider how many of the other events that would be considered as likely to originate in an AGN have an association or not. In particular, there are eight LIGO/Virgo events in O3 with total masses greater than 100 $M_{\odot}$. There is no inherent theoretical basis for this value other than as a fiducial value for hierarchical merger events favored in the AGN channel. Sufficient spatial coverage of the 90\% confidence area of an event and sufficient sampling of the region in the 200 days following the event is required to detect any potential flare. Two events have slightly too low a coverage ($\sim$25\%) to expect a detection, the others all have about 50\% coverage so if all events have a detectable flare then we should expect to get about half of the events. However, there is also an orientation effect to take into account since we expect half the events to be oriented away from us and so not observable. It is thus reasonable to expect two candidate flares associated with high mass mergers in O3 if all high mass mergers originate in the AGN channel. Indeed, we find that two of the events we have identified with associated AGN flares, GW190521 and GW190403\_051519, have total masses greater than 100 $M_\odot$.

\begin{deluxetable}{ccc}
\tablenum{5}
\tablecaption{Parameter values for determining the mean number of associations for an event. The value and uncertainty for each parameter are the mean fraction of events that pass the associated filter and its standard deviation from 1000 simulations of the full O3 LIGO//Virgo run.   \label{tab:parameters}}
\tablewidth{0pt}
\tablehead{
\colhead{Parameter} & \colhead{Value} & \colhead{Uncertainty} \\
& & }
\startdata
$p_{\rm flare}$ & 0.051 & - \\
$f_{\rm prof}$ & 0.006 & 0.017 \\
$f_{\rm FP}$ & 0.288 & 0.277 \\
$f_{\rm WISE}$ & 0.973 & 0.111 \\
$f_{K}$ & 0.962 & 0.154 \\
$f_{\rm CP}$ & 0.026 & 0.190 \\
\hline
Total & $2.12 \times 10^{-6}$ & $7.46 \times 10^{-6}$ \\
\hline
\enddata
\end{deluxetable}

\section{Discussion}
We can make simple inferences about the number of EM counterparts we should expect to observe in a study such as this. First, we should expect that EM counterparts are only in principle detectable for the fraction $f_{\rm AGN}$ of BBH mergers that originate in the AGN channel. Second, only those BBH mergers that occur in the fraction of AGN that are type-1 AGN could possibly have detectable optical signatures (i.e., unobscured AGN showing broad emission lines, corresponding to those sources that are approximately face-on to the observer according to the standard AGN unification paradigm; $f_{\rm type-1} \sim 0.5$). Third, only the fraction BBH mergers that are kicked out of the AGN on the side facing us ($f_{\rm side} \sim 0.5$) could yield a signature that would not be washed out by the optical depth of the accretion disk. Fourth, we require the AGN to lie in that fraction of the sky ($f_{\rm sky} \sim 0.5$) covered by ZTF and fifth, we require the EM counterpart to be detectable against intrinsic variability and false positive signatures, which includes signatures that may emerge on the far side of a face-on disk. 
Now suppose a fraction $f_{\rm AGN}$ of all $N_{\rm BBH}$ LIGO/Virgo-detected BBH mergers occur in AGN disks. Then, the number of EM counterparts potentially detectable by our survey $N_{\rm EM,BBH}$ is

\begin{equation}
    N_{\rm EM,BBH} \approx 3 \left( \frac{N_{\rm BBH}}{83}\right) \left( \frac{f_{\rm AGN}}{0.5}\right)\left( \frac{f_{\rm type-1}}{0.5}\right)\left( \frac{f_{\rm side}}{0.5}\right)\left( \frac{f_{\rm sky}}{0.5}\right)
\end{equation}
\noindent 
In principle, for O4 and beyond, we could increase $f_{\rm sky} \sim 1$ by including a large-sky survey in the Southern Hemisphere, thereby doubling the number of potentially detectable counterparts.

\subsection{What if our sample consists entirely of false positives?}
We have isolated some extreme variability events in AGN that occurred in LIGO/Virgo O3 publicly announced event volumes. These events are not likely intrinsic disk variability according to models of stochastic AGN variability for these (or most) AGN (see \S\ref{sec:discrim_agn}). These events are also not likely known false positives, such as microlensing events, SNe or TDEs (see S\ref{sec:discrim_other}).  However, even if the AGN channel is responsible for most of the mergers LIGO/Virgo observes, it is possible that no EM counterparts will ever be detected from this channel due to muffling of EM counterparts by optically thick disks. In this case, only a statistical approach will reveal the AGN contribution to the observed rate \citep{Imre17,Veronesi22}. If we assume that all the candidate events in Table~\ref{tab:flare_summary} are in fact drawn from a tail of rare disk variability events, then we can conclude that such events occur at a rate of  $\mathcal{O}$($10^{-6}{\rm yr}^{-1} {\rm AGN}^{-1}$). The rate of expected disk-crossing events expected is much higher than the rate here \citep{Fabj20}. The short-timescale nature of the events implies that these events either represent short-lived embedded explosive breakout from within the disk that are not SNe, or explosive events very close to the ISCO.  The possible candidates for such events can be observationally tested. For example, an off-center explosive event not associated with a kicked BBH merger must yield a temporary, asymmetric broad optical line profile on a timescale of days to weeks after the event, depending on the semi-major axis of the flaring event \citep{McK19a}. Unfortunately, we could not test this possibility for any of the  flares in our sample as they were identified long after any such signature might have developed and decayed. Nevertheless, for future GW observing runs, spectral follow-up on interesting candidate AGN GW events with associated AGN flaring can be a powerful technique for identifying off-center flaring events. 

By contrast, explosive events near the ISCO will cause the optical broad lines to reverberate symmetrically and so any change in broad optical lines that is not confidently asymmetric implies a flare origin close to the ISCO. Such flaring could be associated with disk instabilities or magnetic flux explosive release. If our sample of flaring events are due to such extreme effects near the central engine, this allows us to constrain models of magnetic field build-up around the SMBH due to accretion, as well as models of disk instabilities and the frequency of their occurrence. 

\subsection{Implications for EM follow-up in the future: O4 and beyond}
EM follow-up of GW merger events is time and effort expensive. Therefore we suggest that the International Gravitational-Wave Network\footnote{The O4 observing run will employ the LIGO, Virgo, and KAGRA gravitational wave detectors.} (IGWN) update skymaps publicly once parameterization for an individual event has settled down. In particular, since we are searching for EM counterparts that might emerge on timeframes of weeks after individual events, it would be very useful (and presumably low cost) for IGWN to automatically release public skymaps about one month after individual events.

As waveforms used in parameterizing individual events change, the resulting skymaps and error volumes can also change (even by small amounts). It would be very useful if IGWN were to publicly list waveform models used to arrive at particular parameterizations (without necessarily revealing other information about mergers).

In order to optimize follow-up (including spectroscopy) during future IGWN operating runs, it would be useful for IGWN to list `AGN possible' flags in public data releases. We recommend that such a flag correspond to multiple merger parameters including: (i) high mass; i.e., $M_{1}>50M_{\odot}$, or intermediate mass black hole (IMBH) formation events $M_{\rm tot}>100M_{\odot}$; (ii) significantly asymmetric mass ratios ($q<0.3$); and (iii) strongly misaligned spins such that a strong recoil kick ($v_k$) would be expected. By making an `AGN possible' flag multi-parameter and not tied to, e.g., IMBH formation events, it allows us to cross-check EM flare parameters with GW measurements. For example, a 'AGN possible' IGWN flag corresponding to a likely large recoil kick could help rule out a flare that is significantly delayed from the merger time as a false positive. 

We also recommend that coordinating infrastructure for transient followup, such as community alert brokers and Target and Observation Managers (TOMs), may want to maintain watchlists of AGN within the 90\% confidence volumes of IGWN skymaps. Automatic followup, e.g., spectroscopy, could then be triggered for those sources which started to exhibit flaring activity. 



\section{Conclusions}
Our picture of the accretion disk of an AGN is evolving from the simple Sunyaev-Shakirov thin disk model to a dynamic environment with encounters between gas, members of the nuclear star cluster, and clouds of stellar mass BHs in orbit around the central SMBH. In particular, AGN disks are a promising source of the stellar origin compact object mergers being detected by GW observatories. They are also the only BH merger channel where an EM counterpart must occur (whether detectable or not).

We have conducted a systematic search for possible EM counterparts in AGN to BBH object mergers detected by LIGO/Virgo in O3. We filtered out expected false positives, such as SNe and TDEs, and developed a change point algorithm to test whether specific AGN flares are consistent with stochastic variability in their hosts or are more likely to be the result of some other mechanism. We found 7 AGN flares associated with 12 merger events. This is statistically unlikely, with $p \sim 10^{-3}$. Simulation of random LIGO events and our selection procedure confirm the spatial coincidence rate.  

However, our knowledge of the phenomenology of AGN flaring is as yet incomplete: for example, we expect TDEs and SNe embedded in the accretion disk but have no real idea of what these would look like in terms of a detectable signal. We are conducting work on the set of AGN flares detected in ZTF and other large optical time domain surveys to identify categories of events and their respective rates as a way to resolve this unknown false positive issue. Detailed numerical simulations of such events, involving full magnetohydrodynamics, general relativity, and radiative transfer code are also underway by other groups which will aid the search for EM counterparts in AGN.  If we consider that at least one of the associations we have identified is real then this has significant implications for both GW and AGN physics.

\vspace{10mm}
The authors thank Will Farr and Colm Talbot for very useful discussions about LIGO/Virgo waveform choices and parameterization. MJG, BM, and KESF acknowledge the Center for Computational Astrophysics at the Flatiron Institute, New York for their hospitality and support. 

This work was supported in part by the National Science Foundation grants AST-1815034, AST-1831412, and AST-2108402, the NASA grant 16-ADAP16-0232, and Simons Foundation grant 533845. The work of DS was carried out at Jet Propulsion Laboratory, California Institute of Technology, under a contract with NASA. MWC is supported by the National Science Foundation with grant numbers PHY-2010970 and OAC-2117997.
PR acknowledges the support received from the Agence Nationale de la Recherche of the French government through the program ‘‘Investissements d’Avenir’’ (16-IDEX-0001 CAP 20-25).

This work made use of the Million Quasars Catalogue.

Based on observations obtained with the Samuel Oschin Telescope 48-inch and the 60-inch Telescope at the Palomar Observatory as part of the Zwicky Transient Facility (ZTF) project. ZTF is supported by the National Science Foundation under Grant No. AST-1440341 and a collaboration including Caltech, IPAC, the Weizmann Institute for Science, the Oskar Klein Center at Stockholm University, the University of Maryland, the University of Washington, Deutsches Elektronen-Synchrotron and Humboldt University, Los Alamos National Laboratories, the TANGO Consortium of Taiwan, the University of Wisconsin at Milwaukee, and Lawrence Berkeley National Laboratories. Operations are conducted by COO, IPAC, and UW.

The ZTF forced photometry service was funded under the Heising-Simons Foundation grant 12540303 (PI: Graham).


\vspace{5mm}
\facilities{PO:1.2m (Zwicky Transient Facility), Hale (DBSP), Keck:I (LRIS)}

\software{astropy \citep{astropy:2013,astropy:2018},
corner \citep{corner}, ligo.skymap \citep{ligo.skymap}, scikit-learn \citep{scikit-learn}
}

\appendix

\section{Gaussian process change point detection}
\label{gpchange}
A Gaussian process (GP) is a random process produced by a collection of random variables such that any finite set of those variables follows a multivariate Gaussian distribution. A GP is completely specified by a mean function, $m(\cdot) = \mathbb{E}[f(\cdot)]$, and a kernel covariance function, $k = Cov(f(\cdot), f(\cdot))$. These are parameterized with vectors ${\bm \theta_m}$ and ${\bm \theta_k}$, respectively, with ${\bm \theta} = ({\bm \theta_m}, {\bm \theta_k})$ denoting a vector of hyperparameters for a given GP.

A time series, ${\bm y}_{1:N} = \{y_{t_i}\}_{i=1}^N$, consists of a set of $N$ observations at times ${\bm t}_{1:N} = \{t_i\}_{i=1}^N$, which we model with a GP, $f$: $y_{t} = f(t) + \epsilon_{t}$, where $\epsilon_t \sim {\cal N}(\epsilon_t \mid 0, \sigma^2)$ is white Gaussian noise with a zero mean and variance $\sigma^2$. 

For a given set of $N$  we can compute the posterior distributions of function values and observations: ${\bf f}_{1:N} | {\bm t}_{1:N}, {\bm \theta} \sim {\cal N} ({\bf f}_{1:N} | {\bm \mu}, {\bf K})$ where ${\bf f}_{1:N} = f({\bm t}_{1:N}) = \{f(t_i)\}^N_{i=1}$ are function $f$ values at the given input times, ${\bm \mu} = \{\mu_i\}_{i=1}^N = \{m(t_i)\}^N_{i=1}$ are realizations of the GP mean function at the input times; 
${\bm K} = \{{\bm K}_{i,j}\}_{i,j=1}^N = \{k(t_i, t_j)\}^N_{i,j=1}$
are realizations of the GP covariance function at the input times, and
 \[ {\bf y}_{1:N}| t_{1:N}, {\bm \theta} \sim {\cal N}({\bm y}_{1:N} | {\bm \mu}, {\bm K} + \sigma^2 {\bm I}), \]
 
\noindent
where ${\bm I}$ is the identity matrix. The marginal log likelihood function of observed data is given by
\[
\log p ({\bm y}_{1:N} | {\bm t}_{1:N}, {\bm \theta} ) = -\frac{1}{2} ({\bm y}_{1:N} - {\bm \mu})^T ({\bm K} + \sigma^2 {\bm I})^{-1} ({\bm y}_{1:N} - {\bm \mu})  - \frac{1}{2} \log \mathrm{det} ({\bm K} + \sigma^2 {\bm I}) - \frac{N}{2} \log 2\pi.
\]
  
\noindent 
Estimates of the hyperparameters can be obtained by maximizing the marginal likelihood ${\bm \hat{\theta}_a} = \mathrm{\arg \max_{\theta_a}} \log p({\bm y} | {\bm t}, {\bm \theta_a})$.

A {\em change point} represents a transition between different states in a process that generates the time series, i.e., a change in the latent probability distribution of observed data. For a time series described by a GP, this can mean a change at some $t = t_{\star}$ in hyperparameter values or even a change in the mean and/or covariance functions
\begin{eqnarray}
\arraycolsep=5pt
\begin{array}{ccccc}
y_t  =  f_0(t) + \epsilon_{t}^0, & {\bm \theta} = {\bm \theta_0}, & {\bm \mu} = m_0(t), & {\bm K} = k_0 (t_i, t_j), &  t < t_\star,  \\
y_t =  f_1(t) + \epsilon_{t}^1, & {\bm \theta} = {\bm \theta_1}, & {\bm \mu} = m_1(t),  & {\bm K} = k_1 (t_i, t_j),  & t \ge t_\star. \\
\end{array}
\end{eqnarray}
 
\noindent
It can be shown that when ${\bm x}$ is a random vector distributed as a multivariate Gaussian and ${\bm A}$ is an arbitrary symmetric matrix, the quadratic form ${\bm x}^T {\bm A}^{-1} {\bm x}$ has a generalized chi-squared distribution. We can therefore consider a test statistic for change point detection, $\lambda$, assuming a null hypothesis, ${\cal H}_0$, that the GP remains unchanged during the whole observation period and an alternative claim, ${\cal H}_1$, that there exists some window, ${\bm t}_L = \{t_\star \le t < t_\star + L\}$, over which the GP has different hyperparameters or a different functional form
\begin{eqnarray}
\lambda & = &  ({\bm y}_{L} - {\bm \mu}_0)^T ({\bm K}_0 + \sigma^2{\bm I})^{-1} ({\bm y}_{L} - {\bm \mu}_0) \\
 & = & -2 \log p({\bm y}_{L} | {\bm t}_{L}, {\bm \hat{\theta}}_0) - \log \mathrm{det}({\bm K}_{0} + \sigma^2 {\bm I}) - N \log 2\pi,
\end{eqnarray}

\noindent 
evaluated for the observations ${\bm y}_L $ at times ${\bm t}_L$ within the window and where the hyperparameters ${\bm \hat{\theta}}_0$ are evaluated for the time series excluding the window range, i.e., $\{t \le t_\star, t > t_\star + L\}$. There is no closed-form expression for the distribution of $\lambda$ but we can estimate a significance level for any measured value from the distribution of $\lambda$ associated with sample observations in the window range drawn from the GP posterior distribution
\[{\bm \hat{y}}_L  | t_L , {\bm \hat{\theta}}_0 \sim {\cal N}({\bm y}_{L} | {\bm \mu}'_0, {\bm \Sigma}'_0 ) \]

\noindent
where
\begin{eqnarray}
{\bm \mu}'_0 & = & \mu_0({\bm t}_L) + {\bm k}_L^T ({\bm K}_0 + \sigma^2 {\bm I})^{-1} ({\bm y} - {\bm \mu}_0), \\
{\bm \Sigma}' _0& = &  k_0({\bm t}_L, {\bm t}_L) - {\bm k}_L^T ({\bm K}_0 + \sigma^2 {\bm I})^{-1} {\bm k}_L,\\
{\bm k}_L & =  & k_0({\bm t}_L, {\bm t}.) 
\end{eqnarray}

\noindent
We simulate an AGN light curve via a damped random walk (DRW) process characterized by a timescale $\tau$ and an amplitude $\sigma^2$. A (zero centered) data point $m_{i+1}$ at time $t_{i+1}$ is given by
\[ m_{i+1} = m_i e^{-\Delta t / \tau} + G \left[\sigma^2 (1 - e^{-2\Delta t/\tau}) \right] \]

\noindent 
where $G(s^2)$ is a Gaussian deviate with variance $s^2$ and $\Delta t = t_{i+1} - t_i$. We can add a flare of amplitude $A$ peaking at time $t = t_0$ 
with rise and decay times, $t_g$ and $t_e$, respectively to this 

\begin{eqnarray}
m(t) & = m_{DRW}(t) +  A \exp\left(-\frac{(t - t_0)^2}{2 t_g^2}\right), & t \le t_0, \\
& = m_{DRW}(t) + A \exp\left(-\frac{(t - t_0)}{t_e}\right),& t > t_0. 
\end{eqnarray}

\begin{figure*}
   \centering
   \includegraphics[width = \textwidth]{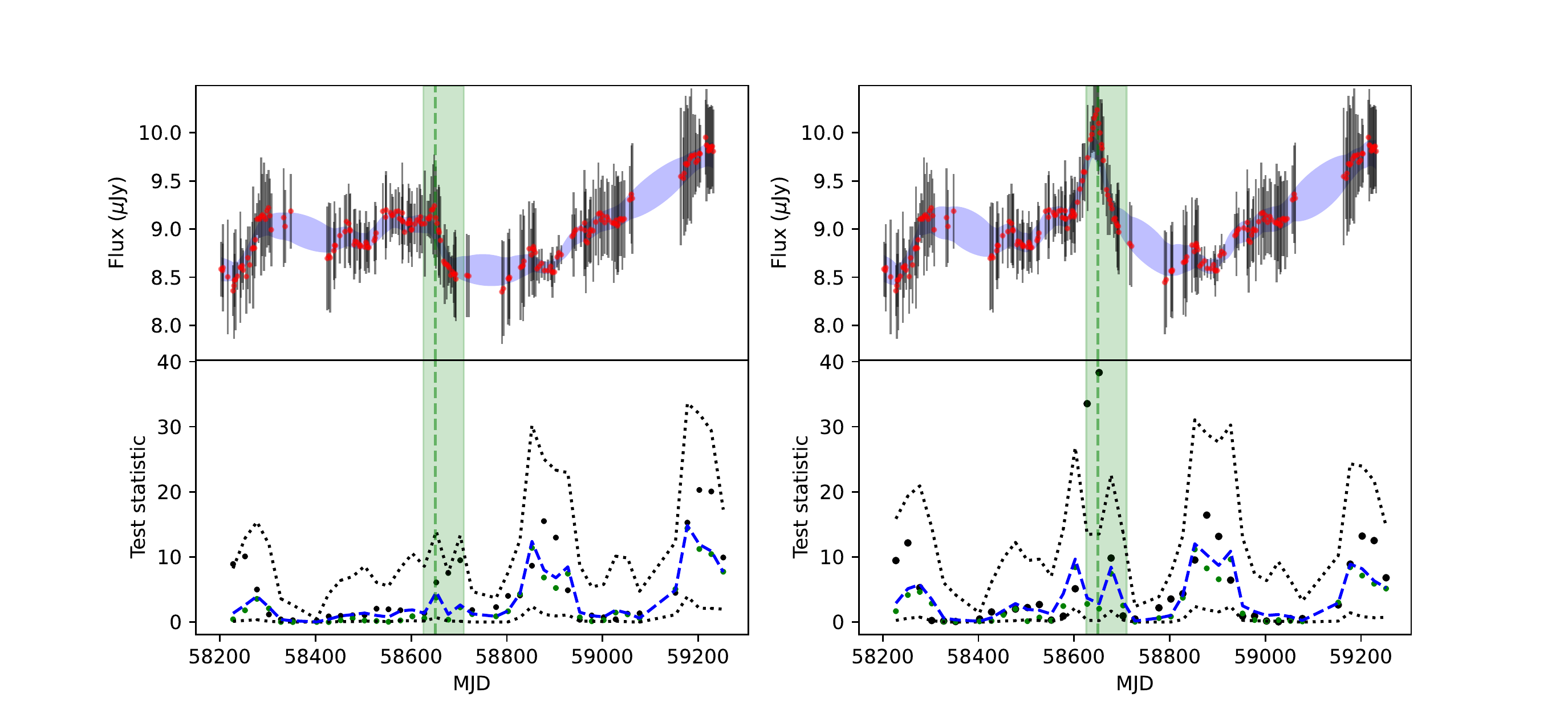}
   \caption{(left) A simulated ZTF AGN light curve from a DRW process and the change point test statistic (black points in the lower plot) as a function of time. The full GP fit to the data and predicted uncertainties (blue shaded region) are shown. In the lower plot, the dashed blue line indicates the median value of the test statistic from 1000 posterior samples at each location and the dotted lines the $0.5^{\mathrm{th}}$ and $99.5^{\mathrm{th}}$ percentiles respectively. (right) The same light curve with a flare peaking at $t = 58650$ and with a rise time of 25 days, a decay time of 60 days, and an amplitude of 10\% the median flux of the DRW model. The peak and duration of the flare are indicated in both plots by the dashed green line and shaded green region respectively. The test statistic indicates that the flare is statistically inconsistent with the DRW model.}
\label{fig:changepoint}
\end{figure*}

\noindent
Fig.~\ref{fig:changepoint} shows the test statistic for both a plain DRW model generated with observation times taken from ZTF and the same model plus a flare with an amplitude equal to 10\% of the median flux of the DRW process. A window with a width of 50 days was used and significance levels for the test statistic determined from 1000 samples drawn from the posterior at each window location. The test statistic in the vicinity of the flare peak ($t \sim t0$) is 
a statistically significant deviate indicating that this region of the light curve is not consistent with being generated by the same process as the rest of the light curve.

\noindent
Note: we assume a particular kernel for the null hypothesis; however, it may be that a given time series is not well described by that model and so the observed values of the quadratic statistic do not generally match those sampled from the posterior. The hyperparameter estimates in the window generally match those for the full time series. It is also possible that the noise modelling is insufficient.

\section{Spectra of candidate EMGW-associated AGN}
\label{app:spectra}

\begin{figure*}
   \centering
   \includegraphics[width = 0.497\textwidth]{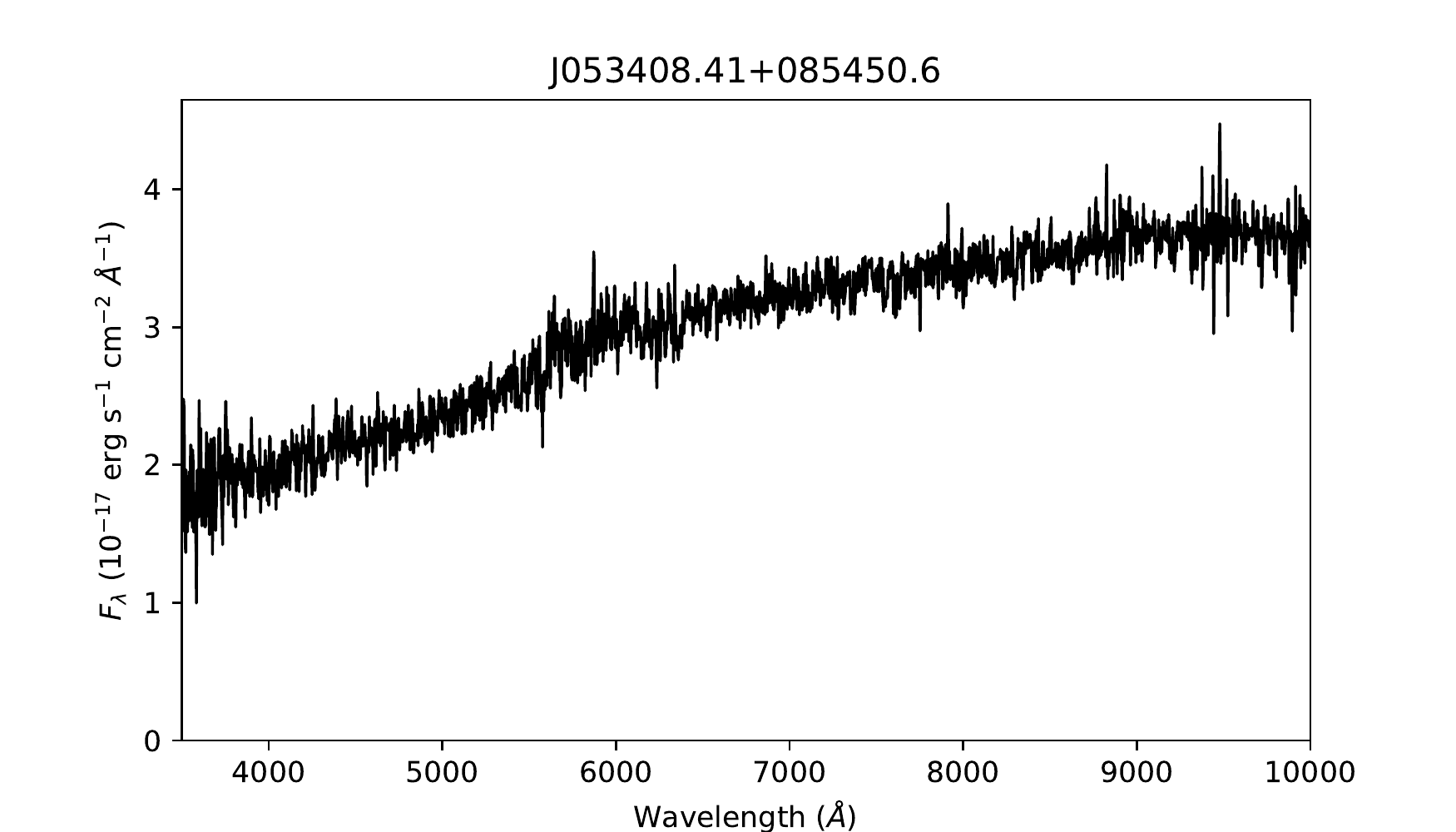}
   \includegraphics[width = 0.497\textwidth]{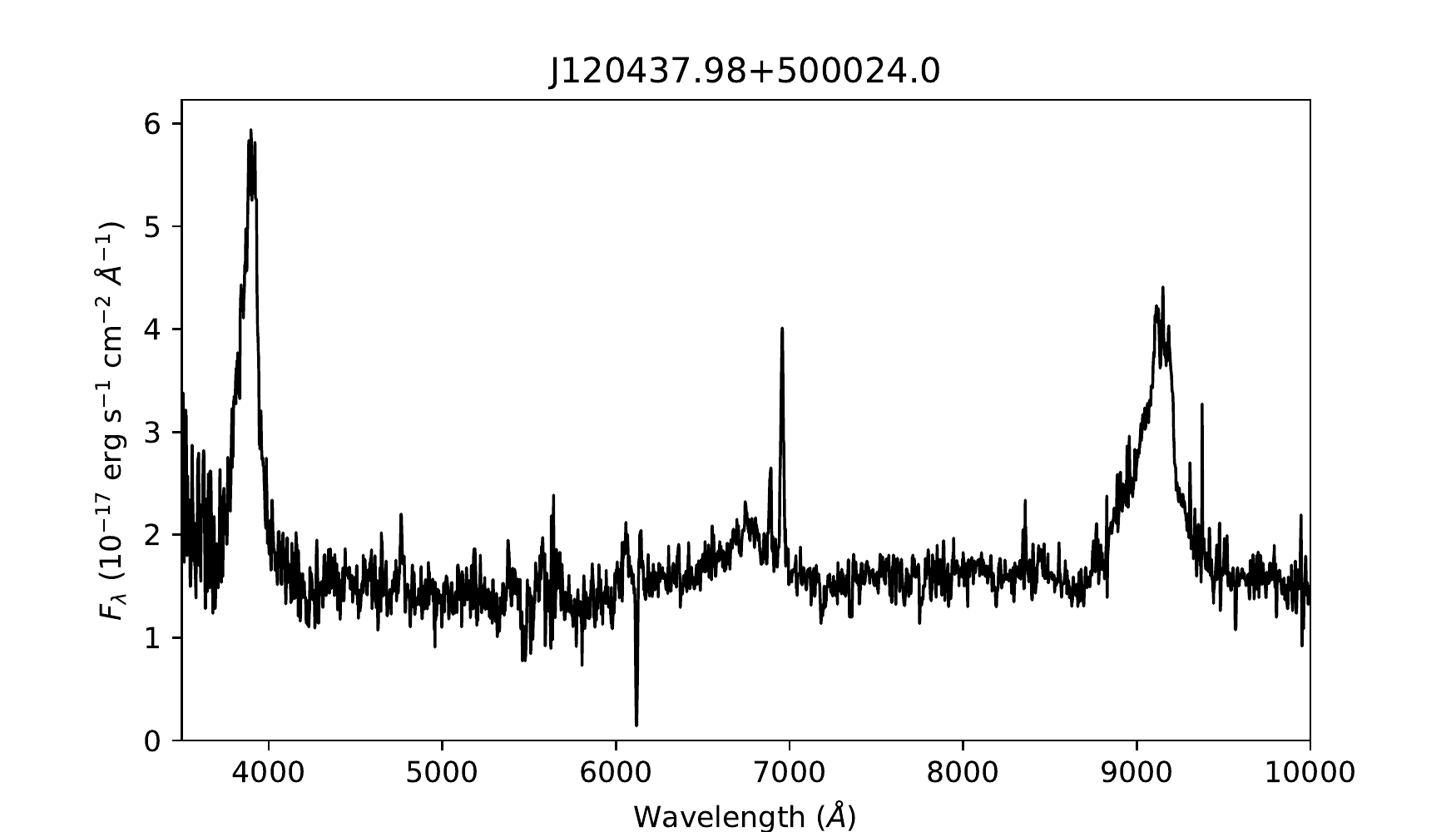}
   \includegraphics[width = 0.497\textwidth]{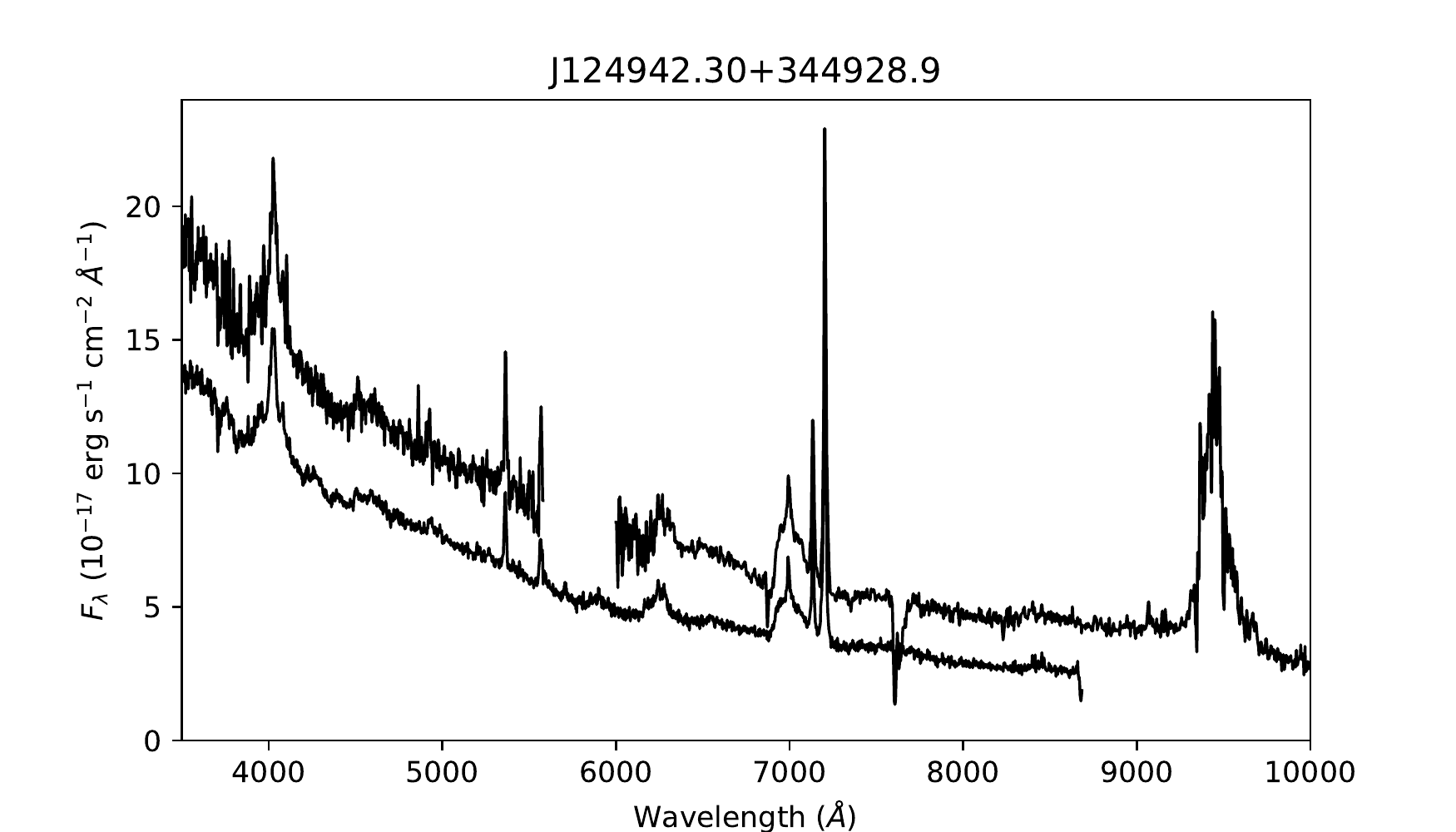}
   \includegraphics[width = 0.497\textwidth]{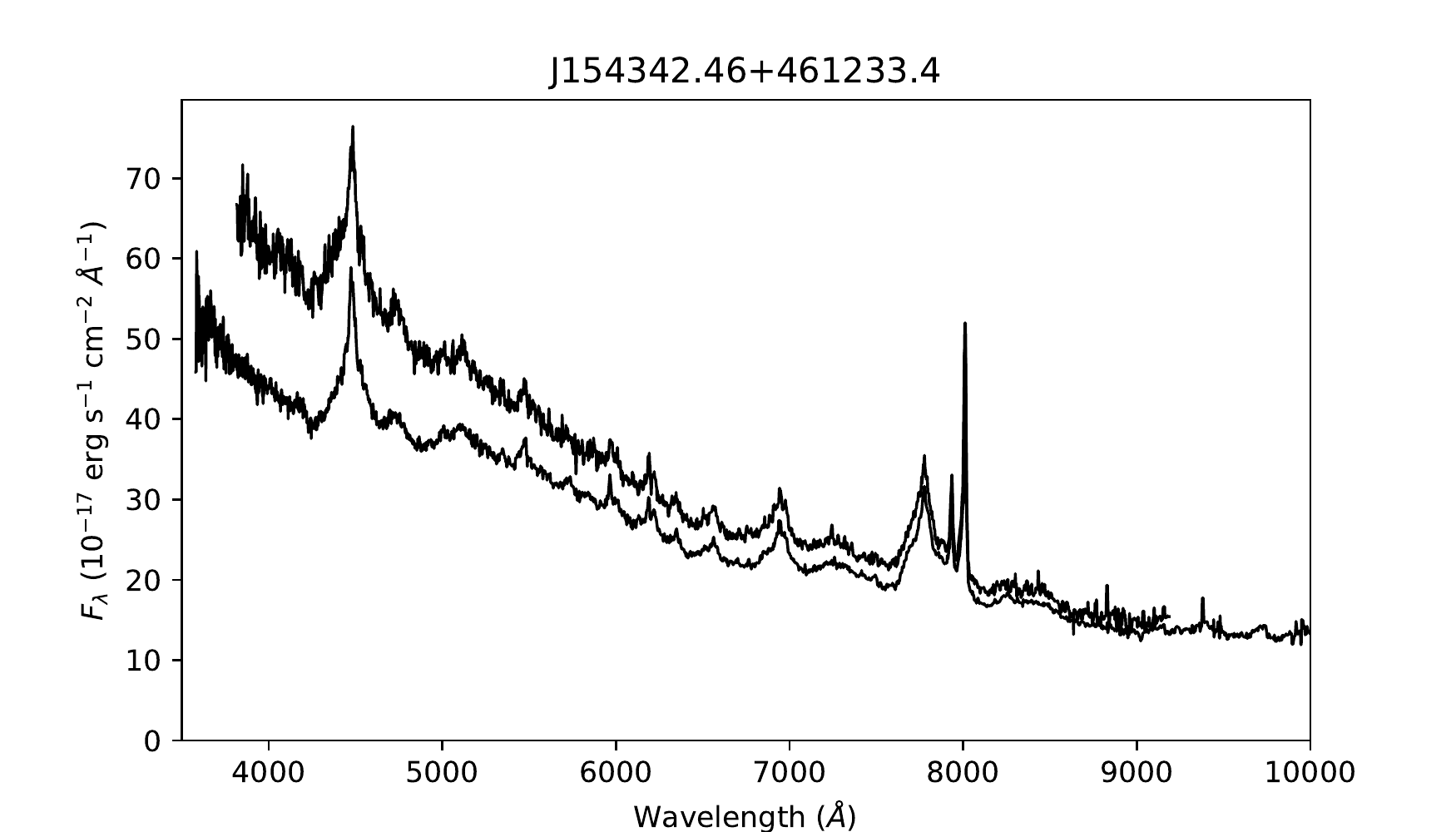}
   \includegraphics[width = 0.497\textwidth]{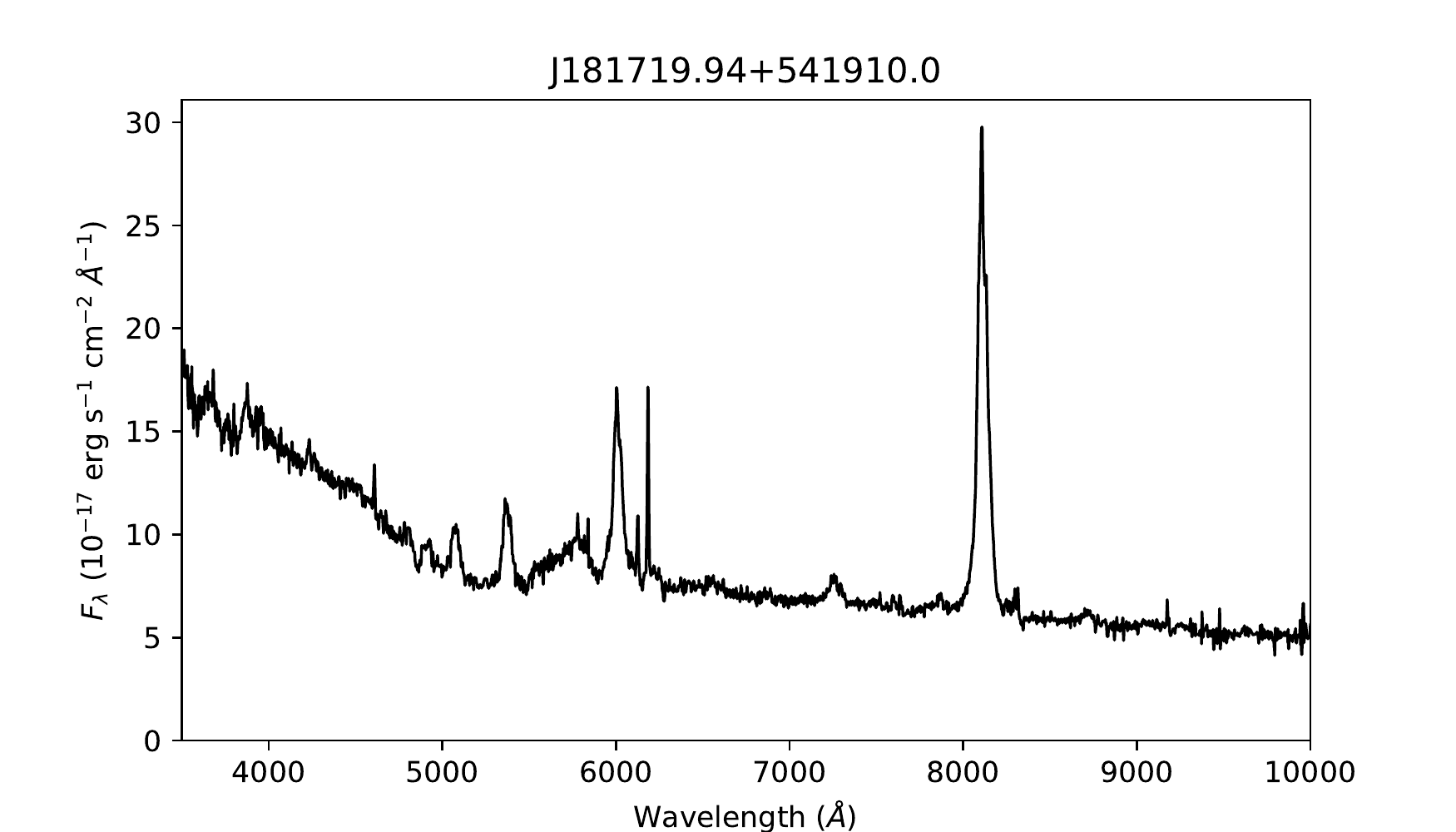}
   \includegraphics[width = 0.497\textwidth]{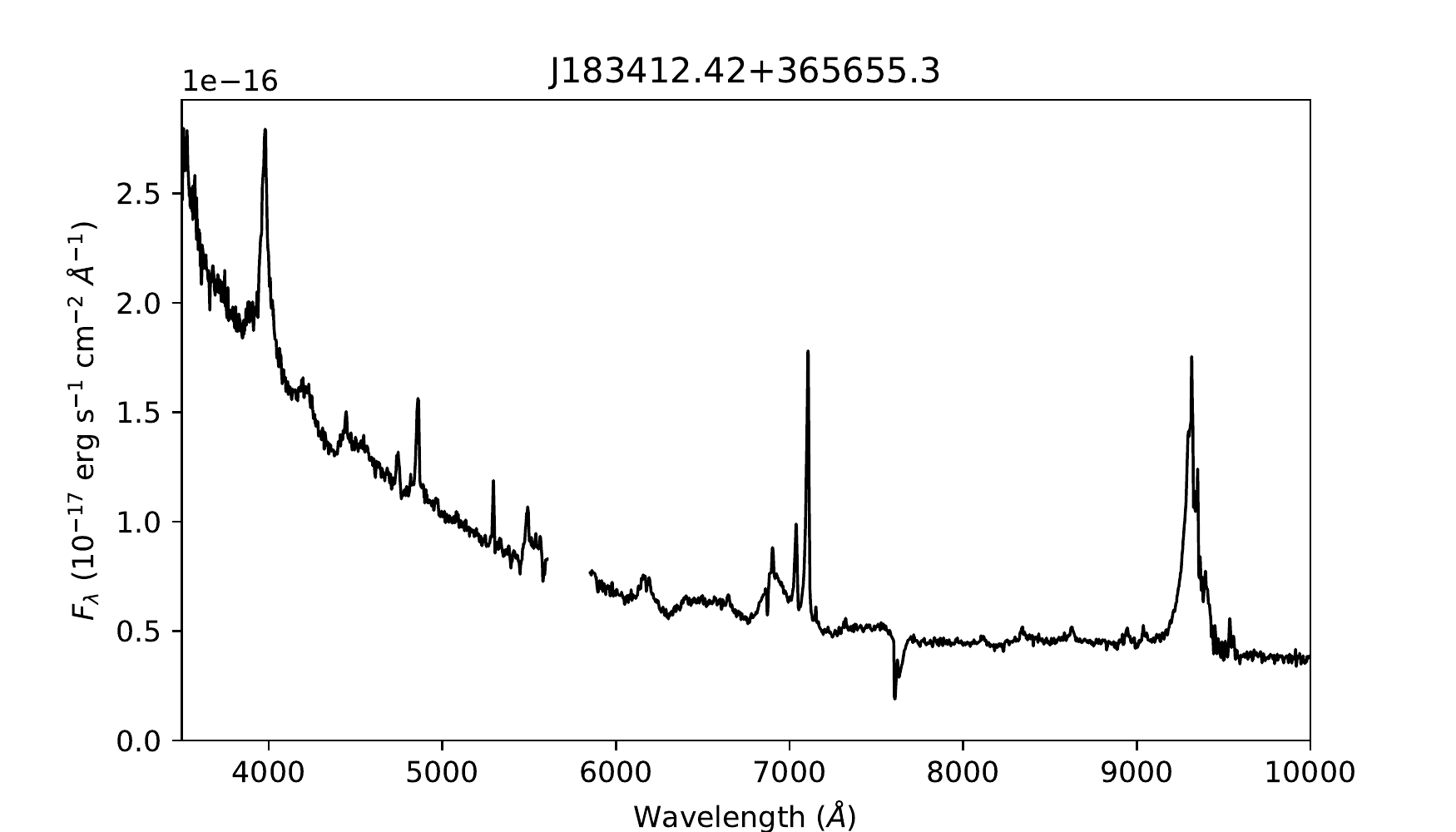}
   \includegraphics[width = 0.497\textwidth]{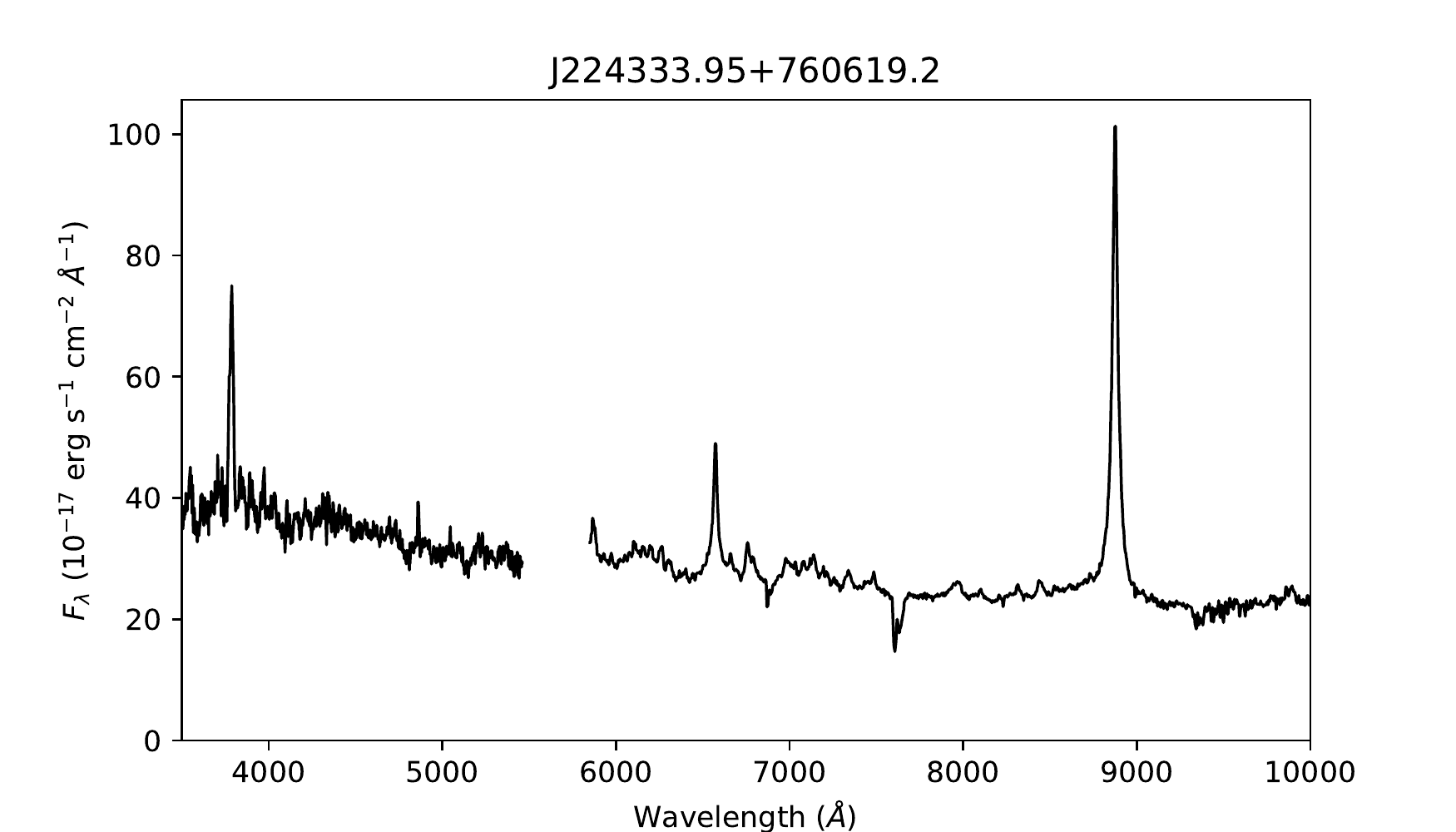}
   \caption{The spectra for the AGN associated with LIGO/Virgo events. They have been smoothed with a 5\AA median filter.}
\label{fig:spectra}
\end{figure*}

Spectroscopic observations for all candidate EMGW-associated AGN (except for J154342.46+461233.4 where there were two existing SDSS spectra) were obtained with the Low Resolution Imaging Spectrometer \citep[LRIS; ][]{Oke1995}
on the 10-m Keck {\sc I} telescope and the Double Spectrograph (DBSP) instrument on the 200-inch Palomar Hale telescope (see Table~\ref{tab:observing}). The data were reduced with standard pipelines for both instruments. The reduced spectra are shown in Fig.~\ref{fig:spectra}.

\begin{deluxetable}{ccccc}
\tablenum{6}
\tablecaption{Observing log for candidate AGN associated with LIGO/Virgo events. \label{tab:observing}}
\tablewidth{0pt}
\tablehead{
\colhead{Name} & \colhead{Date} & \colhead{Telescope} & \colhead{Instrument} & Exposure \\
&  (UT) & & & (s)}
\startdata
J053408.41$+$085450.6 & 2022 February 25 & Keck-I & LRIS & 600 \\
J120437.98$+$500024.0 & 2022 February 25 & Keck-I & LRIS & 600\\
J124942.30$+$344928.9 & 2020 January 25 & Keck-I & LRIS & 600 \\
    & 2022 May 27 & P200 & DBSP & 900 \\
J154342.46$+$461233.4 &  2003 April 02 & SDSS & \nodata & \nodata \\
 & 2017 May 05 & SDSS & \nodata & \nodata \\
J181719.94$+$541910.0 & 2022 February 25 & Keck-I & LRIS & 600 \\
J183412.42$+$365655.3 & 2021 September 10 & Keck-I & LRIS & 600 \\
J224333.95$+$760619.2 & 2022 April 28 & P200 & DBSP & 900 \\
\hline
\enddata
\end{deluxetable}

\section{Disk Exit Time Derivation}
\label{app:derivation}
We assume the binary center of mass is initially orbiting in the midplane of the AGN disk with Keplerian velocity, $v_{\rm orb}$.
At merger, the remnant experiences a kick velocity, $v_{k}$, in an arbitrary direction; the most rapid exit time will occur if the kick is directed perpendicular to the midplane of the disk. The time, $t$, to reach a height $z$ above the disk midplane will be $t=z/v_{k}$.  We assume a Gaussian atmosphere for the gas density away from the midplane, i.e. $\rho=\rho_{0} \exp(-z^2/(2H^2))$, where $\rho$ is the gas density, $\rho_{0}$ is the gas density at the midplane, $z$ is the height above the midplane and $H$ is the scale height of the atmosphere. Rearranging, we find $z=H \sqrt{-2 \ln (\rho/\rho_0)}$. 
If we say the height at which the remnant exits the disk is the height where the optical depth, $\tau$, is unity, and use the relation that $\tau \propto \rho$, the vertical distance the remnant traverses to exit is $z_{\rm exit}=H \sqrt{-2 \ln (1/\tau_{\rm mp})}$, where $\tau_{\rm mp}$ is the midplane optical depth.
Thus to find the time for the remnant to exit, we have
\begin{eqnarray}
    t_{\rm exit}&=&\frac{H \sqrt{2 \ln(\tau_{mp})}}{v_k},
\end{eqnarray}
as in Eqn.~\ref{eqn:t_exit}.

\bibliography{main}{}

\begin{thebibliography}{}
\expandafter\ifx\csname natexlab\endcsname\relax\def\natexlab#1{#1}\fi
\providecommand{\url}[1]{\href{#1}{#1}}
\providecommand{\dodoi}[1]{doi:~\href{http://doi.org/#1}{\nolinkurl{#1}}}
\providecommand{\doeprint}[1]{\href{http://ascl.net/#1}{\nolinkurl{http://ascl.net/#1}}}
\providecommand{\doarXiv}[1]{\href{https://arxiv.org/abs/#1}{\nolinkurl{https://arxiv.org/abs/#1}}}

\bibitem[{{Aasi} {et~al.}(2015)}]{2015CQGra..32g4001L}
{Aasi}, J., {et~al.} 2015, Classical and Quantum Gravity, 32, 074001,
  \dodoi{10.1088/0264-9381/32/7/074001}

\bibitem[{{Acernese} {et~al.}(2015)}]{2015CQGra..32b4001A}
{Acernese}, F., {et~al.} 2015, Classical and Quantum Gravity, 32, 024001,
  \dodoi{10.1088/0264-9381/32/2/024001}

\bibitem[{{Anand} {et~al.}(2021){Anand}, {Coughlin}, {Kasliwal}, {Bulla},
  {Ahumada}, {Sagu{\'e}s Carracedo}, {Almualla}, {Andreoni}, {Stein},
  {Foucart}, {Singer}, {Sollerman}, {Bellm}, {Bolin}, {Caballero-Garc{\'\i}a},
  {Castro-Tirado}, {Cenko}, {De}, {Dekany}, {Duev}, {Feeney}, {Fremling},
  {Goldstein}, {Golkhou}, {Graham}, {Guessoum}, {Hankins}, {Hu}, {Kong},
  {Kool}, {Kulkarni}, {Kumar}, {Laher}, {Masci}, {Mr{\'o}z}, {Nissanke},
  {Porter}, {Reusch}, {Riddle}, {Rosnet}, {Rusholme}, {Serabyn},
  {S{\'a}nchez-Ram{\'\i}rez}, {Rigault}, {Shupe}, {Smith}, {Soumagnac},
  {Walters}, \& {Valeev}}]{anand21}
{Anand}, S., {Coughlin}, M.~W., {Kasliwal}, M.~M., {et~al.} 2021, Nature
  Astronomy, 5, 46, \dodoi{10.1038/s41550-020-1183-3}

\bibitem[{{Antoni} {et~al.}(2019){Antoni}, {MacLeod}, \&
  {Ramirez-Ruiz}}]{Antoni19}
{Antoni}, A., {MacLeod}, M., \& {Ramirez-Ruiz}, E. 2019, \apj, 884,
  \dodoi{10.3847/1538-4357/ab3466}

\bibitem[{{Antonini}(2014)}]{Antonini14}
{Antonini}, F. 2014, \apj, 794, 106, \dodoi{10.1088/0004-637X/794/2/106}

\bibitem[{{Antonini} \& {Rasio}(2016)}]{AntoniniRasio16}
{Antonini}, F., \& {Rasio}, F.~A. 2016, \apj, 831, 187,
  \dodoi{10.3847/0004-637X/831/2/187}

\bibitem[{{Ashton} {et~al.}(2021){Ashton}, {Ackley}, {Hernandez}, \&
  {Piotrzkowski}}]{Ashton21}
{Ashton}, G., {Ackley}, K., {Hernandez}, I.~M., \& {Piotrzkowski}, B. 2021,
  Classical and Quantum Gravity, 38, 235004, \dodoi{10.1088/1361-6382/ac33bb}

\bibitem[{{Assef} {et~al.}(2018){Assef}, {Prieto}, {Stern}, {Cutri},
  {Eisenhardt}, {Graham}, {Jun}, {Rest}, {Flewelling}, {Kaiser}, {Kudritzki},
  \& {Waters}}]{Assef18}
{Assef}, R.~J., {Prieto}, J.~L., {Stern}, D., {et~al.} 2018, \apj, 866, 26,
  \dodoi{10.3847/1538-4357/aaddf7}

\bibitem[{{Astropy Collaboration} {et~al.}(2013){Astropy Collaboration},
  {Robitaille}, {Tollerud}, {Greenfield}, {Droettboom}, {Bray}, {Aldcroft},
  {Davis}, {Ginsburg}, {Price-Whelan}, {Kerzendorf}, {Conley}, {Crighton},
  {Barbary}, {Muna}, {Ferguson}, {Grollier}, {Parikh}, {Nair}, {Unther},
  {Deil}, {Woillez}, {Conseil}, {Kramer}, {Turner}, {Singer}, {Fox}, {Weaver},
  {Zabalza}, {Edwards}, {Azalee Bostroem}, {Burke}, {Casey}, {Crawford},
  {Dencheva}, {Ely}, {Jenness}, {Labrie}, {Lim}, {Pierfederici}, {Pontzen},
  {Ptak}, {Refsdal}, {Servillat}, \& {Streicher}}]{astropy:2013}
{Astropy Collaboration}, {Robitaille}, T.~P., {Tollerud}, E.~J., {et~al.} 2013,
  \aap, 558, A33, \dodoi{10.1051/0004-6361/201322068}

\bibitem[{{Astropy Collaboration} {et~al.}(2018){Astropy Collaboration},
  {Price-Whelan}, {Sip{\H{o}}cz}, {G{\"u}nther}, {Lim}, {Crawford}, {Conseil},
  {Shupe}, {Craig}, {Dencheva}, {Ginsburg}, {Vand erPlas}, {Bradley},
  {P{\'e}rez-Su{\'a}rez}, {de Val-Borro}, {Aldcroft}, {Cruz}, {Robitaille},
  {Tollerud}, {Ardelean}, {Babej}, {Bach}, {Bachetti}, {Bakanov}, {Bamford},
  {Barentsen}, {Barmby}, {Baumbach}, {Berry}, {Biscani}, {Boquien}, {Bostroem},
  {Bouma}, {Brammer}, {Bray}, {Breytenbach}, {Buddelmeijer}, {Burke},
  {Calderone}, {Cano Rodr{\'\i}guez}, {Cara}, {Cardoso}, {Cheedella}, {Copin},
  {Corrales}, {Crichton}, {D'Avella}, {Deil}, {Depagne}, {Dietrich}, {Donath},
  {Droettboom}, {Earl}, {Erben}, {Fabbro}, {Ferreira}, {Finethy}, {Fox},
  {Garrison}, {Gibbons}, {Goldstein}, {Gommers}, {Greco}, {Greenfield},
  {Groener}, {Grollier}, {Hagen}, {Hirst}, {Homeier}, {Horton}, {Hosseinzadeh},
  {Hu}, {Hunkeler}, {Ivezi{\'c}}, {Jain}, {Jenness}, {Kanarek}, {Kendrew},
  {Kern}, {Kerzendorf}, {Khvalko}, {King}, {Kirkby}, {Kulkarni}, {Kumar},
  {Lee}, {Lenz}, {Littlefair}, {Ma}, {Macleod}, {Mastropietro}, {McCully},
  {Montagnac}, {Morris}, {Mueller}, {Mumford}, {Muna}, {Murphy}, {Nelson},
  {Nguyen}, {Ninan}, {N{\"o}the}, {Ogaz}, {Oh}, {Parejko}, {Parley}, {Pascual},
  {Patil}, {Patil}, {Plunkett}, {Prochaska}, {Rastogi}, {Reddy Janga},
  {Sabater}, {Sakurikar}, {Seifert}, {Sherbert}, {Sherwood-Taylor}, {Shih},
  {Sick}, {Silbiger}, {Singanamalla}, {Singer}, {Sladen}, {Sooley},
  {Sornarajah}, {Streicher}, {Teuben}, {Thomas}, {Tremblay}, {Turner},
  {Terr{\'o}n}, {van Kerkwijk}, {de la Vega}, {Watkins}, {Weaver}, {Whitmore},
  {Woillez}, {Zabalza}, \& {Astropy Contributors}}]{astropy:2018}
{Astropy Collaboration}, {Price-Whelan}, A.~M., {Sip{\H{o}}cz}, B.~M., {et~al.}
  2018, \aj, 156, 123, \dodoi{10.3847/1538-3881/aabc4f}

\bibitem[{{Bailer-Jones}(2011)}]{BailerJones2011}
{Bailer-Jones}, C.~A.~L. 2011, \mnras, 411, 435,
  \dodoi{10.1111/j.1365-2966.2010.17699.x}

\bibitem[{{Bartos} {et~al.}(2017{\natexlab{a}}){Bartos}, {Haiman}, {Marka},
  {Metzger}, {Stone}, \& {Marka}}]{Imre17}
{Bartos}, I., {Haiman}, Z., {Marka}, Z., {et~al.} 2017{\natexlab{a}}, Nature
  Communications, 8, 831, \dodoi{10.1038/s41467-017-00851-7}

\bibitem[{{Bartos} {et~al.}(2017{\natexlab{b}}){Bartos}, {Kocsis}, {Haiman}, \&
  {M{\'a}rka}}]{Bartos17}
{Bartos}, I., {Kocsis}, B., {Haiman}, Z., \& {M{\'a}rka}, S.
  2017{\natexlab{b}}, \apj, 835, 165, \dodoi{10.3847/1538-4357/835/2/165}

\bibitem[{{Belczynski} {et~al.}(2010){Belczynski}, {Bulik}, {Fryer},
  {et~al.}}]{Belczynski10}
{Belczynski}, K., {Bulik}, T., {Fryer}, C.~L., {et~al.} 2010, \apj, 714, 1217,
  \dodoi{10.1088/0004-637X/714/2/1217}

\bibitem[{{Bellm} {et~al.}(2019{\natexlab{a}}){Bellm}, {Kulkarni}, {Graham},
  {et~al.}}]{Bellm19}
{Bellm}, E.~C., {Kulkarni}, S.~R., {Graham}, M.~J., {et~al.}
  2019{\natexlab{a}}, \pasp, 131, 018002, \dodoi{10.1088/1538-3873/aaecbe}

\bibitem[{{Bellm} {et~al.}(2019{\natexlab{b}}){Bellm}, {Kulkarni}, {Barlow},
  {Feindt}, {Graham}, {Goobar}, {Kupfer}, {Ngeow}, {Nugent}, {Ofek}, {Prince},
  {Riddle}, {Walters}, \& {Ye}}]{Bellm19b}
{Bellm}, E.~C., {Kulkarni}, S.~R., {Barlow}, T., {et~al.} 2019{\natexlab{b}},
  \pasp, 131, 068003, \dodoi{10.1088/1538-3873/ab0c2a}

\bibitem[{{Bellovary} {et~al.}(2016){Bellovary}, {Mac Low}, {McKernan}, \&
  {Ford}}]{Bellovary16}
{Bellovary}, J.~M., {Mac Low}, M.-M., {McKernan}, B., \& {Ford}, K.~E.~S. 2016,
  \apjl, 819, L17, \dodoi{10.3847/2041-8205/819/2/L17}

\bibitem[{{Berry} {et~al.}(2015){Berry}, {Mandel}, {Middleton},
  {et~al.}}]{Berry15}
{Berry}, C. P.~L., {Mandel}, I., {Middleton}, H., {et~al.} 2015, \apj, 804,
  114, \dodoi{10.1088/0004-637X/804/2/114}

\bibitem[{{Boone}(2019)}]{Boone19}
{Boone}, K. 2019, \aj, 158, 257, \dodoi{10.3847/1538-3881/ab5182}

\bibitem[{{Calder{\'o}n Bustillo} {et~al.}(2021){Calder{\'o}n Bustillo},
  {Leong}, {Chandra}, {McKernan}, \& {Ford}}]{Juan21}
{Calder{\'o}n Bustillo}, J., {Leong}, S. H.~W., {Chandra}, K., {McKernan}, B.,
  \& {Ford}, K.~E.~S. 2021, arXiv e-prints, arXiv:2112.12481.
\newblock \doarXiv{2112.12481}

\bibitem[{{Callister} {et~al.}(2021){Callister}, {Haster}, {Ng}, {Vitale}, \&
  {Farr}}]{Callister21}
{Callister}, T.~A., {Haster}, C.-J., {Ng}, K. K.~Y., {Vitale}, S., \& {Farr},
  W.~M. 2021, \apjl, 922, L5, \dodoi{10.3847/2041-8213/ac2ccc}

\bibitem[{{Cannizzaro} {et~al.}(2020){Cannizzaro}, {Fraser}, {Jonker},
  {et~al.}}]{Cannizzaro20}
{Cannizzaro}, G., {Fraser}, M., {Jonker}, P.~G., {et~al.} 2020, \mnras, 180,
  \dodoi{10.1093/mnras/staa186}

\bibitem[{{Chambers} {et~al.}(2016){Chambers}, {Magnier}, {Metcalfe},
  {Flewelling}, {Huber}, {Waters}, {Denneau}, {Draper}, {Farrow}, {Finkbeiner},
  {Holmberg}, {Koppenhoefer}, {Price}, {Rest}, {Saglia}, {Schlafly}, {Smartt},
  {Sweeney}, {Wainscoat}, {Burgett}, {Chastel}, {Grav}, {Heasley}, {Hodapp},
  {Jedicke}, {Kaiser}, {Kudritzki}, {Luppino}, {Lupton}, {Monet}, {Morgan},
  {Onaka}, {Shiao}, {Stubbs}, {Tonry}, {White}, {Ba{\~n}ados}, {Bell},
  {Bender}, {Bernard}, {Boegner}, {Boffi}, {Botticella}, {Calamida},
  {Casertano}, {Chen}, {Chen}, {Cole}, {Deacon}, {Frenk}, {Fitzsimmons},
  {Gezari}, {Gibbs}, {Goessl}, {Goggia}, {Gourgue}, {Goldman}, {Grant},
  {Grebel}, {Hambly}, {Hasinger}, {Heavens}, {Heckman}, {Henderson}, {Henning},
  {Holman}, {Hopp}, {Ip}, {Isani}, {Jackson}, {Keyes}, {Koekemoer}, {Kotak},
  {Le}, {Liska}, {Long}, {Lucey}, {Liu}, {Martin}, {Masci}, {McLean}, {Mindel},
  {Misra}, {Morganson}, {Murphy}, {Obaika}, {Narayan}, {Nieto-Santisteban},
  {Norberg}, {Peacock}, {Pier}, {Postman}, {Primak}, {Rae}, {Rai}, {Riess},
  {Riffeser}, {Rix}, {R{\"o}ser}, {Russel}, {Rutz}, {Schilbach}, {Schultz},
  {Scolnic}, {Strolger}, {Szalay}, {Seitz}, {Small}, {Smith}, {Soderblom},
  {Taylor}, {Thomson}, {Taylor}, {Thakar}, {Thiel}, {Thilker}, {Unger},
  {Urata}, {Valenti}, {Wagner}, {Walder}, {Walter}, {Watters}, {Werner},
  {Wood-Vasey}, \& {Wyse}}]{Chambers16}
{Chambers}, K.~C., {Magnier}, E.~A., {Metcalfe}, N., {et~al.} 2016, arXiv
  e-prints, arXiv:1612.05560.
\newblock \doarXiv{1612.05560}

\bibitem[{{Chan} {et~al.}(2019){Chan}, {Piran}, {Krolik}, \& {Saban}}]{Chan19}
{Chan}, C.-H., {Piran}, T., {Krolik}, J.~H., \& {Saban}, D. 2019, \apj, 881,
  113, \dodoi{10.3847/1538-4357/ab2b40}

\bibitem[{{Chen} {et~al.}(2022){Chen}, {Haster}, {Vitale}, {Farr}, \&
  {Isi}}]{Chen22}
{Chen}, H.-Y., {Haster}, C.-J., {Vitale}, S., {Farr}, W.~M., \& {Isi}, M. 2022,
  \mnras, 513, 2152, \dodoi{10.1093/mnras/stac989}

\bibitem[{{Condon} {et~al.}(1998){Condon}, {Cotton}, {Greisen}, {Yin},
  {Perley}, {Taylor}, \& {Broderick}}]{Condon1998}
{Condon}, J.~J., {Cotton}, W.~D., {Greisen}, E.~W., {et~al.} 1998, \aj, 115,
  1693, \dodoi{10.1086/300337}

\bibitem[{{Coughlin} {et~al.}(2019){Coughlin}, {Ahumada}, {Anand}, {De},
  {Hankins}, {Kasliwal}, {Singer}, {Bellm}, {Andreoni}, {Cenko}, {Cooke},
  {Copperwheat}, {Dugas}, {Jencson}, {Perley}, {Yu}, {Bhalerao}, {Kumar},
  {Bloom}, {Anupama}, {Ashley}, {Bagdasaryan}, {Biswas}, {Buckley}, {Burdge},
  {Cook}, {Cromer}, {Cunningham}, {D'A{\`\i}}, {Dekany}, {Delacroix},
  {Dichiara}, {Duev}, {Dutta}, {Feeney}, {Frederick}, {Gatkine}, {Ghosh},
  {Goldstein}, {Golkhou}, {Goobar}, {Graham}, {Hanayama}, {Horiuchi}, {Hung},
  {Jha}, {Kong}, {Giomi}, {Kaplan}, {Karambelkar}, {Kowalski}, {Kulkarni},
  {Kupfer}, {Masci}, {Mazzali}, {Moore}, {Mogotsi}, {Neill}, {Ngeow},
  {Mart{\'\i}nez-Palomera}, {La Parola}, {Pavana}, {Ofek}, {Patil}, {Riddle},
  {Rigault}, {Rusholme}, {Serabyn}, {Shupe}, {Sharma}, {Singh}, {Sollerman},
  {Soon}, {Staats}, {Taggart}, {Tan}, {Travouillon}, {Troja}, {Waratkar}, \&
  {Yatsu}}]{coughlin19}
{Coughlin}, M.~W., {Ahumada}, T., {Anand}, S., {et~al.} 2019, \apjl, 885, L19,
  \dodoi{10.3847/2041-8213/ab4ad8}

\bibitem[{{de Mink} \& {Mandel}(2016)}]{deMink16}
{de Mink}, S.~E., \& {Mandel}, I. 2016, \mnras, 460, 3545,
  \dodoi{10.1093/mnras/stw1219}

\bibitem[{{Fabj} {et~al.}(2020){Fabj}, {Nasim}, {Caban}, {Ford}, {McKernan}, \&
  {Bellovary}}]{Fabj20}
{Fabj}, G., {Nasim}, S.~S., {Caban}, F., {et~al.} 2020, \mnras, 499, 2608,
  \dodoi{10.1093/mnras/staa3004}

\bibitem[{{Finn} \& {Chernoff}(1993)}]{FinnChernoff93}
{Finn}, L.~S., \& {Chernoff}, D.~F. 1993, \prd, 47, 2198,
  \dodoi{10.1103/PhysRevD.47.2198}

\bibitem[{{Flesch}(2019)}]{Flesch19}
{Flesch}, E.~W. 2019, arXiv e-prints, arXiv:1912.05614.
\newblock \doarXiv{1912.05614}

\bibitem[{{Foley} {et~al.}(2011){Foley}, {Sanders}, \& {Kirshner}}]{Foley11}
{Foley}, R.~J., {Sanders}, N.~E., \& {Kirshner}, R.~P. 2011, \apj, 742, 89,
  \dodoi{10.1088/0004-637X/742/2/89}

\bibitem[{{Ford} \& {McKernan}(2021)}]{FMcK22}
{Ford}, K.~E.~S., \& {McKernan}, B. 2021, arXiv e-prints, arXiv:2109.03212.
\newblock \doarXiv{2109.03212}

\bibitem[{Foreman-Mackey(2016)}]{corner}
Foreman-Mackey, D. 2016, The Journal of Open Source Software, 1, 24,
  \dodoi{10.21105/joss.00024}

\bibitem[{{Fragione} {et~al.}(2019){Fragione}, {Leigh}, \&
  {Perna}}]{Fragione19}
{Fragione}, G., {Leigh}, N. W.~C., \& {Perna}, R. 2019, \mnras, 488, 2825,
  \dodoi{10.1093/mnras/stz1803}

\bibitem[{{Fremling} {et~al.}(2020){Fremling}, {Miller}, {Sharma}, {Dugas},
  {Perley}, {Taggart}, {Sollerman}, {Goobar}, {Graham}, {Neill}, {Nordin},
  {Rigault}, {Walters}, {Andreoni}, {Bagdasaryan}, {Belicki}, {Cannella},
  {Bellm}, {Cenko}, {De}, {Dekany}, {Frederick}, {Golkhou}, {Graham}, {Helou},
  {Ho}, {Kasliwal}, {Kupfer}, {Laher}, {Mahabal}, {Masci}, {Riddle},
  {Rusholme}, {Schulze}, {Shupe}, {Smith}, {van Velzen}, {Yan}, {Yao},
  {Zhuang}, \& {Kulkarni}}]{Fremling20}
{Fremling}, C., {Miller}, A.~A., {Sharma}, Y., {et~al.} 2020, \apj, 895, 32,
  \dodoi{10.3847/1538-4357/ab8943}

\bibitem[{{Gaia Collaboration} {et~al.}(2022){Gaia Collaboration}, {Vallenari},
  {Brown}, {Prusti}, {de Bruijne}, {Arenou}, {Babusiaux}, {Biermann},
  {Creevey}, {Ducourant}, {Evans}, {Eyer}, {Guerra}, {Hutton}, {Jordi},
  {Klioner}, {Lammers}, {Lindegren}, {Luri}, {Mignard}, {Panem}, {Pourbaix},
  {Randich}, {Sartoretti}, {Soubiran}, {Tanga}, {Walton}, {Bailer-Jones},
  {Bastian}, {Drimmel}, {Jansen}, {Katz}, {Lattanzi}, {van Leeuwen}, {Bakker},
  {Cacciari}, {Casta{\~n}eda}, {De Angeli}, {Fabricius}, {Fouesneau},
  {Fr{\'e}mat}, {Galluccio}, {Guerrier}, {Heiter}, {Masana}, {Messineo},
  {Mowlavi}, {Nicolas}, {Nienartowicz}, {Pailler}, {Panuzzo}, {Riclet}, {Roux},
  {Seabroke}, {Sordo{\o}rcit}, {Th{\'e}venin}, {Gracia-Abril}, {Portell},
  {Teyssier}, {Altmann}, {Andrae}, {Audard}, {Bellas-Velidis}, {Benson},
  {Berthier}, {Blomme}, {Burgess}, {Busonero}, {Busso}, {C{\'a}novas}, {Carry},
  {Cellino}, {Cheek}, {Clementini}, {Damerdji}, {Davidson}, {de Teodoro},
  {Nu{\~n}ez Campos}, {Delchambre}, {Dell'Oro}, {Esquej},
  {Fern{\'a}ndez-Hern{\'a}ndez}, {Fraile}, {Garabato}, {Garc{\'\i}a-Lario},
  {Gosset}, {Haigron}, {Halbwachs}, {Hambly}, {Harrison}, {Hern{\'a}ndez},
  {Hestroffer}, {Hodgkin}, {Holl}, {Jan{\ss}en}, {Jevardat de Fombelle},
  {Jordan}, {Krone-Martins}, {Lanzafame}, {L{\"o}ffler}, {Marchal}, {Marrese},
  {Moitinho}, {Muinonen}, {Osborne}, {Pancino}, {Pauwels}, {Recio-Blanco},
  {Reyl{\'e}}, {Riello}, {Rimoldini}, {Roegiers}, {Rybizki}, {Sarro}, {Siopis},
  {Smith}, {Sozzetti}, {Utrilla}, {van Leeuwen}, {Abbas}, {{\'A}brah{\'a}m},
  {Abreu Aramburu}, {Aerts}, {Aguado}, {Ajaj}, {Aldea-Montero}, {Altavilla},
  {{\'A}lvarez}, {Alves}, {Anders}, {Anderson}, {Anglada Varela}, {Antoja},
  {Baines}, {Baker}, {Balaguer-N{\'u}{\~n}ez}, {Balbinot}, {Balog}, {Barache},
  {Barbato}, {Barros}, {Barstow}, {Bartolom{\'e}}, {Bassilana}, {Bauchet},
  {Becciani}, {Bellazzini}, {Berihuete}, {Bernet}, {Bertone}, {Bianchi},
  {Binnenfeld}, {Blanco-Cuaresma}, {Blazere}, {Boch}, {Bombrun}, {Bossini},
  {Bouquillon}, {Bragaglia}, {Bramante}, {Breedt}, {Bressan}, {Brouillet},
  {Brugaletta}, {Bucciarelli}, {Burlacu}, {Butkevich}, {Buzzi}, {Caffau},
  {Cancelliere}, {Cantat-Gaudin}, {Carballo}, {Carlucci}, {Carnerero},
  {Carrasco}, {Casamiquela}, {Castellani}, {Castro-Ginard}, {Chaoul},
  {Charlot}, {Chemin}, {Chiaramida}, {Chiavassa}, {Chornay}, {Comoretto},
  {Contursi}, {Cooper}, {Cornez}, {Cowell}, {Crifo}, {Cropper}, {Crosta},
  {Crowley}, {Dafonte}, {Dapergolas}, {David}, {David}, {de Laverny}, {De
  Luise}, {De March}, {De Ridder}, {de Souza}, {de Torres}, {del Peloso}, {del
  Pozo}, {Delbo}, {Delgado}, {Delisle}, {Demouchy}, {Dharmawardena}, {Di
  Matteo}, {Diakite}, {Diener}, {Distefano}, {Dolding}, {Edvardsson}, {Enke},
  {Fabre}, {Fabrizio}, {Faigler}, {Fedorets}, {Fernique}, {Fienga}, {Figueras},
  {Fournier}, {Fouron}, {Fragkoudi}, {Gai}, {Garcia-Gutierrez},
  {Garcia-Reinaldos}, {Garc{\'\i}a-Torres}, {Garofalo}, {Gavel}, {Gavras},
  {Gerlach}, {Geyer}, {Giacobbe}, {Gilmore}, {Girona}, {Giuffrida}, {Gomel},
  {Gomez}, {Gonz{\'a}lez-N{\'u}{\~n}ez}, {Gonz{\'a}lez-Santamar{\'\i}a},
  {Gonz{\'a}lez-Vidal}, {Granvik}, {Guillout}, {Guiraud},
  {Guti{\'e}rrez-S{\'a}nchez}, {Guy}, {Hatzidimitriou}, {Hauser}, {Haywood},
  {Helmer}, {Helmi}, {Sarmiento}, {Hidalgo}, {Hilger}, {H{\l}adczuk}, {Hobbs},
  {Holland}, {Huckle}, {Jardine}, {Jasniewicz}, {Jean-Antoine Piccolo},
  {Jim{\'e}nez-Arranz}, {Jorissen}, {Juaristi Campillo}, {Julbe}, {Karbevska},
  {Kervella}, {Khanna}, {Kontizas}, {Kordopatis}, {Korn}, {K{\'o}sp{\'a}l},
  {Kostrzewa-Rutkowska}, {Kruszy{\'n}ska}, {Kun}, {Laizeau}, {Lambert},
  {Lanza}, {Lasne}, {Le Campion}, {Lebreton}, {Lebzelter}, {Leccia}, {Leclerc},
  {Lecoeur-Taibi}, {Liao}, {Licata}, {Lindstr{\o}m}, {Lister}, {Livanou},
  {Lobel}, {Lorca}, {Loup}, {Madrero Pardo}, {Magdaleno Romeo}, {Managau},
  {Mann}, {Manteiga}, {Marchant}, {Marconi}, {Marcos}, {Marcos Santos},
  {Mar{\'\i}n Pina}, {Marinoni}, {Marocco}, {Marshall}, {Polo},
  {Mart{\'\i}n-Fleitas}, {Marton}, {Mary}, {Masip}, {Massari},
  {Mastrobuono-Battisti}, {Mazeh}, {McMillan}, {Messina}, {Michalik}, {Millar},
  {Mints}, {Molina}, {Molinaro}, {Moln{\'a}r}, {Monari}, {Mongui{\'o}},
  {Montegriffo}, {Montero}, {Mor}, {Mora}, {Morbidelli}, {Morel}, {Morris},
  {Muraveva}, {Murphy}, {Musella}, {Nagy}, {Noval}, {Oca{\~n}a}, {Ogden},
  {Ordenovic}, {Osinde}, {Pagani}, {Pagano}, {Palaversa}, {Palicio},
  {Pallas-Quintela}, {Panahi}, {Payne-Wardenaar}, {Pe{\~n}alosa Esteller},
  {Penttil{\"a}}, {Pichon}, {Piersimoni}, {Pineau}, {Plachy}, {Plum}, {Poggio},
  {Pr{\v{s}}a}, {Pulone}, {Racero}, {Ragaini}, {Rainer}, {Raiteri}, {Rambaux},
  {Ramos}, {Ramos-Lerate}, {Re Fiorentin}, {Regibo}, {Richards}, {Rios Diaz},
  {Ripepi}, {Riva}, {Rix}, {Rixon}, {Robichon}, {Robin}, {Robin}, {Roelens},
  {Rogues}, {Rohrbasser}, {Romero-G{\'o}mez}, {Rowell}, {Royer}, {Ruz Mieres},
  {Rybicki}, {Sadowski}, {S{\'a}ez N{\'u}{\~n}ez}, {Sagrist{\`a} Sell{\'e}s},
  {Sahlmann}, {Salguero}, {Samaras}, {Sanchez Gimenez}, {Sanna},
  {Santove{\~n}a}, {Sarasso}, {Schultheis}, {Sciacca}, {Segol}, {Segovia},
  {S{\'e}gransan}, {Semeux}, {Shahaf}, {Siddiqui}, {Siebert}, {Siltala},
  {Silvelo}, {Slezak}, {Slezak}, {Smart}, {Snaith}, {Solano}, {Solitro},
  {Souami}, {Souchay}, {Spagna}, {Spina}, {Spoto}, {Steele},
  {Steidelm{\"u}ller}, {Stephenson}, {S{\"u}veges}, {Surdej}, {Szabados},
  {Szegedi-Elek}, {Taris}, {Taylo}, {Teixeira}, {Tolomei}, {Tonello}, {Torra},
  {Torra}, {Torralba Elipe}, {Trabucchi}, {Tsounis}, {Turon}, {Ulla}, {Unger},
  {Vaillant}, {van Dillen}, {van Reeven}, {Vanel}, {Vecchiato}, {Viala},
  {Vicente}, {Voutsinas}, {Weiler}, {Wevers}, {Wyrzykowski}, {Yoldas}, {Yvard},
  {Zhao}, {Zorec}, {Zucker}, \& {Zwitter}}]{GaiaDR3}
{Gaia Collaboration}, {Vallenari}, A., {Brown}, A.~G.~A., {et~al.} 2022, arXiv
  e-prints, arXiv:2208.00211.
\newblock \doarXiv{2208.00211}

\bibitem[{{Gerosa} \& {Berti}(2019)}]{Gerosa19}
{Gerosa}, D., \& {Berti}, E. 2019, \prd, 100, 041301,
  \dodoi{10.1103/PhysRevD.100.041301}

\bibitem[{{Gerosa} \& {Fishbach}(2021)}]{GerosaFishbach21}
{Gerosa}, D., \& {Fishbach}, M. 2021, Nature Astronomy, 5, 749,
  \dodoi{10.1038/s41550-021-01398-w}

\bibitem[{{Graham} {et~al.}(2017){Graham}, {Djorgovski}, {Drake}, {Stern},
  {Mahabal}, {Glikman}, {Larson}, \& {Christensen}}]{Graham17}
{Graham}, M.~J., {Djorgovski}, S.~G., {Drake}, A.~J., {et~al.} 2017, \mnras,
  470, 4112, \dodoi{10.1093/mnras/stx1456}

\bibitem[{{Graham} {et~al.}(2019){Graham}, {Kulkarni}, {Bellm},
  {et~al.}}]{Graham19}
{Graham}, M.~J., {Kulkarni}, S.~R., {Bellm}, E.~C., {et~al.} 2019, \pasp, 131,
  078001, \dodoi{10.1088/1538-3873/ab006c}

\bibitem[{{Graham} {et~al.}(2020){Graham}, {Ford}, {McKernan}, {Ross}, {Stern},
  {Burdge}, {Coughlin}, {Djorgovski}, {Drake}, {Duev}, {Kasliwal}, {Mahabal},
  {van Velzen}, {Belecki}, {Bellm}, {Burruss}, {Cenko}, {Cunningham}, {Helou},
  {Kulkarni}, {Masci}, {Prince}, {Reiley}, {Rodriguez}, {Rusholme}, {Smith}, \&
  {Soumagnac}}]{Graham20}
{Graham}, M.~J., {Ford}, K.~E.~S., {McKernan}, B., {et~al.} 2020, \prl, 124,
  251102, \dodoi{10.1103/PhysRevLett.124.251102}

\bibitem[{{Guo} {et~al.}(2018){Guo}, {Shen}, \& {Wang}}]{PyQSOSFit}
{Guo}, H., {Shen}, Y., \& {Wang}, S. 2018, {PyQSOFit: Python code to fit the
  spectrum of quasars}, Astrophysics Source Code Library.
\newblock \doeprint{1809.008}

\bibitem[{{Hammerstein} {et~al.}(2021){Hammerstein}, {Gezari}, {van Velzen},
  {Cenko}, {Roth}, {Ward}, {Frederick}, {Hung}, {Graham}, {Foley}, {Bellm},
  {Cannella}, {Drake}, {Kupfer}, {Laher}, {Mahabal}, {Masci}, {Riddle},
  {Rojas-Bravo}, \& {Smith}}]{Hammerstein21}
{Hammerstein}, E., {Gezari}, S., {van Velzen}, S., {et~al.} 2021, \apjl, 908,
  L20, \dodoi{10.3847/2041-8213/abdcb4}

\bibitem[{{Ho} \& {Kim}(2015)}]{Ho15}
{Ho}, L.~C., \& {Kim}, M. 2015, \apj, 809, 123,
  \dodoi{10.1088/0004-637X/809/2/123}

\bibitem[{{Jiang} {et~al.}(2019){Jiang}, {Stone}, \& {Davis}}]{YanFei19}
{Jiang}, Y.-F., {Stone}, J.~M., \& {Davis}, S.~W. 2019, \apj, 880, 67,
  \dodoi{10.3847/1538-4357/ab29ff}

\bibitem[{{Kasen} \& {Bildsten}(2010)}]{Kasen10}
{Kasen}, D., \& {Bildsten}, L. 2010, \apj, 717, 245,
  \dodoi{10.1088/0004-637X/717/1/245}

\bibitem[{{Kasliwal} {et~al.}(2020){Kasliwal}, {Anand}, {Ahumada}, {Stein},
  {Carracedo}, {Andreoni}, {Coughlin}, {Singer}, {Kool}, {De}, {Kumar},
  {AlMualla}, {Yao}, {Bulla}, {Dobie}, {Reusch}, {Perley}, {Cenko}, {Bhalerao},
  {Kaplan}, {Sollerman}, {Goobar}, {Copperwheat}, {Bellm}, {Anupama}, {Corsi},
  {Nissanke}, {Agudo}, {Bagdasaryan}, {Barway}, {Belicki}, {Bloom}, {Bolin},
  {Buckley}, {Burdge}, {Burruss}, {Caballero-Garc{\'\i}a}, {Cannella},
  {Castro-Tirado}, {Cook}, {Cooke}, {Cunningham}, {Dahiwale}, {Deshmukh},
  {Dichiara}, {Duev}, {Dutta}, {Feeney}, {Franckowiak}, {Frederick},
  {Fremling}, {Gal-Yam}, {Gatkine}, {Ghosh}, {Goldstein}, {Golkhou}, {Graham},
  {Graham}, {Hankins}, {Helou}, {Hu}, {Ip}, {Jaodand}, {Karambelkar}, {Kong},
  {Kowalski}, {Khandagale}, {Kulkarni}, {Kumar}, {Laher}, {Li}, {Mahabal},
  {Masci}, {Miller}, {Mogotsi}, {Mohite}, {Mooley}, {Mroz}, {Newman}, {Ngeow},
  {Oates}, {Patil}, {Pandey}, {Pavana}, {Pian}, {Riddle},
  {S{\'a}nchez-Ram{\'\i}rez}, {Sharma}, {Singh}, {Smith}, {Soumagnac},
  {Taggart}, {Tan}, {Tzanidakis}, {Troja}, {Valeev}, {Walters}, {Waratkar},
  {Webb}, {Yu}, {Zhang}, {Zhou}, \& {Zolkower}}]{kasliwal20}
{Kasliwal}, M.~M., {Anand}, S., {Ahumada}, T., {et~al.} 2020, \apj, 905, 145,
  \dodoi{10.3847/1538-4357/abc335}

\bibitem[{{Kasliwal} {et~al.}(2015){Kasliwal}, {Vogeley}, \&
  {Richards}}]{Kasliwal15}
{Kasliwal}, V.~P., {Vogeley}, M.~S., \& {Richards}, G.~T. 2015, \mnras, 451,
  4328, \dodoi{10.1093/mnras/stv1230}

\bibitem[{{Kimura} {et~al.}(2021){Kimura}, {Murase}, \& {Bartos}}]{Kimura21}
{Kimura}, S.~S., {Murase}, K., \& {Bartos}, I. 2021, \apj, 916, 111,
  \dodoi{10.3847/1538-4357/ac0535}

\bibitem[{{Krolik}(1999)}]{Krolik99}
{Krolik}, J.~H. 1999, {Active galactic nuclei : from the central black hole to
  the galactic environment}

\bibitem[{{Lawrence} {et~al.}(2016){Lawrence}, {Bruce}, {MacLeod}, {Gezari},
  {Elvis}, {Ward}, {Smartt}, {Smith}, {Wright}, {Fraser}, {Marshall}, {Kaiser},
  {Burgett}, {Magnier}, {Tonry}, {Chambers}, {Wainscoat}, {Waters}, {Price},
  {Metcalfe}, {Valenti}, {Kotak}, {Mead}, {Inserra}, {Chen}, \&
  {Soderberg}}]{Lawrence16}
{Lawrence}, A., {Bruce}, A.~G., {MacLeod}, C., {et~al.} 2016, \mnras, 463, 296,
  \dodoi{10.1093/mnras/stw1963}

\bibitem[{{LIGO Scientific Collaboration} \& {Virgo
  Collaboration}(2019)}]{Abbott19}
{LIGO Scientific Collaboration}, \& {Virgo Collaboration}. 2019, \apjl, 882,
  L24, \dodoi{10.3847/2041-8213/ab3800}

\bibitem[{{Mapelli}(2021)}]{Mapelli21}
{Mapelli}, M. 2021, in Handbook of Gravitational Wave Astronomy, 16,
  \dodoi{10.1007/978-981-15-4702-7\_16-1}

\bibitem[{{Masci} {et~al.}(2019){Masci}, {Laher}, {Rusholme}, {Shupe}, {Groom},
  {Surace}, {Jackson}, {Monkewitz}, {Beck}, {Flynn}, {Terek}, {Landry},
  {Hacopians}, {Desai}, {Howell}, {Brooke}, {Imel}, {Wachter}, {Ye}, {Lin},
  {Cenko}, {Cunningham}, {Rebbapragada}, {Bue}, {Miller}, {Mahabal}, {Bellm},
  {Patterson}, {Juri{\'c}}, {Golkhou}, {Ofek}, {Walters}, {Graham}, {Kasliwal},
  {Dekany}, {Kupfer}, {Burdge}, {Cannella}, {Barlow}, {Van Sistine}, {Giomi},
  {Fremling}, {Blagorodnova}, {Levitan}, {Riddle}, {Smith}, {Helou}, {Prince},
  \& {Kulkarni}}]{Masci19}
{Masci}, F.~J., {Laher}, R.~R., {Rusholme}, B., {et~al.} 2019, \pasp, 131,
  018003, \dodoi{10.1088/1538-3873/aae8ac}

\bibitem[{{Matthews} \& {Sandage}(1963)}]{Matthews63}
{Matthews}, T.~A., \& {Sandage}, A.~R. 1963, \apj, 138, 30,
  \dodoi{10.1086/147615}

\bibitem[{{McKernan} {et~al.}(2019){McKernan}, {Ford}, {Bartos},
  {et~al.}}]{McK19a}
{McKernan}, B., {Ford}, K.~E.~S., {Bartos}, I., {et~al.} 2019, \apjl, 884, L50,
  \dodoi{10.3847/2041-8213/ab4886}

\bibitem[{{McKernan} {et~al.}(2018){McKernan}, {Ford}, {Bellovary},
  {et~al.}}]{McK18}
{McKernan}, B., {Ford}, K.~E.~S., {Bellovary}, J., {et~al.} 2018, \apj, 866,
  66, \dodoi{10.3847/1538-4357/aadae5}

\bibitem[{{McKernan} {et~al.}(2022{\natexlab{a}}){McKernan}, {Ford},
  {Callister}, {Farr}, {O'Shaughnessy}, {Smith}, {Thrane}, \& {Vajpeyi}}]{qX21}
{McKernan}, B., {Ford}, K.~E.~S., {Callister}, T., {et~al.} 2022{\natexlab{a}},
  \mnras, 514, 3886, \dodoi{10.1093/mnras/stac1570}

\bibitem[{{McKernan} {et~al.}(2022{\natexlab{b}}){McKernan}, {Ford},
  {Cantiello}, {Graham}, {Jermyn}, {Leigh}, {Ryu}, \& {Stern}}]{Starfall21}
{McKernan}, B., {Ford}, K.~E.~S., {Cantiello}, M., {et~al.} 2022{\natexlab{b}},
  \mnras, 514, 4102, \dodoi{10.1093/mnras/stac1310}

\bibitem[{{McKernan} {et~al.}(2014){McKernan}, {Ford}, {Kocsis}, {Lyra}, \&
  {Winter}}]{McK14}
{McKernan}, B., {Ford}, K.~E.~S., {Kocsis}, B., {Lyra}, W., \& {Winter}, L.~M.
  2014, \mnras, 441, 900, \dodoi{10.1093/mnras/stu553}

\bibitem[{{McKernan} {et~al.}(2012){McKernan}, {Ford}, {Lyra}, \&
  {Perets}}]{McK12}
{McKernan}, B., {Ford}, K.~E.~S., {Lyra}, W., \& {Perets}, H.~B. 2012, \mnras,
  425, 460, \dodoi{10.1111/j.1365-2966.2012.21486.x}

\bibitem[{{McKernan} {et~al.}(2020){McKernan}, {Ford}, \&
  {O'Shaughnessy}}]{McK20R}
{McKernan}, B., {Ford}, K.~E.~S., \& {O'Shaughnessy}, R. 2020, \mnras, 498,
  4088, \dodoi{10.1093/mnras/staa2681}

\bibitem[{{Meyer} {et~al.}(2019){Meyer}, {Scargle}, \& {Blandford}}]{meyer2019}
{Meyer}, M., {Scargle}, J.~D., \& {Blandford}, R.~D. 2019, \apj, 877, 39,
  \dodoi{10.3847/1538-4357/ab1651}

\bibitem[{{Moreno} {et~al.}(2019){Moreno}, {Vogeley}, {Richards}, \&
  {Yu}}]{Moreno19}
{Moreno}, J., {Vogeley}, M.~S., {Richards}, G.~T., \& {Yu}, W. 2019, \pasp,
  131, 063001, \dodoi{10.1088/1538-3873/ab1597}

\bibitem[{{Mukherjee} {et~al.}(2020){Mukherjee}, {Ghosh}, {Graham},
  {Karathanasis}, {Kasliwal}, {Maga{\~n}a Hernandez}, {Nissanke}, {Silvestri},
  \& {Wandelt}}]{Mukherjee20}
{Mukherjee}, S., {Ghosh}, A., {Graham}, M.~J., {et~al.} 2020, arXiv e-prints,
  arXiv:2009.14199.
\newblock \doarXiv{2009.14199}

\bibitem[{{Oke} {et~al.}(1995){Oke}, {Cohen}, {Carr}, {Cromer}, {Dingizian},
  {Harris}, {Labrecque}, {Lucinio}, {Schaal}, {Epps}, \& {Miller}}]{Oke1995}
{Oke}, J.~B., {Cohen}, J.~G., {Carr}, M., {et~al.} 1995, \pasp, 107, 375,
  \dodoi{10.1086/133562}

\bibitem[{{Ostriker}(1999)}]{Ostriker99}
{Ostriker}, E.~C. 1999, \apj, 513, 252, \dodoi{10.1086/306858}

\bibitem[{{Palmese} {et~al.}(2021){Palmese}, {Fishbach}, {Burke}, {Annis}, \&
  {Liu}}]{Palmese21}
{Palmese}, A., {Fishbach}, M., {Burke}, C.~J., {Annis}, J., \& {Liu}, X. 2021,
  \apjl, 914, L34, \dodoi{10.3847/2041-8213/ac0883}

\bibitem[{{Pan} \& {Yang}(2021)}]{Pan21}
{Pan}, Z., \& {Yang}, H. 2021, \apj, 923, 173, \dodoi{10.3847/1538-4357/ac249c}

\bibitem[{{Patterson} {et~al.}(2019){Patterson}, {Bellm}, {Rusholme}, {Masci},
  {Juric}, {Krughoff}, {Golkhou}, {Graham}, {Kulkarni}, {Helou}, \& {Zwicky
  Transient Facility Collaboration}}]{Patterson19}
{Patterson}, M.~T., {Bellm}, E.~C., {Rusholme}, B., {et~al.} 2019, \pasp, 131,
  018001, \dodoi{10.1088/1538-3873/aae904}

\bibitem[{Pedregosa {et~al.}(2011)Pedregosa, Varoquaux, Gramfort, Michel,
  Thirion, Grisel, Blondel, Prettenhofer, Weiss, Dubourg, Vanderplas, Passos,
  Cournapeau, Brucher, Perrot, \& Duchesnay}]{scikit-learn}
Pedregosa, F., Varoquaux, G., Gramfort, A., {et~al.} 2011, Journal of Machine
  Learning Research, 12, 2825

\bibitem[{{Perley} {et~al.}(2020){Perley}, {Fremling}, {Sollerman}, {Miller},
  {Dahiwale}, {Sharma}, {Bellm}, {Biswas}, {Brink}, {Bruch}, {De}, {Dekany},
  {Drake}, {Duev}, {Filippenko}, {Gal-Yam}, {Goobar}, {Graham}, {Graham}, {Ho},
  {Irani}, {Kasliwal}, {Kim}, {Kulkarni}, {Mahabal}, {Masci}, {Modak}, {Neill},
  {Nordin}, {Riddle}, {Soumagnac}, {Strotjohann}, {Schulze}, {Taggart},
  {Tzanidakis}, {Walters}, \& {Yan}}]{Perley20}
{Perley}, D.~A., {Fremling}, C., {Sollerman}, J., {et~al.} 2020, \apj, 904, 35,
  \dodoi{10.3847/1538-4357/abbd98}

\bibitem[{{Perna} {et~al.}(2021){Perna}, {Lazzati}, \& {Cantiello}}]{Perna21}
{Perna}, R., {Lazzati}, D., \& {Cantiello}, M. 2021, \apjl, 906, L7,
  \dodoi{10.3847/2041-8213/abd319}

\bibitem[{{Petrov} {et~al.}(2022){Petrov}, {Singer}, {Coughlin}, {Kumar},
  {Almualla}, {Anand}, {Bulla}, {Dietrich}, {Foucart}, \&
  {Guessoum}}]{Petrov21}
{Petrov}, P., {Singer}, L.~P., {Coughlin}, M.~W., {et~al.} 2022, \apj, 924, 54,
  \dodoi{10.3847/1538-4357/ac366d}

\bibitem[{{Planck Collaboration} {et~al.}(2016){Planck Collaboration}, {Ade},
  {Aghanim}, {Arnaud}, {Ashdown}, {Aumont}, {Baccigalupi}, {Banday},
  {Barreiro}, {Bartlett}, {Bartolo}, {Battaner}, {Battye}, {Benabed},
  {Beno{\^\i}t}, {Benoit-L{\'e}vy}, {Bernard}, {Bersanelli}, {Bielewicz},
  {Bock}, {Bonaldi}, {Bonavera}, {Bond}, {Borrill}, {Bouchet}, {Boulanger},
  {Bucher}, {Burigana}, {Butler}, {Calabrese}, {Cardoso}, {Catalano},
  {Challinor}, {Chamballu}, {Chary}, {Chiang}, {Chluba}, {Christensen},
  {Church}, {Clements}, {Colombi}, {Colombo}, {Combet}, {Coulais}, {Crill},
  {Curto}, {Cuttaia}, {Danese}, {Davies}, {Davis}, {de Bernardis}, {de Rosa},
  {de Zotti}, {Delabrouille}, {D{\'e}sert}, {Di Valentino}, {Dickinson},
  {Diego}, {Dolag}, {Dole}, {Donzelli}, {Dor{\'e}}, {Douspis}, {Ducout},
  {Dunkley}, {Dupac}, {Efstathiou}, {Elsner}, {En{\ss}lin}, {Eriksen},
  {Farhang}, {Fergusson}, {Finelli}, {Forni}, {Frailis}, {Fraisse},
  {Franceschi}, {Frejsel}, {Galeotta}, {Galli}, {Ganga}, {Gauthier}, {Gerbino},
  {Ghosh}, {Giard}, {Giraud-H{\'e}raud}, {Giusarma}, {Gjerl{\o}w},
  {Gonz{\'a}lez-Nuevo}, {G{\'o}rski}, {Gratton}, {Gregorio}, {Gruppuso},
  {Gudmundsson}, {Hamann}, {Hansen}, {Hanson}, {Harrison}, {Helou},
  {Henrot-Versill{\'e}}, {Hern{\'a}ndez-Monteagudo}, {Herranz}, {Hildebrandt},
  {Hivon}, {Hobson}, {Holmes}, {Hornstrup}, {Hovest}, {Huang}, {Huffenberger},
  {Hurier}, {Jaffe}, {Jaffe}, {Jones}, {Juvela}, {Keih{\"a}nen}, {Keskitalo},
  {Kisner}, {Kneissl}, {Knoche}, {Knox}, {Kunz}, {Kurki-Suonio}, {Lagache},
  {L{\"a}hteenm{\"a}ki}, {Lamarre}, {Lasenby}, {Lattanzi}, {Lawrence}, {Leahy},
  {Leonardi}, {Lesgourgues}, {Levrier}, {Lewis}, {Liguori}, {Lilje},
  {Linden-V{\o}rnle}, {L{\'o}pez-Caniego}, {Lubin}, {Mac{\'\i}as-P{\'e}rez},
  {Maggio}, {Maino}, {Mandolesi}, {Mangilli}, {Marchini}, {Maris}, {Martin},
  {Martinelli}, {Mart{\'\i}nez-Gonz{\'a}lez}, {Masi}, {Matarrese}, {McGehee},
  {Meinhold}, {Melchiorri}, {Melin}, {Mendes}, {Mennella}, {Migliaccio},
  {Millea}, {Mitra}, {Miville-Desch{\^e}nes}, {Moneti}, {Montier}, {Morgante},
  {Mortlock}, {Moss}, {Munshi}, {Murphy}, {Naselsky}, {Nati}, {Natoli},
  {Netterfield}, {N{\o}rgaard-Nielsen}, {Noviello}, {Novikov}, {Novikov},
  {Oxborrow}, {Paci}, {Pagano}, {Pajot}, {Paladini}, {Paoletti}, {Partridge},
  {Pasian}, {Patanchon}, {Pearson}, {Perdereau}, {Perotto}, {Perrotta},
  {Pettorino}, {Piacentini}, {Piat}, {Pierpaoli}, {Pietrobon}, {Plaszczynski},
  {Pointecouteau}, {Polenta}, {Popa}, {Pratt}, {Pr{\'e}zeau}, {Prunet},
  {Puget}, {Rachen}, {Reach}, {Rebolo}, {Reinecke}, {Remazeilles}, {Renault},
  {Renzi}, {Ristorcelli}, {Rocha}, {Rosset}, {Rossetti}, {Roudier},
  {Rouill{\'e} d'Orfeuil}, {Rowan-Robinson}, {Rubi{\~n}o-Mart{\'\i}n},
  {Rusholme}, {Said}, {Salvatelli}, {Salvati}, {Sandri}, {Santos},
  {Savelainen}, {Savini}, {Scott}, {Seiffert}, {Serra}, {Shellard}, {Spencer},
  {Spinelli}, {Stolyarov}, {Stompor}, {Sudiwala}, {Sunyaev}, {Sutton},
  {Suur-Uski}, {Sygnet}, {Tauber}, {Terenzi}, {Toffolatti}, {Tomasi},
  {Tristram}, {Trombetti}, {Tucci}, {Tuovinen}, {T{\"u}rler}, {Umana},
  {Valenziano}, {Valiviita}, {Van Tent}, {Vielva}, {Villa}, {Wade}, {Wandelt},
  {Wehus}, {White}, {White}, {Wilkinson}, {Yvon}, {Zacchei}, \&
  {Zonca}}]{Planck15}
{Planck Collaboration}, {Ade}, P.~A.~R., {Aghanim}, N., {et~al.} 2016, \aap,
  594, A13, \dodoi{10.1051/0004-6361/201525830}

\bibitem[{{Rees}(1988)}]{rees88}
{Rees}, M.~J. 1988, \nat, 333, 523, \dodoi{10.1038/333523a0}

\bibitem[{{Ricci} {et~al.}(2020){Ricci}, {Kara}, {Loewenstein}, {Trakhtenbrot},
  {Arcavi}, {Remillard}, {Fabian}, {Gendreau}, {Arzoumanian}, {Li}, {Ho},
  {MacLeod}, {Cackett}, {Altamirano}, {Gandhi}, {Kosec}, {Pasham}, {Steiner},
  \& {Chan}}]{Ricci20}
{Ricci}, C., {Kara}, E., {Loewenstein}, M., {et~al.} 2020, \apjl, 898, L1,
  \dodoi{10.3847/2041-8213/ab91a1}

\bibitem[{{Rodriguez} {et~al.}(2016{\natexlab{a}}){Rodriguez}, {Chatterjee}, \&
  {Rasio}}]{Rodriquez16b}
{Rodriguez}, C.~L., {Chatterjee}, S., \& {Rasio}, F.~A. 2016{\natexlab{a}},
  \prd, 93, 084029, \dodoi{10.1103/PhysRevD.93.084029}

\bibitem[{{Rodriguez} {et~al.}(2016{\natexlab{b}}){Rodriguez}, {Haster},
  {Chatterjee}, {Kalogera}, \& {Rasio}}]{Rodriquez16a}
{Rodriguez}, C.~L., {Haster}, C.-J., {Chatterjee}, S., {Kalogera}, V., \&
  {Rasio}, F.~A. 2016{\natexlab{b}}, \apjl, 824, L8,
  \dodoi{10.3847/2041-8205/824/1/L8}

\bibitem[{{Ross} {et~al.}(2018){Ross}, {Ford}, {Graham}, {et~al.}}]{Ross18}
{Ross}, N.~P., {Ford}, K.~E.~S., {Graham}, M., {et~al.} 2018, \mnras, 480,
  4468, \dodoi{10.1093/mnras/sty2002}

\bibitem[{{Rosswog} {et~al.}(2009){Rosswog}, {Ramirez-Ruiz}, \&
  {Hix}}]{rosswog09}
{Rosswog}, S., {Ramirez-Ruiz}, E., \& {Hix}, W.~R. 2009, \apj, 695, 404,
  \dodoi{10.1088/0004-637X/695/1/404}

\bibitem[{{Ryu} {et~al.}(2020){Ryu}, {Krolik}, {Piran}, \& {Noble}}]{Taeho20}
{Ryu}, T., {Krolik}, J., {Piran}, T., \& {Noble}, S.~C. 2020, \apj, 904, 98,
  \dodoi{10.3847/1538-4357/abb3cf}

\bibitem[{{Samsing} {et~al.}(2022){Samsing}, {Bartos}, {D'Orazio}, {Haiman},
  {Kocsis}, {Leigh}, {Liu}, {Pessah}, \& {Tagawa}}]{Samsing22}
{Samsing}, J., {Bartos}, I., {D'Orazio}, D.~J., {et~al.} 2022, \nat, 603, 237,
  \dodoi{10.1038/s41586-021-04333-1}

\bibitem[{{Scargle} {et~al.}(2013){Scargle}, {Norris}, {Jackson}, \&
  {Chiang}}]{scargle13}
{Scargle}, J.~D., {Norris}, J.~P., {Jackson}, B., \& {Chiang}, J. 2013, \apj,
  764, 167, \dodoi{10.1088/0004-637X/764/2/167}

\bibitem[{{Secunda} {et~al.}(2019){Secunda}, {Bellovary}, {Mac Low},
  {et~al.}}]{Secunda19}
{Secunda}, A., {Bellovary}, J., {Mac Low}, M.-M., {et~al.} 2019, \apj, 878, 85,
  \dodoi{10.3847/1538-4357/ab20ca}

\bibitem[{{Secunda} {et~al.}(2020){Secunda}, {Bellovary}, {Mac Low}, {Ford},
  {McKernan}, {Leigh}, {Lyra}, {Sandor}, \& {Adorno}}]{Secunda20}
---. 2020, arXiv e-prints, arXiv:2004.11936.
\newblock \doarXiv{2004.11936}

\bibitem[{{Shen} \& {Liu}(2012)}]{Shen12}
{Shen}, Y., \& {Liu}, X. 2012, \apj, 753, 125,
  \dodoi{10.1088/0004-637X/753/2/125}

\bibitem[{Singer \& Price(2016)}]{Singer16}
Singer, L.~P., \& Price, L.~R. 2016, Phys. Rev. D, 93, 024013,
  \dodoi{10.1103/PhysRevD.93.024013}

\bibitem[{{Singer} \& {Price}(2016)}]{ligo.skymap}
{Singer}, L.~P., \& {Price}, L.~R. 2016, \prd, 93, 024013,
  \dodoi{10.1103/PhysRevD.93.024013}

\bibitem[{{Singer} {et~al.}(2016){Singer}, {Chen}, {Holz}, {Farr}, {Price},
  {Raymond}, {Cenko}, {Gehrels}, {Cannizzo}, {Kasliwal}, {Nissanke},
  {Coughlin}, {Farr}, {Urban}, {Vitale}, {Veitch}, {Graff}, {Berry},
  {Mohapatra}, \& {Mandel}}]{singer16b}
{Singer}, L.~P., {Chen}, H.-Y., {Holz}, D.~E., {et~al.} 2016, \apjl, 829, L15,
  \dodoi{10.3847/2041-8205/829/1/L15}

\bibitem[{{Sirko} \& {Goodman}(2003)}]{SirkoGoodman03}
{Sirko}, E., \& {Goodman}, J. 2003, \mnras, 341, 501,
  \dodoi{10.1046/j.1365-8711.2003.06431.x}

\bibitem[{{Stern} {et~al.}(2018){Stern}, {McKernan}, {Graham},
  {et~al.}}]{Stern18}
{Stern}, D., {McKernan}, B., {Graham}, M.~J., {et~al.} 2018, \apj, 864, 27,
  \dodoi{10.3847/1538-4357/aac726}

\bibitem[{{Stone} {et~al.}(2017){Stone}, {Metzger}, \& {Haiman}}]{Stone17}
{Stone}, N.~C., {Metzger}, B.~D., \& {Haiman}, Z. 2017, \mnras, 464, 946,
  \dodoi{10.1093/mnras/stw2260}

\bibitem[{{Tagawa} {et~al.}(2021){Tagawa}, {Haiman}, {Bartos}, {Kocsis}, \&
  {Omukai}}]{Tagawa21}
{Tagawa}, H., {Haiman}, Z., {Bartos}, I., {Kocsis}, B., \& {Omukai}, K. 2021,
  \mnras, 507, 3362, \dodoi{10.1093/mnras/stab2315}

\bibitem[{{Tagawa} {et~al.}(2019){Tagawa}, {Haiman}, \& {Kocsis}}]{Tagawa19}
{Tagawa}, H., {Haiman}, Z., \& {Kocsis}, B. 2019, arXiv e-prints,
  arXiv:1912.08218.
\newblock \doarXiv{1912.08218}

\bibitem[{{Thompson} {et~al.}(2005){Thompson}, {Quataert}, \&
  {Murray}}]{Thompson05}
{Thompson}, T.~A., {Quataert}, E., \& {Murray}, N. 2005, \apj, 630, 167,
  \dodoi{10.1086/431923}

\bibitem[{{Vajpeyi} {et~al.}(2022){Vajpeyi}, {Thrane}, {Smith}, {McKernan}, \&
  {Saavik Ford}}]{Vajpeyi22}
{Vajpeyi}, A., {Thrane}, E., {Smith}, R., {McKernan}, B., \& {Saavik Ford},
  K.~E. 2022, \apj, 931, 82, \dodoi{10.3847/1538-4357/ac6180}

\bibitem[{{van Velzen} {et~al.}(2021){van Velzen}, {Gezari}, {Hammerstein},
  {Roth}, {Frederick}, {Ward}, {Hung}, {Cenko}, {Stein}, {Perley}, {Taggart},
  {Foley}, {Sollerman}, {Blagorodnova}, {Andreoni}, {Bellm}, {Brinnel}, {De},
  {Dekany}, {Feeney}, {Fremling}, {Giomi}, {Golkhou}, {Graham}, {Ho},
  {Kasliwal}, {Kilpatrick}, {Kulkarni}, {Kupfer}, {Laher}, {Mahabal}, {Masci},
  {Miller}, {Nordin}, {Riddle}, {Rusholme}, {van Santen}, {Sharma}, {Shupe}, \&
  {Soumagnac}}]{vanVelzen21}
{van Velzen}, S., {Gezari}, S., {Hammerstein}, E., {et~al.} 2021, \apj, 908, 4,
  \dodoi{10.3847/1538-4357/abc258}

\bibitem[{{Varma} {et~al.}(2022){Varma}, {Biscoveanu}, {Islam}, {Shaik},
  {Haster}, {Isi}, {Farr}, {Field}, \& {Vitale}}]{Varma22}
{Varma}, V., {Biscoveanu}, S., {Islam}, T., {et~al.} 2022, \prl, 128, 191102,
  \dodoi{10.1103/PhysRevLett.128.191102}

\bibitem[{{Veitch} {et~al.}(2015){Veitch}, {Raymond}, {Farr}, {Farr}, {Graff},
  {Vitale}, {Aylott}, {Blackburn}, {Christensen}, {Coughlin}, {Del Pozzo},
  {Feroz}, {Gair}, {Haster}, {Kalogera}, {Littenberg}, {Mandel},
  {O'Shaughnessy}, {Pitkin}, {Rodriguez}, {R{\"o}ver}, {Sidery}, {Smith}, {Van
  Der Sluys}, {Vecchio}, {Vousden}, \& {Wade}}]{Veitch15}
{Veitch}, J., {Raymond}, V., {Farr}, B., {et~al.} 2015, \prd, 91, 042003,
  \dodoi{10.1103/PhysRevD.91.042003}

\bibitem[{{Veronesi} {et~al.}(2022){Veronesi}, {Rossi}, {van Velzen}, \&
  {Buscicchio}}]{Veronesi22}
{Veronesi}, N., {Rossi}, E.~M., {van Velzen}, S., \& {Buscicchio}, R. 2022,
  \mnras, \dodoi{10.1093/mnras/stac1346}

\bibitem[{{Villar} {et~al.}(2020){Villar}, {Cranmer}, {Contardo}, {Ho}, \&
  {Yao-Yu Lin}}]{Villar20}
{Villar}, V.~A., {Cranmer}, M., {Contardo}, G., {Ho}, S., \& {Yao-Yu Lin}, J.
  2020, arXiv e-prints, arXiv:2010.11194.
\newblock \doarXiv{2010.11194}

\bibitem[{{Wang} {et~al.}(2021{\natexlab{a}}){Wang}, {Liu}, {Ho}, {Li}, \&
  {Du}}]{WangJ21}
{Wang}, J.-M., {Liu}, J.-R., {Ho}, L.~C., {Li}, Y.-R., \& {Du}, P.
  2021{\natexlab{a}}, \apjl, 916, L17, \dodoi{10.3847/2041-8213/ac0b46}

\bibitem[{{Wang} {et~al.}(2021{\natexlab{b}}){Wang}, {McKernan}, {Ford},
  {Perna}, {Leigh}, \& {Low}}]{Yihan21}
{Wang}, Y.-H., {McKernan}, B., {Ford}, S., {et~al.} 2021{\natexlab{b}}, \apjl,
  923, L23, \dodoi{10.3847/2041-8213/ac400a}

\bibitem[{{Woosley}(2017)}]{Woosley17}
{Woosley}, S.~E. 2017, \apj, 836, 244, \dodoi{10.3847/1538-4357/836/2/244}

\bibitem[{{Yang} {et~al.}(2021){Yang}, {Bartos}, {Fragione}, {Haiman},
  {Kowalski}, {Marka}, {Perna}, \& {Tagawa}}]{Yang21}
{Yang}, Y., {Bartos}, I., {Fragione}, G., {et~al.} 2021, arXiv e-prints,
  arXiv:2105.02342.
\newblock \doarXiv{2105.02342}

\bibitem[{{Yang} {et~al.}(2019){Yang}, {Bartos}, {Gayathri}, {et~al.}}]{Yang19}
{Yang}, Y., {Bartos}, I., {Gayathri}, V., {et~al.} 2019, \prl, 123, 181101,
  \dodoi{10.1103/PhysRevLett.123.181101}

\end{thebibliography}
\bibliographystyle{aasjournal}



\end{document}